\documentclass[fleqn,usenatbib]{mnras}
\usepackage{newtxtext,newtxmath}
\usepackage{siunitx}
\usepackage{rotating}
\usepackage{lscape}
\usepackage[T1]{fontenc}
\DeclareRobustCommand{\VAN}[3]{#2}
\let\VANthebibliography\thebibliography
\def\thebibliography{\DeclareRobustCommand{\VAN}[3]{##3}\VANthebibliography}

\usepackage{graphicx}	
\usepackage{amsmath}	
\usepackage[dvipsnames]{xcolor} 
\usepackage{enumitem} 
\usepackage{caption}
\usepackage{subcaption}
\usepackage{gensymb}
\usepackage{adjustbox}

\title[AGN feedback on molecular gas in quasars]{Quasar Feedback Survey: molecular gas affected by central outflows and by $\sim$10\,kpc radio lobes reveal dual feedback effects in ‘radio quiet’ quasars}
\author[A.\,Girdhar, C.\,M.\,Harrison, V.\,Mainieri, et al.]{A.\,Girdhar,$^{1,2,3}$\thanks{E-mail: astroaishgirdhar@gmail.com}
C.\,M.\,Harrison,$^{1}$\thanks{E-mail: christopher.harrison@newcastle.ac.uk}
V.\,Mainieri,$^{3}$
R.\,Fern{\'a}ndez\,Aranda,$^{4,5,3}$ 
D.\,M.\,Alexander,$^{6}$ \newauthor
F.\,Arrigoni\,Battaia,$^{7}$ 
M.\,Bianchin,$^{8,9}$
G.\,Calistro\,Rivera,$^{3}$ 
C.\,Circosta,$^{10,11}$ 
T.\,Costa,$^{7,1}$ 
A.\,C.\,Edge,$^{6}$ \newauthor
E.\,P.\,Farina,$^{12}$ 
D.\,Kakkad,$^{13}$ 
P.\,Kharb,$^{14}$ 
S.\,J.\,Molyneux,$^{15,3}$ 
D.\,Mukherjee,$^{16}$ 
A.\,Njeri,$^{1}$ 
Silpa S.,$^{14,17}$ \newauthor
G.\,Venturi,$^{18,19}$ 
and S.\,R.\,Ward$^{2,3,1}$ 
\\
\\
$^{1}$School of Mathematics, Statistics and Physics, Newcastle University, NE1 7RU, UK\\
$^{2}$Ludwig-Maximilians-Universit{\"a}t, Professor-Huber-Platz 2, D-80539 M{\"u}nchen, Germany\\
$^{3}$European Southern Observatory, Karl--Schwarzschild--Stra{\ss}e 2, D-85748 Garching bei M{\"u}nchen, Germany\\
$^{4}$Department of Physics, University of Crete, 70013, Heraklion, Greece \\
$^{5}$Institute of Astrophysics, Foundation for Research and Technology, Hellas, Voutes, 70013 Heraklion, Greece \\
$^{6}$Centre for Extragalactic Astronomy, Department of Physics, Durham University, South Road, Durham DH1 3LE, UK\\
$^{7}$Max Planck Institut f\"ur Astrophysik, Karl--Schwarzschild--Stra{\ss}e 1, D-85748, Garching bei M\"unchen, Germany \\
$^{8}$Departamento de F{\'i}sica, CCNE, Universidade Federal de Santa Maria, 97105-900 Santa Maria, RS, Brazil \\
$^{9}$Department of Physics and Astronomy, University of California, 4129 Frederick Reines Hall, Irvine, CA 92697, USA \\
$^{10}$European Space Agency (ESA), European Space Astronomy Centre (ESAC), Camino bajo del Castillo s/n, 28692 Villanueva de la Cañada, Madrid, Spain \\
$^{11}$Department of Physics \& Astronomy, University College London, Gower Street, London, WC1E 6BT, UK\\
$^{12}$Gemini Observatory, NSF’s NOIRLab, 670 N A’ohoku Place, Hilo, Hawai'i 96720, USA \\
$^{13}$Space Telescope Science Institute, 3700 San Martin Drive, Baltimore, MD 21218, USA\\
$^{14}$National Centre for Radio Astrophysics - Tata Institute of Fundamental Research, Pune University Campus, Post Bag 3, Ganeshkhind, Pune 411007, India\\ 
$^{15}$Astrophysics Research Institute, Liverpool John Moores University, 146 Brownlow Hill, Liverpool L3 5RF, UK\\ 
$^{16}$Inter-University Centre for Astronomy and Astrophysics, Post Bag 4, Pune - 411007, India\\
$^{17}$Departamento de Astronomía, Universidad de Concepción, Concepción, Chile \\
$^{18}$Scuola Normale Superiore, Piazza dei Cavalieri 7, I-56126 Pisa, Italy \\
$^{19}$INAF - Osservatorio Astrofisico di Arcetri, Largo E. Fermi 5, I-50125 Firenze, Italy
}

\date{Accepted XXX. Received YYY; in original form ZZZ}

\pubyear{2023}

\begin{document}
\label{firstpage}
\pagerange{\pageref{firstpage}--\pageref{lastpage}}
\maketitle

 
\begin{abstract}
We present a study of molecular gas, traced via CO\,(3--2) from ALMA data, of four z$<$0.2, `radio quiet', type 2 quasars (L$_{\mathrm{bol}}\sim10^{45.3-46.2}$\,erg\,s$^{-1}$; L$_{\mathrm{1.4\,GHz}}\sim10^{23.7-24.3}$\,W\,Hz$^{-1}$). Targets were selected to have extended radio lobes ($\geq$\,10\,kpc), and compact, moderate-power jets (1--10\,kpc; P$_{\mathrm{jet}}\sim10^{43.2-43.7}$\,erg\,s$^{-1}$). All targets show evidence of central molecular outflows, or injected turbulence, within the gas disks (traced via high-velocity wing components in CO emission-line profiles). The inferred velocities (V$_{\mathrm{out}}$=250\,--\,440\,km\,s$^{-1}$) and spatial scales (0.6\,--\,1.6\,kpc), are consistent with those of other samples of luminous low-redshift AGN. In two targets, we observe extended molecular gas structures beyond the central disks, containing 9\,--\,53\,\% of the total molecular gas mass. These structures tend to be elongated, extending from the core, and wrap-around (or along) the radio lobes. Their properties are similar to the molecular gas filaments observed around radio lobes of, mostly `radio loud', Brightest Cluster Galaxies. They have: projected distances of 5\,--\,13\,kpc; bulk velocities of 100\,--340\,km\,s$^{-1}$; velocity dispersion of 30\,--\,130\,km\,s$^{-1}$; inferred mass outflow rates of 4\,--\,20\,M$_{\odot}$\,yr$^{-1}$; and estimated kinetic powers of 10$^{40.3-41.7}$\,erg\,s$^{-1}$. Our observations are consistent with simulations that suggest moderate-power jets can have a direct (but modest) impact on molecular gas on small scales, through direct jet-cloud interactions. Then, on larger scales, jet-cocoons can push gas aside. Both processes could contribute to the long-term regulation of star formation.
\end{abstract}




\section{Introduction}

Active Galactic Nuclei (AGN) are observed as sites of growing black holes (\citealt{kormendyHo13}) and are capable of converting the energy from accreted material into intense episodes of emitted energy in the form of radiation, accretion disk winds, and jets of relativistic particles. This energy can be extremely high, also exceeding the binding energy of the galaxy itself (\citealt{cattaneo09,bower12}) and is theoretically capable of affecting the host galaxy through regulation of star-formation (\citealt{mcnamara12, fabian12}). Depending on how the available energy couples to the interstellar medium (ISM), the gas could be driven due to wide-angled accretion disk winds, radiation pressure on dust, and/or due to the acceleration by radio jets \citep[e.g.,][]{sijacki07, fabian12, kingPounds15, ishibashi16, mukherjee16, costa18, costa20, tannerWeaver22,almeida23}. These processes can also influence the fuel available for feeding the black hole itself, thereby giving this process a self-limiting nature, and thus earning the name `AGN feedback'.
 
Direct evidence of the influence of AGN on the ISM comes from observations that have confirmed the presence of galactic-scale AGN outflows over different phases, including ionized, neutral, and molecular forms (e.g., \citealt{morganti05, nesvadba08, feruglio10, alexander10, harrison12, rupke13, liu13, cicone14, villarMartin14,  kingPounds15, fiore17, rupke17, cicone18, chris2018, forsterSchreiber19,davies20, roy21, venturi21, ramosAlmeida22, girdhar22, kakkad22, kakkad23}). While each of these phases are crucial in forming a complete understanding of galaxy evolution, comprehending the impact of AGN on molecular gas is particularly popular because (i) molecular gas is the main reservoir for fuelling star-formation and the growth of supermassive black holes; (ii) most of the mass in galactic outflows is seen to reside in the molecular gas phase (e.g., \citealt{fiore17} compiled literature measurements and found that for AGN with L$_{\mathrm{bol}}\sim\,10^{45-46}$\,erg\,s$^{-1}$, the observed molecular outflows typically have 100\,$\times$ more mass than the ionised outflows). It is hence important to understand the effect of powerful quasars on the molecular gas in their host galaxy \citep[e.g.,][]{feruglio10, mainieri11, alatalo11, cicone14, morganti15, harrison17, fiore17, mainieri21, ward22, oliveira23}. 

For the brightest AGN, with high accretion rates, the dominant feedback mechanism is typically expected to be due to accretion disk winds (which can propagate into the host galaxies) or directly due to radiation pressure. This can lead to the disturbance or removal of inter-stellar gas \citep[e.g.,][]{costa18,costa20}. Many of the observational studies focusing on high accretion rate AGN have looked at starburst and highly luminous quasar targets. Specifically, there is a class of observational work searching for underlying wing components\footnote{The emission-line wing components refer to the presence of any high-velocity wing components, that mark a deviation from a single Gaussian fit to the emission line. Observationally, this high-velocity wing is often attributed to non-gravitational motions and is used to identify gas outflows or turbulence.} in CO emission-line profiles, as a tracer of molecular gas outflows, and then investigating these outflow properties as a function of star formation rates, stellar masses, and AGN luminosities (\citealt{cicone14, fiore17, fluetsch19}). Another class of observational studies have focused on massive, radio-luminous Brightest Cluster Galaxies (BCGs), located in cool-core clusters, revealing molecular gas entrained in filamentary structures along with the large radio lobes and X-ray cavities in the systems (\citealt{salomeCombes04, david14, mcnamara14, tremblay16, vantyghem16, russell17, russell18, tremblay18, russell19, olivares19, tamhane22}). Therefore, these classes of studies appear to investigate different types of feedback effects on the molecular gas, with the former assuming a dominant role of AGN winds/radiation (at least for driving the most powerful molecular outflows) and the latter finding a dominant role of radio jets. 

One might conclude a simple overall picture of two feedback modes on the molecular ISM; one acting on larger scales, beyond the gas disk, and caused by powerful radio jets (e.g., in the BCGs) and one acting within the molecular gas disks, due to the radiative output of high accretion rate AGN. However, these different feedback mechanisms, acting on two scales are typically not investigated within the same objects. For example, potential radio-jet-related processes, are often assumed to be sub-dominant in radiatively luminous AGN  with low to moderate radio powers (such as `radio quiet' quasars). However, recent observational studies have come to highlight the importance of low- and moderate-power radio jets (P$_{\mathrm{jet}}\leq\,10^{45}$\,erg\,s$^{-1}$) in galaxies, which are traditionally classified as `radio quiet' (because their radiative output dominates over that from jets). Low- and moderate-power jets in these systems have been observed to be driving turbulence, outflows, and excitation of the molecular gas (e.g., \citealt{morganti15, rosario19, girdhar22, audibert23, morganti23}), which have an impact which is, at least qualitatively, expected from the jets as seen in simulations (\citealt{mukherjee16,  meenakshi22, tannerWeaver22, morganti23}). This all motivates an observational study to search for the impact on molecular gas, on multiple spatial scales, in systems that contain both radio jets and high luminosity AGN. With this goal in mind, we make use of the spatially resolved, multi-wavelength data from the Quasar Feedback Survey \citep[QFeedS;][]{jarvis21}. \\

QFeedS\footnote{\url{https://blogs.ncl.ac.uk/quasarfeedbacksurvey/}}, includes 42 quasars (L$_{\rm bol}\gtrsim10^{45}$\,erg\,s$^{-1}$) that were selected from the parent population of AGN at z$\leq$0.2 by \cite{jarvis21}. These luminosities are representative of the peak of the luminosity function, L$_*$, at the peak of the cosmic epoch of growth when quasar feedback is also expected to dominate (i.e., $z\sim1$). However, the low redshift provides the advantage to obtain spatially-resolved, sensitive observations of such powerful quasars. While the galaxies studied in this sample are seen to be gas-rich and star-forming, it is a caveat that the conditions of the interstellar medium (ISM) may be different for the host-galaxies of quasars at z\,$\sim$\,1, and are not complete analogs of high redshift AGN. The QFeedS dataset is being used to extract information on the origin of radio emission in `radio quiet' quasars; multi-phase outflows; and the impact of AGN on the host galaxies (\citealt{harrison15,lansbury18,jarvis19,jarvis20,jarvis21,girdhar22,silpa22, molyneux23}). 

One benefit of QFeedS, for exploring different feedback mechanisms, is the availability of sensitive and high-resolution radio imaging provided by the Karl G. Jansky Very Large Array (VLA) \citep[][]{jarvis21}. In this work, we explore the feedback on the molecular gas of these quasar-host galaxies by comparing the radio emission with respect to the spatial distribution and kinematics of the molecular gas, traced with CO (3--2) transition, with data from the Atacama Large Millimeter/submillimeter Array (ALMA). The main focus is to compare with the prior feedback studies performed in BCGs, which look for molecular structures associated with radio lobes (\citealt{russell19, tamhane22}); and to also simultaneously search for the presence of CO emission-line wings (as a tracer of molecular outflows), as performed for a compilation of z$<$\,0.2 AGN and starburst galaxies by \cite{fluetsch19}.

This paper is structured as follows. In Section \ref{sec: data}, we discuss the sample selection and the different observations and their reduction used for this analysis. In Section \ref{sec: methods}, we describe the approach for the emission-line fits to extract the molecular gas kinematics, followed by the stellar kinematics (using data obtained on the Very Large Telescope's Multi Unit Spectroscopic Explorer; VLT/MUSE) and the methods used to extract morphological and kinematic properties of the molecular gas. In Section \ref{sec: results}, we present the results and a discussion of these results, in the context of previous observations and simulation studies. Finally, in Section \ref{sec: conclusions}, we present our conclusions. 

We have adopted the cosmological parameters to be $H_0$ = 70 km\,s$^{-1}$ Mpc$^{-1}$, $\Omega_M$ = 0.3 and $\Omega_{\Lambda}$ = 0.7, throughout. In this cosmology, 1\,arcsec corresponds to 2.47\,kpc for the redshift of $z=$\,0.14 (i.e., the average redshift of the galaxies studied here).  When referred to, we define the radio spectral index, $\alpha$, using the relation, $S_{\nu}\propto \nu^{\alpha}$, where $S_{\nu}$ refers to the flux density at the corresponding frequency $\nu$.  \\


\begin{table*}
\caption{Study targets and their properties: (1) quasar name; (2) redshift; (3) and (4) optical Right Ascension and Declination positions from SDSS (DR7:\,\citealt{sdss_dr7}) in the format hh:mm:ss.ss and dd:mm:ss.s, respectively. Values in (5)-(7) are as follows (from \citealt{jarvis19}): (5) bolometric AGN luminosity; (6) [O\,\textsc{iii}] luminosity; (7) 1.4\,GHz radio luminosity; and (8) largest linear size measured for the radio structures (LLS$_{\rm radio}$; \citealt{jarvis21}).}
\label{tab1: litprops}
\begin{tabular}{l c c c c c c c }
\hline
Quasar & z & RA & Dec & log(L$_{\mathrm{bol}}$) & log(L$_{[\rm O~III]}$) & log(L$_{\rm 1.4GHz}$) & LLS$_{\rm radio}$ \\
 &  (SDSS) & (J2000) & (J2000) & [erg s$^{-1}$] & [erg s$^{-1}$] & [W Hz$^{-1}$]  & [kpc] \\
(1) & (2) & (3) & (4) & (5) & (6) & (7) & (8) \\
\hline
J0945+1737 & 0.128 & 09:45:21.30 & +17:37:53.2 & 45.70 & 42.66 & 24.3 & 11 \\
J1000+1242 &0.148 & 10:00:13.14 & +12:42:26.2 & 45.30 & 42.61 & 24.2 & 21  \\
J1010+1413 & 0.199 & 10:10:22.95 & +14:13:00.9 & 46.20 & 43.13 & 24.0 & 10 \\
J1430+1339 & 0.085 & 14:30:29.88 & +13:39:12.0 & 45.50 & 42.61 & 23.7 & 14 \\
\hline
\end{tabular}
\end{table*}

\section{Targets, Observations and Ancillary Data}\label{sec: data}

We select our targets for this work from QFeedS (\citealt{jarvis21}), a survey of 42 sources that were originally selected from the parent population of emission-line AGN at z$\leq$0.2 (\citealt{mullaney2013}), with quasar-like [O\,\textsc{iii}]$\,\lambda$\,5007\,\AA\, luminosities (L$_{\rm [O~III]}>10^{42.1}$\,erg\,s$^{-1}$). A moderate radio luminosity criteria of L$_{\rm 1.4GHz} > 10^{23.45}$\,W Hz$^{-1}$ is also applied to obtain the QFeedS sample; however, the sample still consists of 88\,\% `radio quiet' sources, based on the criteria of \cite{xu99} (see \citealt{jarvis21} for full details). This is consistent with the ‘radio quiet’ fraction of the overall quasar population (i.e. $\sim$90\,$\%$; \citealt{zakamska04}). 

\begin{figure}
	\includegraphics[width=\columnwidth]{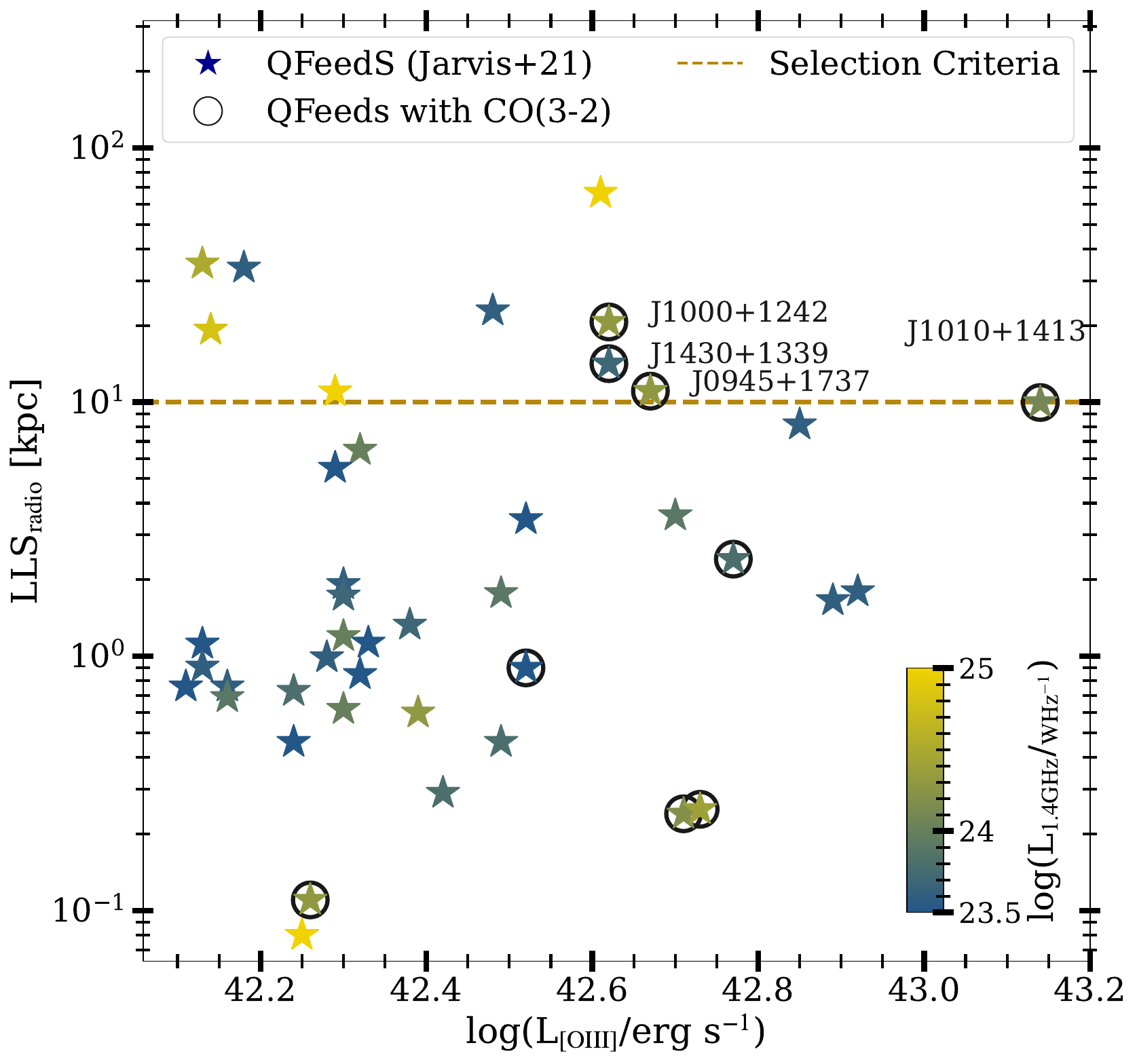}
    \caption{Largest linear size of radio structures (LLS$_{\rm radio}$) versus [O\,\textsc{iii}] luminosity for the parent sample of 42 QFeedS targets, colour-coded by their 1.4\,GHz radio luminosity (\citealt{jarvis21}). The 9/42 QFeedS targets with the required CO(3--2) ALMA data are highlighted with a black circle. A selection criteria of LLS$_{\rm radio}\geq$\,10\,kpc was then applied (dashed line) to select the four targets for this work (labeled with their names, see Section \ref{sec: samplSelect}).}
    \label{fig1: paramplot}
\end{figure}

Figure~\ref{fig1: paramplot} shows the [O\,\textsc{iii}] luminosities and the projected largest linear radio sizes, LLS$_{\rm radio}$, of the 42 quasars from QFeedS. The LLS$_{\rm radio}$ measurements were calculated in \cite{jarvis21} from a set of 1.5--6\,GHz VLA images, with a resolution ranging from 0.3--1\,arcsec. LLS$_{\rm radio}$ is defined as the distance between the farthest radio emission peaks in the lowest resolution image where the source shows radio structures. If the source shows no morphological features beyond the core in any image,  LLS$_{\rm radio}$ is defined as the beam de-convolved size of the core. The sample exhibits a wide range of radio sizes ($\sim$0.1\,kpc to $\sim$60\,kpc). Section~\ref{sec: samplSelect} gives an overview of the sources selected from QFeedS for this study, followed by a description of the data used from ALMA, VLA, and VLT/MUSE (Section~\ref{sec: datared_alma}\,--\,\ref{sec: stelKin}).

\subsection{Sample Selection}\label{sec: samplSelect}
 
\begin{figure*}
\centering
\begin{adjustbox}{addcode={\begin{minipage}{\width}}{\caption{The {\em \textbf{Panels a1, b1, c1, d1:}} show CO (3--2) flux (moment\,0) maps for the four targets selected following Section \ref{sec: samplSelect}. The CO(3--2) images are created by integrating over the velocity ranges annotated in the top-right of each panel. The green contours show CO emission corresponding to levels indicated in the individual legends, and the dashed green contours show the $-$3\,$\sigma$ level of CO emission. The red and the white contours show the radio emission from the high- and low-resolution 6\,GHz images, respectively. The contour levels are mentioned in the individual panel legends and are chosen to highlight the important structures in each target. Zoom-ins of the central emission are shown at the top for J1430+1339 and J0945+1737 (where the e-MERLIN image at 1.5\,GHz, is used and shown through blue contours). A 5\,kpc scale bar is shown in each panel at the bottom-right and the ALMA beams are shown at the bottom-left of the panel. The dashed-grey boxes show the region over which the galaxy-integrated spectra are extracted and are shown in the {\em \textbf{Panels a2, b2, c2, d2:}} with the properties listed in Table \ref{tab2: galprops}. The dashed purple boxes show the central outflow regions with the spectra shown in {\em \textbf{Panels a3, b3, c3, d3:}} and properties listed in Table \ref{tab5: nucoutprops}. The purple curves shows the combined fits while the blue, and orange dashed curves represent the individual Gaussian components. In the case of central spectra, we use the orange curves as the ``broad'' Gaussian component.}
\label{fig2: all4targets}
\end{minipage}},rotate=90,center}
\begin{tabular}{cc}
{\includegraphics[width=0.65\textwidth]{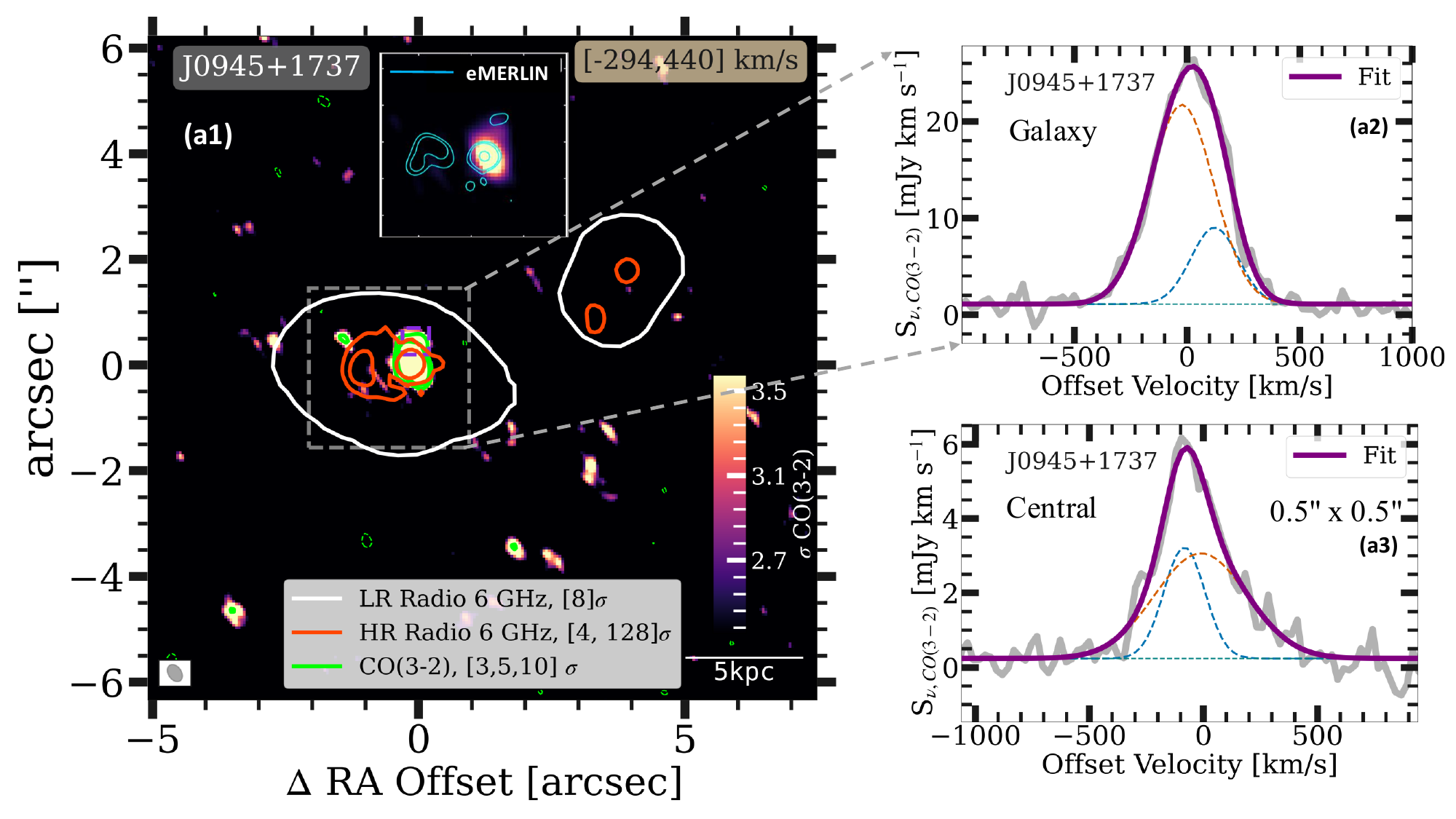}} &
{\includegraphics[width=0.65\textwidth]{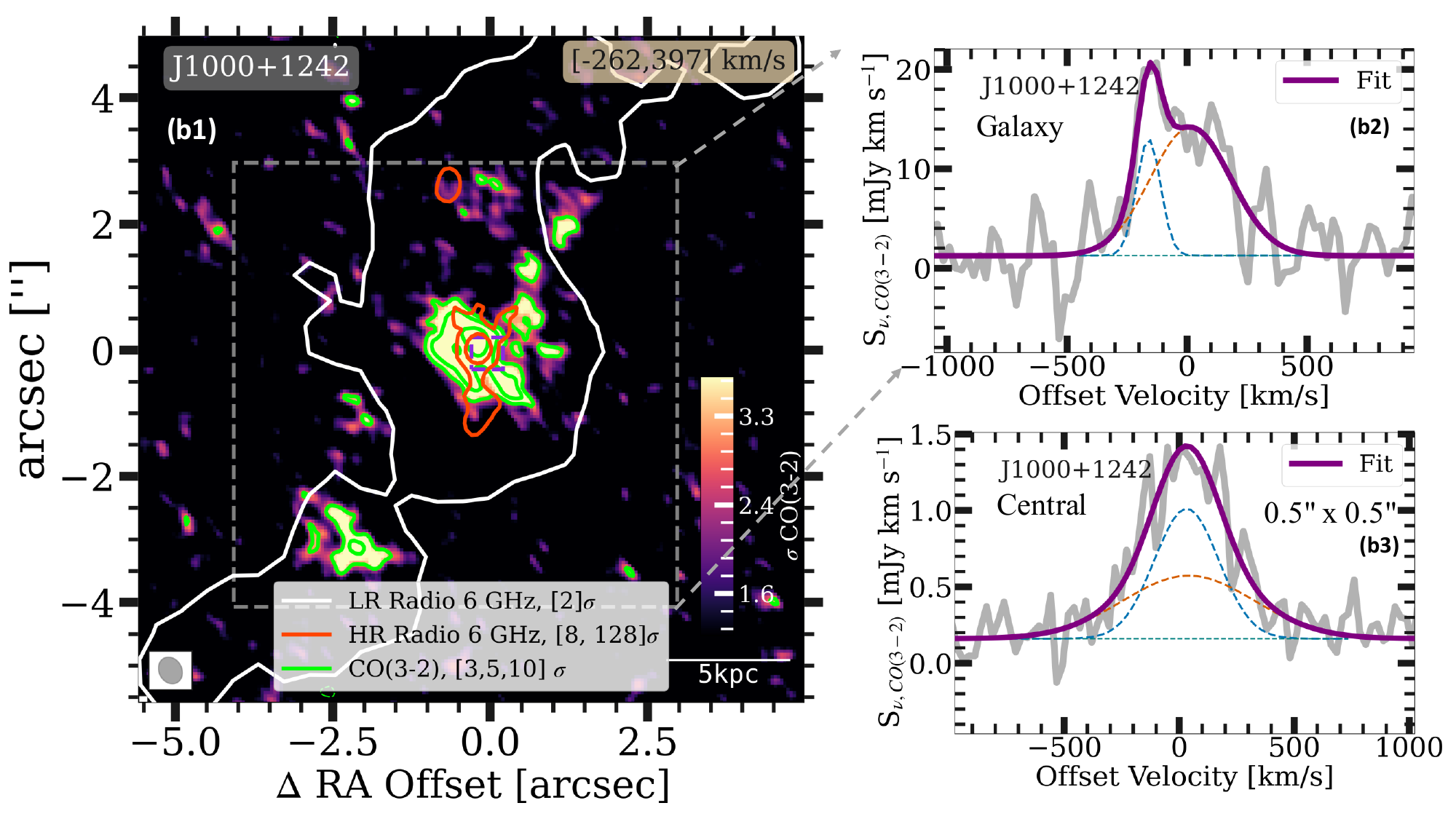}} \\
{\includegraphics[width=0.65\textwidth]{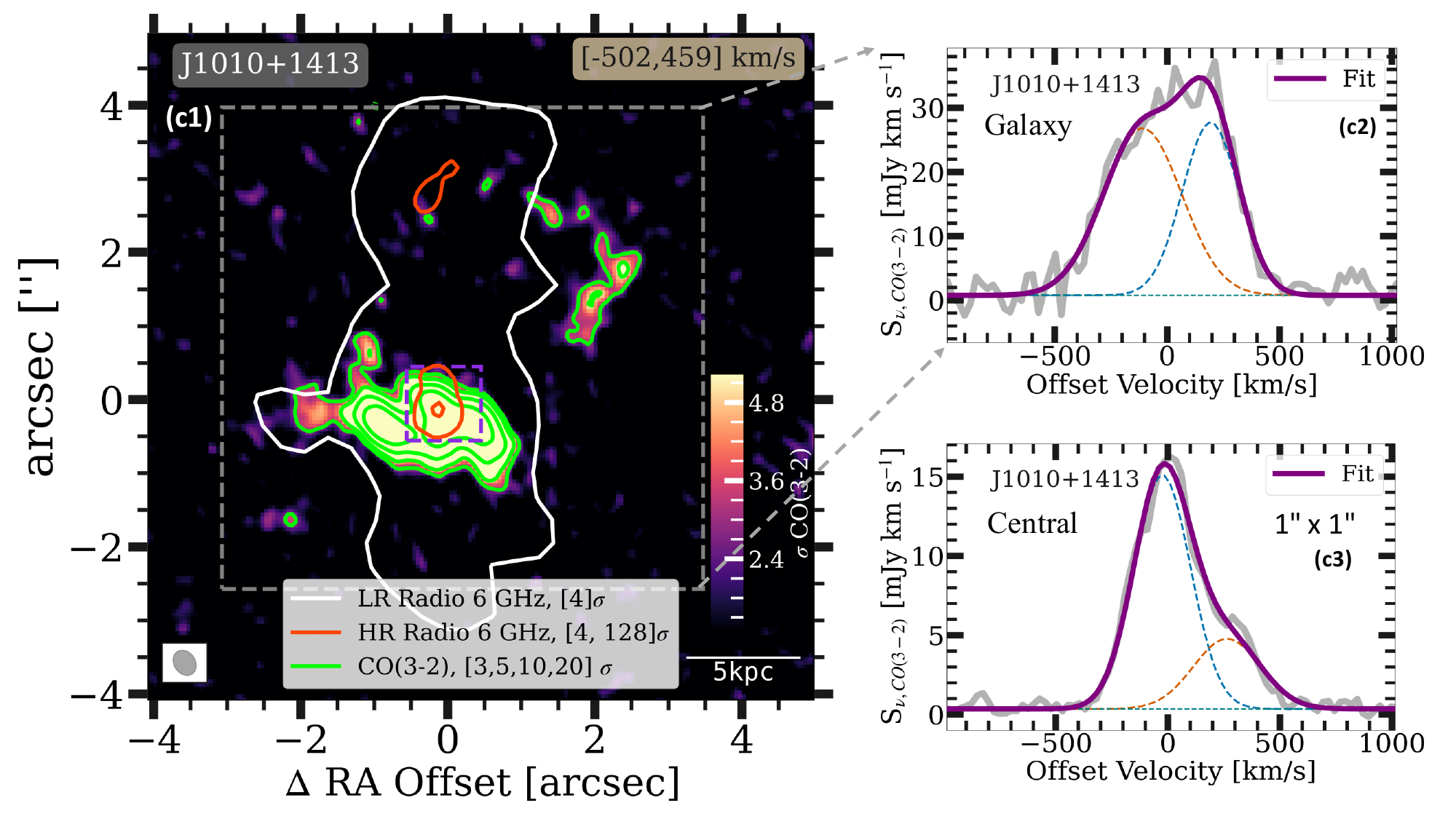}} &
{\includegraphics[width=0.65\textwidth]{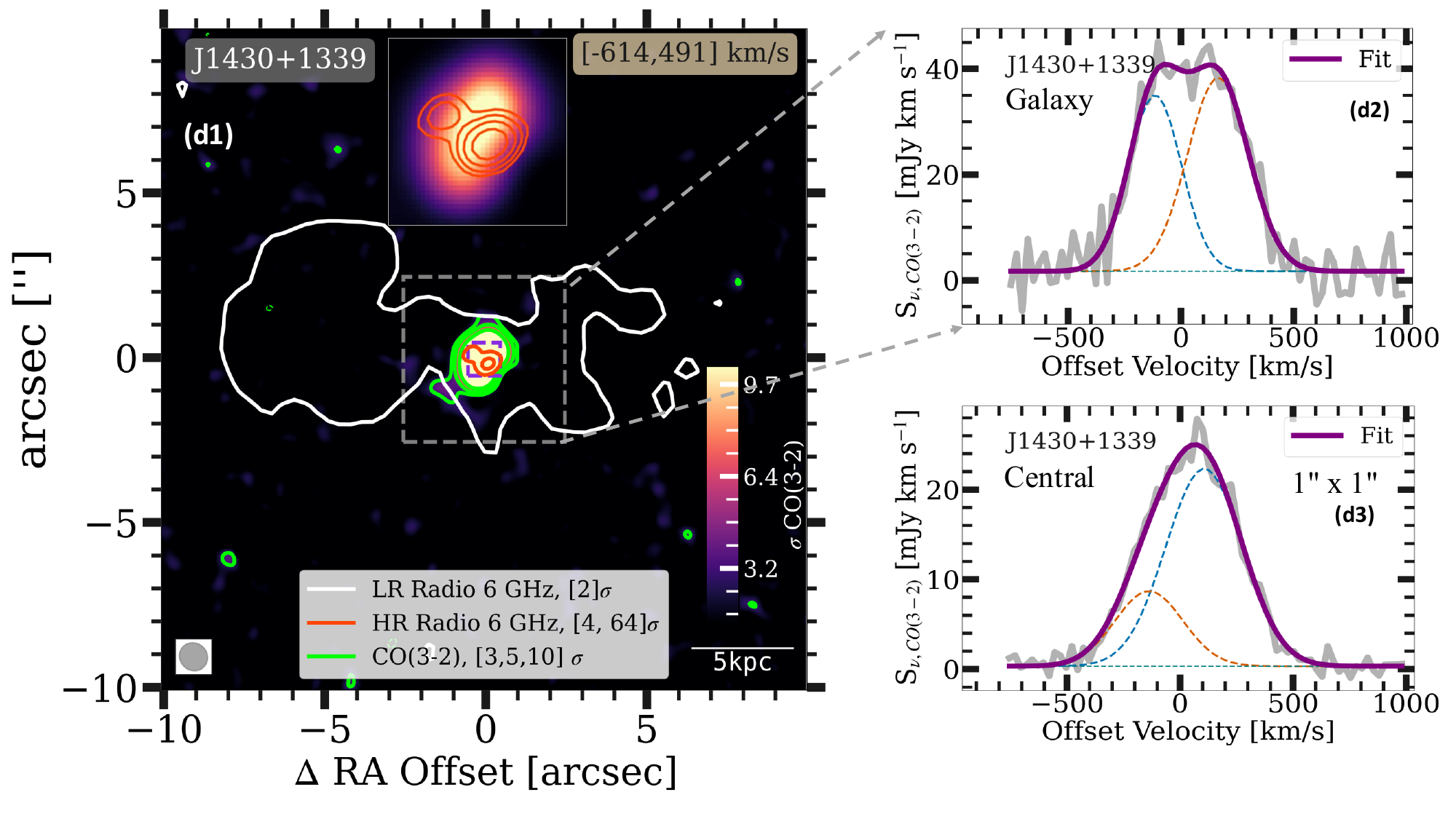}} \\
\end{tabular}
\end{adjustbox}
\end{figure*}

Following our goal to analyze the molecular gas, we identified 9/42 targets from the QFeedS sample which have available CO 12-m array ALMA data (highlighted by the empty black circles in Figure \ref{fig1: paramplot}). For all 9 of these, CO(3--2) data is available in ALMA Band\,7 observations; therefore, we decided to use this as our tracer of molecular gas for this work. We note 2/9 also have CO(2--1) data and 1/9 has CO(1--0) data  which are presented in \cite{ramosAlmeida22,audibert23} and \cite{sun14}. To clearly separate CO emission related to galaxy disks from any extended emission outside the disks, which could be associated with extended radio lobes\footnote{For this work, we refer to radio lobes as the diffuse, and loosely collimated, radio emission that is seen to extend beyond the galaxy disk.}, we further only selected the sources with LLS$_{\rm radio}\geq$10\,kpc (from VLA data, see Section \ref{sec: datared_rad}). As shown in Figure \ref{fig1: paramplot}, 4/9 sources with ALMA data meet this criteria; namely, J0945+1439, J1000+1242, J1010+1413, and J1430+1339. Table \ref{tab1: litprops} lists the basic properties of these four targets. The CO maps and radio images are shown for these four targets in Figure \ref{fig2: all4targets} (center-panel; see Section~\ref{sec: datared_alma} and Section~\ref{sec: datared_rad} for details). 

All four targets are bright AGN with high, quasar-like bolometric luminosities of $\log$\,(L$_{\rm bol}$/erg s$^{-1}$)\,=\,45.3\,--\,46.2 (from the fitting of the spectral energy distributions; \citealt{jarvis19}). These four targets are all classified as `radio quiet' based on the L$_{\rm 1.4GHz}$ versus L$_{\rm [O III]}$ criteria of \cite{xu99}. However, despite their modest radio luminosities of $\log(\mathrm{L}_{\rm 1.4GHz}/$W Hz$^{-1}$)\,=\,23.7\,--\,24.3, all four of these targets have been confirmed to have an excess of radio emission over that expected from star-formation from their radio imaging (see \citealt{jarvis19,jarvis21}). High-resolution VLA data at 1.4 GHz (see \citealt{jarvis21}) reveals collimated structures along with the presence of hotspots consistent with a jet morphology. Furthermore, the imaging and study of the polarization data of these four galaxies suggests a jet origin of the radio emission (see \citealt{silpa22}). 

These targets also have known central AGN-driven outflows and/or high levels of turbulence identified in ionized gas, as traced via broad emission line widths ($\geq$\,600\,km\,s$^{-1}$) of the [O\,\textsc{iii}] emission, extending over the central few kiloparsecs. In all cases, the interactions of radio jets with the ISM seem to be a significant driver with possible contributions from disk winds \citep[][]{harrison14, jarvis19, venturi23}. Near-infrared spectroscopy of J1430+1339 and J0945+1737 further reveals multiple ionized outflow components through different gas tracers  \citep[][]{ramosAlmeida17, speranza22}. Furthermore, for J1430+1339, there is evidence that the small scale inner $\sim$1\,kpc jet (\citealt{harrison15}) influences both the kinematics and excitation state of the cold molecular gas, as traced with CO\,(2--1) kinematics and CO\,(3--2)/CO\,(2--1) emission-line ratios \citep[][]{ramosAlmeida22,audibert23}. 

In summary, these targets are well aligned with our goal to understand the impact of radio jets on the molecular gas on both small scales, within the molecular gas disks, ($\sim$1\,kpc) and on large scales ($\gtrsim$10\,kpc), extended beyond the molecular gas disks, in `radio quiet' quasars.

\subsection{Observation and reduction of the ALMA data}\label{sec: datared_alma}
We use 12-m array ALMA Band 7 observations to obtain the spatially-resolved molecular gas emission, traced by the CO(3--2) transition. Three of the four targets (J0945+1737, J1000+1242, and J1010+1413) were observed in three, one-hour epochs under the ALMA project 2018.1.01767.S (PI: A.P. Thomson); using the C\,43\,-\,4 array configuration. The chosen correlator setup comprises three spectral windows, with one spectral window covering the central frequency $\nu_{\mathrm{obs}}$\,=\,345.795990\,GHz, corresponding to the CO (3--2) line, and the other two windows partially overlapping the former spectral window for an enhanced signal. The fourth target, J1430+1339 was observed under the program code 2016.1.01535.S (PI: G. Lansbury) in the C40-3 configuration. The observation has a single pointing on-source integration time of 30.3 minutes. The spectral window has a bandwidth of 1.875 GHz and was centered at the CO\,(3--2) line, with the same frequency as mentioned above. The resulting angular resolutions of the observations are $\mathrm{\theta_{res}}\sim0.33\,-\,0.65$\,arcsec (corresponding to linear scales of 0.75\,--\,1.09\,kpc for the respective source redshifts) and a largest angular scale $\mathrm{\theta}_\mathrm{{LAS}}\sim$\,4\,arcsec (i.e., 4.6\,--\,6.4\,kpc) for the former three and $\sim$\,19\,arcsec\,($\sim$\,30\,kpc), for J1430+1339. 

The data for the four galaxies were reduced and calibrated using the Common Astronomy Software Applications (CASA; \citealt{McMullin07}). Using CASA v6.4.3, the imaging of the cubes was made with the task \texttt{tclean}. The cleaning was performed in a mask centered in the peak luminosity pixel of each galaxy, with a radius varying between 5\,arcsec and 8\,arcsec to make sure all the resolved emission was included. A channel width of 25 km\,s$^{-1}$ was selected, and a pixel scale of 0.05” was used to sample all the synthesized beams. The \texttt{H{\"o}gbom CLEAN} algorithm was run to a flux density threshold of 2 times the root mean square of each of the cubes. Different weightings of the Briggs mode were compared: a robustness = 2.0 (close to natural weighting), a robustness = 0.5 (between uniform and natural weighting), and applying a tapering of the visibilities in the u-v plane. We decided to use the robustness = 2.0 to maximize the recovery of the extended emission without losing significant flux, or the central source small-scale structures. For this work, we use cubes with the continuum and allow a line component to fit the continuum (see Section \ref{sec: emLineProcedure}). The final beam sizes of the observations were an average of 0.37\,arcsec\,$\times$\,0.28\,arcsec for the first three targets and  0.7\,arcsec\,$\times$\,0.6\,arcsec for J1430+1339.

\begin{figure*}
\centering
\begin{tabular}{cc}
{\includegraphics[width=.45\textwidth]{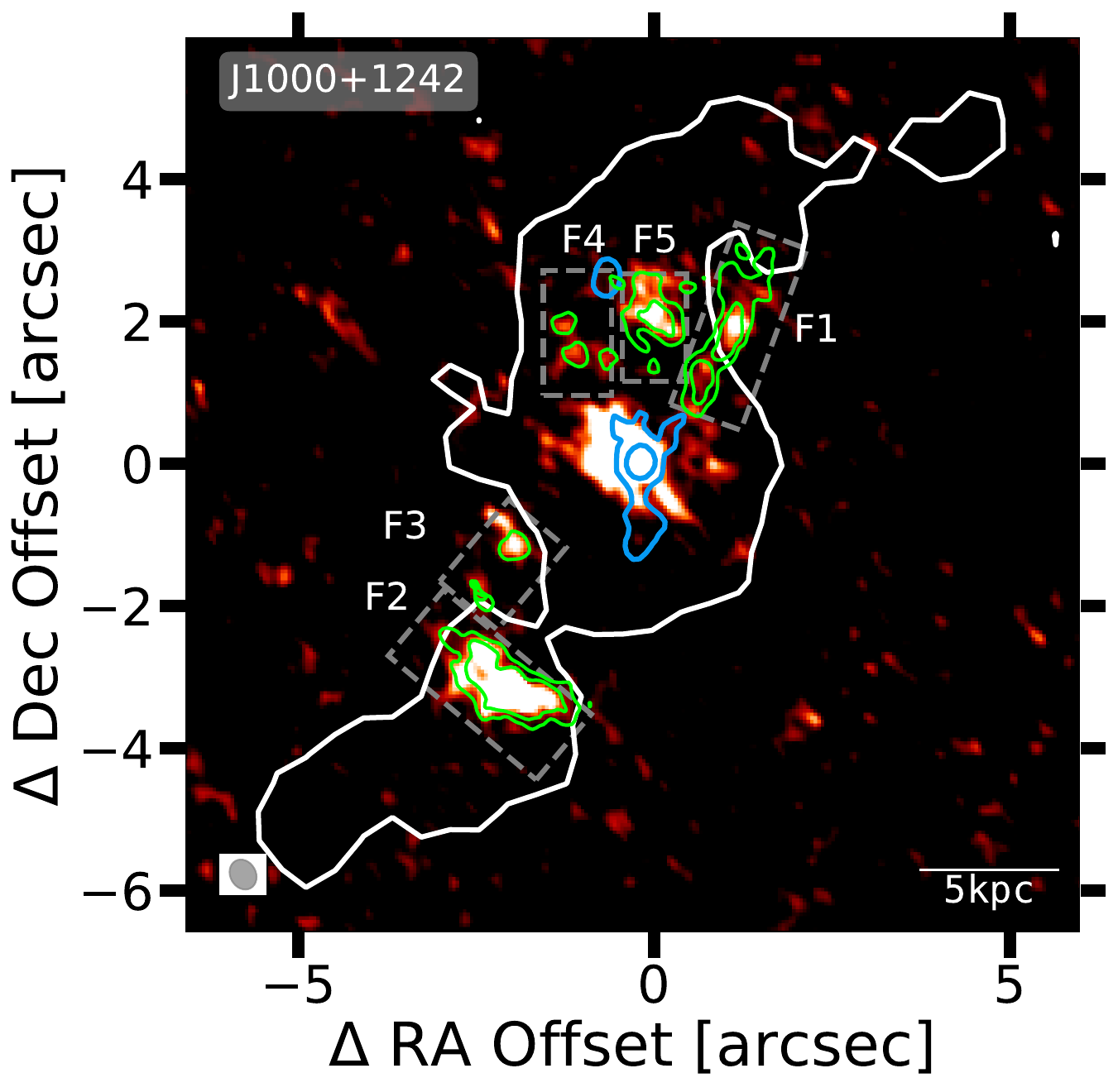}} & {\includegraphics[width=.45\textwidth]{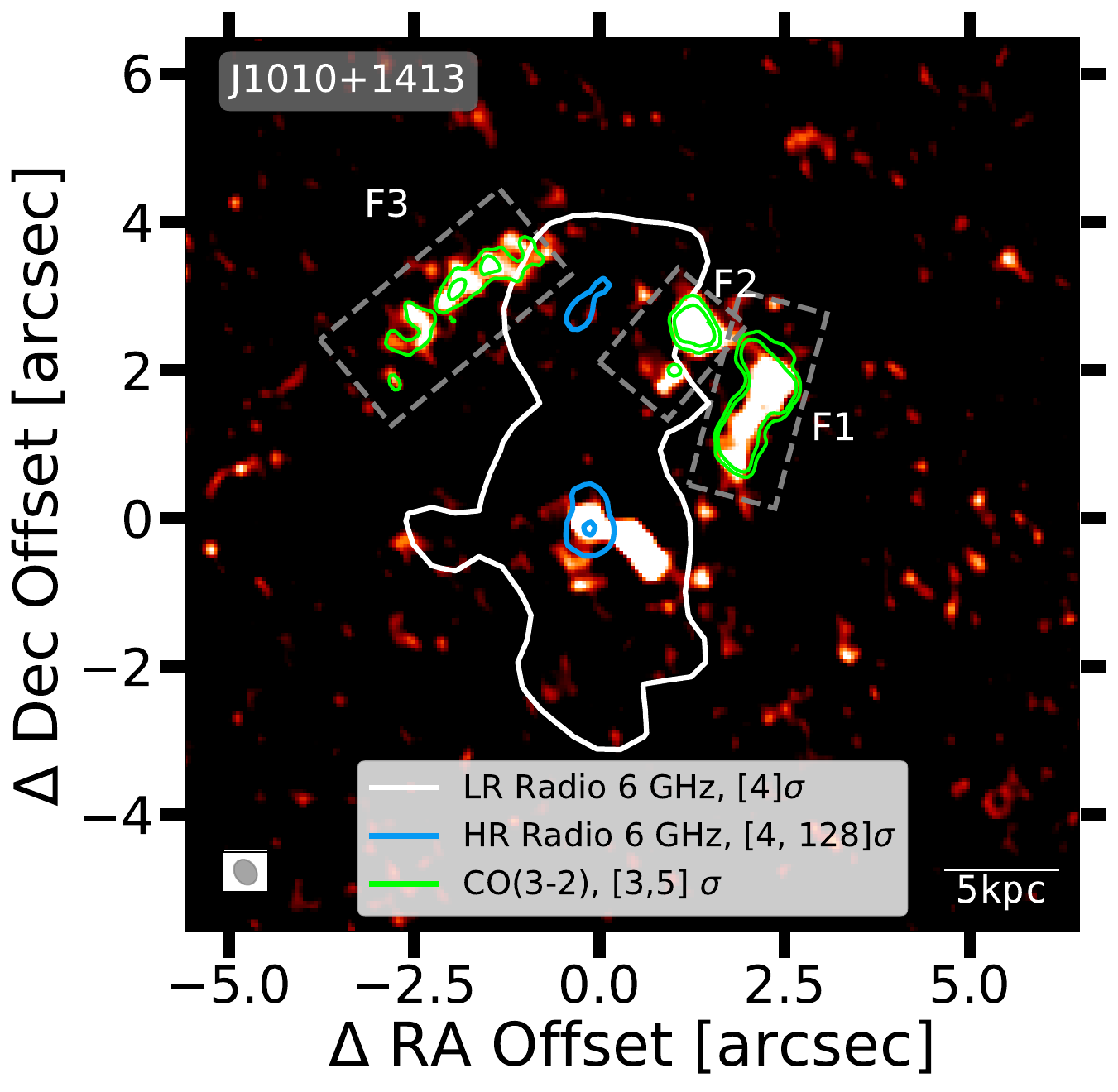}} \\
\end{tabular}
\caption{CO(3--2) emission-line maps, produced using a weighted combination of the individual narrow velocity-range images, highlighting each filament for J1000+1242 and J1010+1413, as shown in Figure~\ref{appfig1: filNarrow} (see Section~\ref{sec: filDefine}). The individual velocity ranges over which the filaments were detected are listed in Table \ref{tab3: filmprops}. The identified filament regions are highlighted with the dashed-grey boxes \textbf{and are labeled respectively as F1\,--\,5}. We overlay as green contours, within each box, the CO(3--2) emission at the 3$\sigma$ and 5$\sigma$ levels from the associated images presented in Figure~\ref{appfig1: filNarrow}. The radio contours are the same as in Figure \ref{fig2: all4targets} but shown here in blue and white colours for high- and low-resolution, respectively. A legend is shown in the right panel.}
\label{fig3: filaments_overview}
\end{figure*}
\subsection{Summary of the radio images}\label{sec: datared_rad}
For our investigation of the relationship between the CO emission-line properties and radio morphology, we use the 6 GHz (C-band) VLA radio images from \cite{jarvis19}. We use both the `low-resolution' (LR) and `high-resolution' (HR), images described from \cite{jarvis19}, which are constructed from a combination of A- and B-configuration VLA data. The 6 GHz (C-band) VLA radio images were obtained by optimizing the imaging parameters and weighting schemes to enhance the extended morphological features for each source (for example,  using uniform, natural or Briggs parameter=0.5 weighting; see Table~3 in \citealt{jarvis19}). In each of the figures in this work, we use these 6\,GHz (C-band) VLA radio images from \cite{jarvis19}, to show the range of radio structures seen on the different spatial scales (e.g., Figure~\ref{fig2: all4targets}). We note that this is why we prefer these over the radio images from \cite{jarvis21}, where a simpler, but consistent, set of imaging parameters were applied to the whole QFeedS sample (but not optimised to show all morpohlogical structures). The LR images have major axis beam sizes of $\sim$1.0\,--\,1.2\,arcsec (i.e., $\sim$\,2 kpc resolution for z\,=\,0.1) , whilst the HR images have beam sizes of $\sim$0.2\,--\,0.3\,arcsec (i.e, $\sim$\,0.5 kpc resolution for z\,=\,0.1). For J0945+1737, we also make use of the 1.5\,GHz e-MERLIN image from \cite{jarvis19}, which has a beam size of 0.2\,--\,0.3\,arcsec. This image highlights the $\sim$\,2.1\,kpc bent jet-like structure in this source.

\subsection{Stellar velocities and velocity dispersion from MUSE data}\label{sec: stelKin}
For this analysis, we are interested in measuring the stellar redshift (z$_*$) and stellar velocity dispersion ($\sigma_*$), integrated over the spatial extent of the galaxies. To do this, we use the available MUSE data for these targets and follow the procedure outlined in \cite{girdhar22} for another QFeedS target, making use of spectral fitting of stellar templates. Full details of the MUSE data and its reduction for these, and other QFeedS targets, are deferred to future works (e.g., \citealt{venturi23}). Therefore, we only provide brief details here.

The four targets have been observed with MUSE, in wide-field mode. This provides a field of view of 1$\times$1 arcmin and a pixel sampling of 0.2\,arcsec. Observations of these targets were taken under proposal IDs 0103.B-0071 (PI: C. Harrison), 0102.B-107 (PI: Sartori) and 0104.B-0476 (PI: G. Venturi). We combine the data from these multiple programs to construct deep final stacked cubes, following the data reduction steps described in \cite{girdhar22}. \\

We obtained stellar kinematics by employing the GIST pipeline (\citealt{bittner19}), following the detailed methodology and parameters described in \cite{girdhar22}. GIST is a framework that inputs fully reduced MUSE cubes and prepares them for stellar continuum fitting to finally provide the stellar kinematics as per the following steps. Firstly, GIST performs a Voronoi tessellation routine (\citealt{cappellariCopin03}) to divide the galaxy into regions with a minimum signal-to-noise-ratio (SNR) in the continuum. We used a threshold of SNR\,=\,30. Following this, for each Voronoi region, GIST obtains the best fit to the stellar continuum, exploiting the pPXF routine (\citealt{cappellariEmsellem04, cappellari17}), with a combination of stellar templates from XSL Library (\citealt{arentsen19,gonneau20}). This results in an accurate measure of the stellar velocity and stellar velocity dispersion in each Voronoi bin. We used flux-weighted averaging over all the Voronoi bins for the systemic redshift (z$_*$) and the stellar velocity dispersion ($\sigma_*$) of the galaxy. These systemic redshifts are used to shift all the molecular gas emission profiles to the rest frame (see Section~\ref{sec: emLineProcedure}). The errors in the values are determined as a median of the formal errors across all the Voronoi bins for each target, where formal errors are 1$\sigma$ uncertainties as obtained by pPXF fitting. For the four targets, we obtain stellar velocity dispersion values between 170\,--\,270\,km\,s$^{-1}$. All values for z$_*$ and $\sigma_*$ are listed in Table~\ref{tab2: galprops}.


\begin{table*}
\caption{Measured global galaxy properties using the spectra shown in Figure \ref{fig2: all4targets} (panels a2\,--\,d2) and extracted from the dashed-grey regions (shown in the panels a1\,--\,d1): (1) quasar name;  (2-3) stellar redshift and stellar velocity dispersion, respectively, measured from the stellar kinematics following Section \ref{sec: stelKin}; (4-7) measurements following Section \ref{sec: emLineProcedure}, namely, (4) mean velocity (V$_{50}$); (5) velocity width (W$_{80}$); (6) velocity dispersion estimated as $\sigma\,=\,W_{80}/2.56$; (7) flux; (8) molecular gas mass; 
(9) jet kinetic power obtained from the median of the total low-resolution flux density and core flux density at 5.2 GHz (\citealt{jarvis19}).}
\label{tab2: galprops}
\begin{tabular}{l c c c c c c c c } 
\hline
Quasar & z$_*$ & $\sigma_*$ & V$_{\mathrm{50,CO,gal}}$ & W$_{\mathrm{80,CO,gal}}$ & ${\sigma}_{\mathrm{CO,gal}}$ & S$_{\mathrm{CO(3-2),gal}}$ & log$({\mathrm{M}}_{\mathrm{mol,gal}}$) & log\,(P$_{\mathrm{jet}}$)\\
 & & [km\,s$^{-1}$]  & [km\,s$^{-1}$]  & [km\,s$^{-1}$]  & [km\,s$^{-1}$] & [Jy km\,s$^{-1}$] & /\,[M$_{\odot}$] & [erg\,s$^{-1}$]\\
(1) & (2) & (3) & (4) & (5) & (6) & (7) & (8) & (9) \\
\hline
J0945+1737 & 0.12840 & 171\,$\pm$\,30 & 0\,$\pm$\,5 & 398\,$\pm$\,13 & 156\,$\pm$\,5 & 9.39\,$\pm$\,1.29 & 9.6 & 43.55\,$\pm$\,0.33 \\
J1000+1242 & 0.14787 & 174\,$\pm$\,18 & $-$54\,$\pm$\,26 & 438\,$\pm$\,102 & 171\,$\pm$\,40 & 6.89\,$\pm$\,0.29 & 9.6 & 43.67\,$\pm$\,0.08 \\
J1010+1413 & 0.19877 & 272\,$\pm$\,20 & 22\,$\pm$\,10 & 583\,$\pm$\,18 & 228\,$\pm$\,7 & 19.86\,$\pm$\,0.17 & 10.4 & 43.18\,$\pm$\,0.15 \\
J1430+1339 & 0.08507 & 182\,$\pm$\,29 & 29\,$\pm$\,13 & 511\,$\pm$\,39 & 200\,$\pm$\,15 & 22.20\,$\pm$\,0.19 & 9.3 & 43.29\,$\pm$\,0.43 \\
\hline
\end{tabular}
\end{table*}

\begin{table*}
\caption{Measured properties for the filamentary molecular gas structures (see Section \ref{sec: filProps}) extracted from within the dashed-grey regions shown in Figure \ref{fig3: filaments_overview}. The individual emission-line profiles are shown in appendix Figures \ref{appfig2: filamentsZoom1} and \ref{appfig3: filamentsZoom2}. The properties listed here are namely: (1) quasar name; (2) filament name (see Figure~\ref{fig3: filaments_overview}); (3) velocity range used to identify the filaments (see Figure \ref{fig3: filaments_overview} and Section \ref{sec: filDefine}); (4) filament axis ratio; (5) radial extent of the filament (R$_{\mathrm{fil}}$); (6) filament velocity (V$_{\mathrm{50,fil}}$); (7) filament velocity width (W$_{\mathrm{80,fil}}$); (8) filament velocity dispersion (${\sigma}_{\mathrm{fil}}$); (9) flux.}
\label{tab3: filmprops}
\begin{tabular}{l c c c c c c c c }
\hline
& & & & & & & & \\
Quasar & Filament & Velocity range & Axis ratio & R$_{\mathrm{fil}}$ & V$_{\mathrm{50,fil}}$ & W$_{\mathrm{80,fil}}$ & ${\sigma}_{\mathrm{fil}}$ & S$_{\mathrm{CO(3-2);fil}}$ \\
 &  & [km\,s$^{-1}$] & & [kpc] & [km\,s$^{-1}$] & [km\,s$^{-1}$] & [km\,s$^{-1}$] & [Jy km\,s$^{-1}$] \\
(1) & (2) & (3) & (4) & (5) & (6) & (7) & (8) & (9) \\
\hline
J1000+1242 & 1 & [93,220] & 6.51\,$\pm$\,0.12 & 10.43\,$\pm$\,0.16 & 134\,$\pm$\,13 & 242\,$\pm$\,25 & 95\,$\pm$\,10 & 0.93\,$\pm$\,0.32 \\
 & 2 & [-262,-109] & 2.39\,$\pm$\,0.03 & 10.97\,$\pm$\,0.16 & $-$177\,$\pm$\,12 & 239\,$\pm$\,24 & 94\,$\pm$\,9 & 1.16\,$\pm$\,0.72 \\
 & 3 & [-211,-160] & 2.96\,$\pm$\,0.11 & 5.76\,$\pm$\,0.16  & $-$193\,$\pm$\,15 & 82\,$\pm$\,12 & 32\,$\pm$\,5 & 0.27\,$\pm$\,0.06 \\
 & 4 & [-185,-109] & 2.04\,$\pm$\,0.08 & 9.12\,$\pm$\,0.16  & $-$189\,$\pm$\,42 & 340\,$\pm$\,190 & 133\,$\pm$\,74 & 0.63\,$\pm$\,0.20 \\
 & 5 & [-160,-33] & 1.15\,$\pm$\,0.03 & 8.15\,$\pm$\,0.16  & $-$96\,$\pm$\,19 &  170\,$\pm$\,28 & 67\,$\pm$\,11 & 0.68\,$\pm$\,0.11 \\
 \hline
J1010+1413 & 1 & [-401,-274] & 3.02\,$\pm$\,0.06 & 13.24\,$\pm$\,0.13 & $-$228\,$\pm$\,6 &  143\,$\pm$\,10 & 56\,$\pm$\,4 & 0.96\,$\pm$\,0.08 \\
 & 2 & [-477,-401] & 1.25\,$\pm$\,0.02 & 12.45\,$\pm$\,0.13 & $-$332\,$\pm$\,9 & 223\,$\pm$\,22 & 87\,$\pm$\,9 & 0.51\,$\pm$\,0.17 \\
 & 3 & [-502,-426] & 3.47\,$\pm$\,0.07 & 13.17\,$\pm$\,0.13 & $-$342\,$\pm$\,14 &  119\,$\pm$\,15 & 47\,$\pm$\,6 & 0.30\,$\pm$\,0.05 \\
\hline
\end{tabular}
\end{table*}


\begin{table*}
\caption{Derived properties for the filamentary molecular gas structures(see Section \ref{sec: filProps}) extracted from integrated spectra within dashed-grey regions shown in Figure \ref{fig3: filaments_overview}. The properties listed here are namely: (1) quasar name; (2) filament number; (3) molecular gas mass in filaments; (4) mass outflow rates; (5) kinetic power; (6) jet kinetic power; (7) jet coupling efficiency; (8) radiative coupling efficiency.}.
\label{tab4: fildprops}
\begin{tabular}{l c c c c c c c}
\hline
Quasar & Filament & log$(\mathrm{M}_{\mathrm{fil}}$) & log$(\dot{\mathrm{M}}_{\mathrm{mol,fil}}$) & log$(\dot{\mathrm{E}}_{\mathrm{kin}}$) & log(P$_{\mathrm{jet}}$) & $\eta\,_{\mathrm{jet}}$ & $\eta\,_{\mathrm{radiative}}$ \\
 &  & /\,[M$_{\odot}$] & /\,[M$_{\odot}$\,yr$^{-1}$] & /\,[erg\,s$^{-1}$]  & /\,[erg\,s$^{-1}$] & \textbf{\,=\,$\dot{\mathrm{E}}_{\mathrm{kin}}$/P$_{\mathrm{jet}}$} & \textbf{\,=\,$\dot{\mathrm{E}}_{\mathrm{kin}}$/L$_{\mathrm{bol}}$} \\
(1) & (2) & (3) & (4) & (5) & (6)  & (7) & (8)\\
\hline
J1000+1242 & 1 & 8.77 & 7.77 & 40.67 & 43.67 & 1$\times$10$^{-3}$ & 3$\times$10$^{-6}$ \\
J1000+1242 & 2 & 8.86 & 10.98 & 40.97 & 43.67 & 2$\times$10$^{-3}$ & 7$\times$10$^{-6}$ \\
J1000+1242 & 3 & 8.22 & 3.64 & 40.26 & 43.67 & 0.4$\times$10$^{-3}$ &  1$\times$10$^{-6}$ \\
J1000+1242 & 4 & 8.60 & 7.92 & 40.92 & 43.67 & 2$\times$10$^{-3}$ & 6$\times$10$^{-6}$ \\
J1000+1242 & 5 & 8.63 & 5.46 & 40.28 & 43.67 & 0.5$\times$10$^{-3}$ &  1$\times$10$^{-6}$ \\
\hline
J1010+1413 & 1 & 9.04 & 19.58 & 41.53 & 43.18 & 2$\times$10$^{-2}$ &  7$\times$10$^{-6}$ \\
J1010+1413 & 2 & 8.77 & 15.49 & 41.72 & 43.18 & 4$\times$10$^{-2}$ &  11$\times$10$^{-6}$ \\
J1010+1413 & 3 & 8.54 & 9.10 & 41.53 & 43.18 & 2$\times$10$^{-2}$ &  7$\times$10$^{-6}$ \\
\hline
\end{tabular}
\end{table*}

\begin{table}
\caption{Measured central outflow properties in CO(3--2) following Section \ref{sec: nucOutMap} and spectra from the region shown through dashed-purple boxes in Figure \ref{fig2: all4targets} (panels a3\,--\,d3): (1) quasar name; (2) velocity (V$_{\mathrm{out}}$); and (3) radial extent (R$_{\mathrm{out}}$) of the observed central outflow.}
\label{tab5: nucoutprops}
\centering
\begin{tabular}{l c c c c}
\hline
Quasar & V$_{\mathrm{out}}$ & R$_{\mathrm{out}}$ \\
 & [km\,s$^{-1}$]  & [kpc] \\
(1) & (2) & (3) \\
\hline
J0945+1737 & 249\,$\pm$\,27 & 0.60\,$\pm$\,0.13 \\
J1000+1242 & 376\,$\pm$\,236 & 0.65\,$\pm$\,0.15 \\
J1010+1413 & 441\,$\pm$\,57 & 1.64\,$\pm$\,0.13 \\
J1430+1339 & 313\,$\pm$\,146 & 0.80\,$\pm$\,0.23 \\
\hline
\end{tabular}
\end{table}

\section{Analysis of the CO emission}\label{sec: methods}

In this Section, we present our analysis steps to obtain the observed and derived properties of the molecular gas on different scales; for the whole galaxy, and for various spatially-resolved scales. In Section~\ref{sec: filDefine}, we formalize our approach to identify any molecular gas structures outside of the central galaxy disks (i.e., the extended molecular gas structures). In Section~\ref{sec: emLineProcedure} we describe our emission-line fitting procedure to characterize the kinematics of the CO emission. In Section~\ref{sec: filProps} we evaluate the properties of the molecular gas structures (velocity, velocity dispersion, projected extent, and estimated masses). Finally, in Section~\ref{sec: nucOutMap}, we present a brief analysis of the broad CO emission-line wing components, as a tracer of central molecular outflows.

\subsection{Identification of extended molecular gas structures}\label{sec: filDefine}

In Figure \ref{fig2: all4targets}, we show CO (3--2) emission-line images, collapsed over the full observed velocity width of the emission-line profiles (the velocity limits are indicated at the top-right of each panel). Overlaid on these images are contours of the 6\,GHz radio emission for both the low-resolution and high-resolution images (shown as white and red contours, respectively; see Section~\ref{sec: datared_rad}). In two of the galaxies, J0945+1737 (\textbf{panel a1}) and J1430+1339 (\textbf{panel d1}), we see that the CO emission is only observed in a central, contiguous region. However, for the other two galaxies, J1000+1242 \textbf{(panel b1)} and J1010+1413 \textbf{(panel c1)}, in addition to the central molecular gas, we see molecular CO structures outside of the central regions. 

To discern extended molecular gas structures in a systematic way, we formalized the following procedure, which is motivated by the qualitative methods used by \cite{russell19} and \cite{tamhane22} to search for extended molecular gas structures around BCGs. These two works will serve as our primary comparison sample (discussed in Section~\ref{sec: compFilProps}). Firstly, we define the central molecular gas disks as central, contiguous CO structures with smooth velocity gradients centered on the systemic galaxy velocities. 

We used a visual inspection of the narrow-band images and individual velocity slices, as well as the kinematic maps, to identify extended molecular gas structures based on the following criteria.
\begin{enumerate}[leftmargin=0.4in]
    \item Emission with a clear morphological and/or kinematic separation from the central molecular gas disk.
    \item Emission detected at a $\geq$\,5$\sigma$ significance level in individual velocity channels of the data cube, but also seen in more than one consecutive velocity channel.
    \item Structures extending to $\geq$\,1\,kpc in projected size.
\end{enumerate}

Using this systematic approach, for J0945+1737 and J1430+1339, we do not identify any extended molecular gas structures away from the central emission. This confirms the observation made from the total CO emission-line images shown in Figure \ref{fig2: all4targets}. These two quasars are hence not included in our analysis pertaining to the extended CO gas structures. 

For J1000+1242 and J1010+1413, we identified 5 and 3 molecular gas structures, respectively, using the above method. We created narrow velocity slice CO images by collapsing over the consecutive velocity channels where any emission $\geq\,5\,\sigma$ was seen associated with these structures. These are shown in Figure~\ref{appfig1: filNarrow}. A combined overview of the molecular gas structures is shown in Figure \ref{fig3: filaments_overview}. For this figure, to distinctly visualize each of these molecular gas structures, we performed a weighted combination of the narrow velocity slices of each filament shown in Figure \ref{appfig1: filNarrow}, with higher weights to the structures with lower surface brightness. Each filament is highlighted with a surrounding dashed grey box,\textbf{ and labeled following the notation as F1\,--\,5, respectively}. These boxes cover the full observed extent (at $\ge3\,\sigma$) of the structures. In Section \ref{sec: filProps}, we evaluate the properties of each of these structures. 

For all 8 identified gas structures, we estimated an axis ratio by fitting a 2D Gaussian over the surface brightness images of each. The uncertainty in the axis ratio was obtained by using the errors in the measurements of the major and minor axis of the 2-D fitted Gaussian, and then propagating the errors to obtain the uncertainty on the axis ratio. These values are listed in Table \ref{tab3: filmprops}. The axis ratios range from 1.2\,--\,6.5, with a median of 2.7, and all but two have an axis ratio $>$2. Therefore, we refer to each of the identified molecular gas structures as `filaments'. This follows the terminology adopted for the molecular gas structures seen in BCGs, which show a similar range in morphology \cite[e.g.,][]{russell19,tamhane22}.

We note that our approach of selecting these structures may not uniquely select physically distinct `filaments'. For example, the identified structures may be part of a larger connected `flow', and/or each `filament' can contain sub-structures. Indeed, the contours in Figure~\ref{fig3: filaments_overview} do show sub-structure. However, we have used the systematic approach described above to identify these structures, without laying much emphasis on the sub-structures. We note that this does not affect our scientific goals, which are primarily to compare the properties of these overall structures to those seen in BCGs. Further, the filaments analyzed in the BCGs were also identified using a very similar approach and definitions. We discuss the origin and properties of these structures in Section~\ref{sec: results}. 

\subsection{Emission-line fitting procedure}\label{sec: emLineProcedure}

We evaluate the molecular gas velocity and velocity dispersion by performing fitting to the CO emission-line profiles. As described below, we studied the properties on various spatial scales from the datacubes:  (1) in individual spatial pixels to produce maps; (2) from a region covering the entire CO emitting region for the full galaxies (shown through dashed-grey boxes in Figure \ref{fig2: all4targets}); (3) regions covering each of the individual filaments (shown through dashed-white boxes in Figure \ref{fig3: filaments_overview}); and (4) central `outflow' regions where we identify broad CO emission-line wing components (shown through dashed-purple boxes in Figure \ref{fig2: all4targets}). 

We used the \textsc{scipy curve fit} routine (\citealt{scipypaper}) to obtain the best fit to the data, within the velocity range of $\pm$\,700\,km\,s$^{-1}$. We modeled the emission-line profiles using one and two Gaussian components; in addition to a linear component for characterizing any underlying continuum emission. To statistically select the best fit to the data, we used the difference in the BIC values (Bayesian Information Criterion; \citealt{schwarz78}), i.e., the model with the lowest BIC value was selected. 

While we use multiple Gaussian components, to characterize the emission-line profiles,  we adopt a non-parametric approach for most of our analysis \citep[following e.g.,][]{harrison14,girdhar22}. The bulk velocity is measured in terms of the median velocity of the line profile, V$_{50}$; and the velocity dispersion is measured through velocity width in terms of W$_{80}$, i.e., the width containing 80\% of the emission-line flux. For a single Gaussian, W$_{80}$ is approximately related to the full-width-at-half-maxima (FWHM) as W$_{80}\,=\,1.088\,\times$\,FWHM; where the FWHM itself can be related to $\sigma$ as FWHM\,=\,2.35\,$\sigma$. 

All line profiles and velocity maps presented in this work are shifted from the observed to the rest-frame using the stellar systemic redshift as listed in Table \ref{tab2: galprops} (following Section \ref{sec: stelKin}). The application of this emission-line fitting process to different spatial scales is explained in detail below.

\begin{figure*}
\centering
\begin{tabular}{cc}
\subcaptionbox{Molecular Gas Velocity (V$_{50}$; km\,s$^{-1}$)}{\includegraphics[width=.45\textwidth]{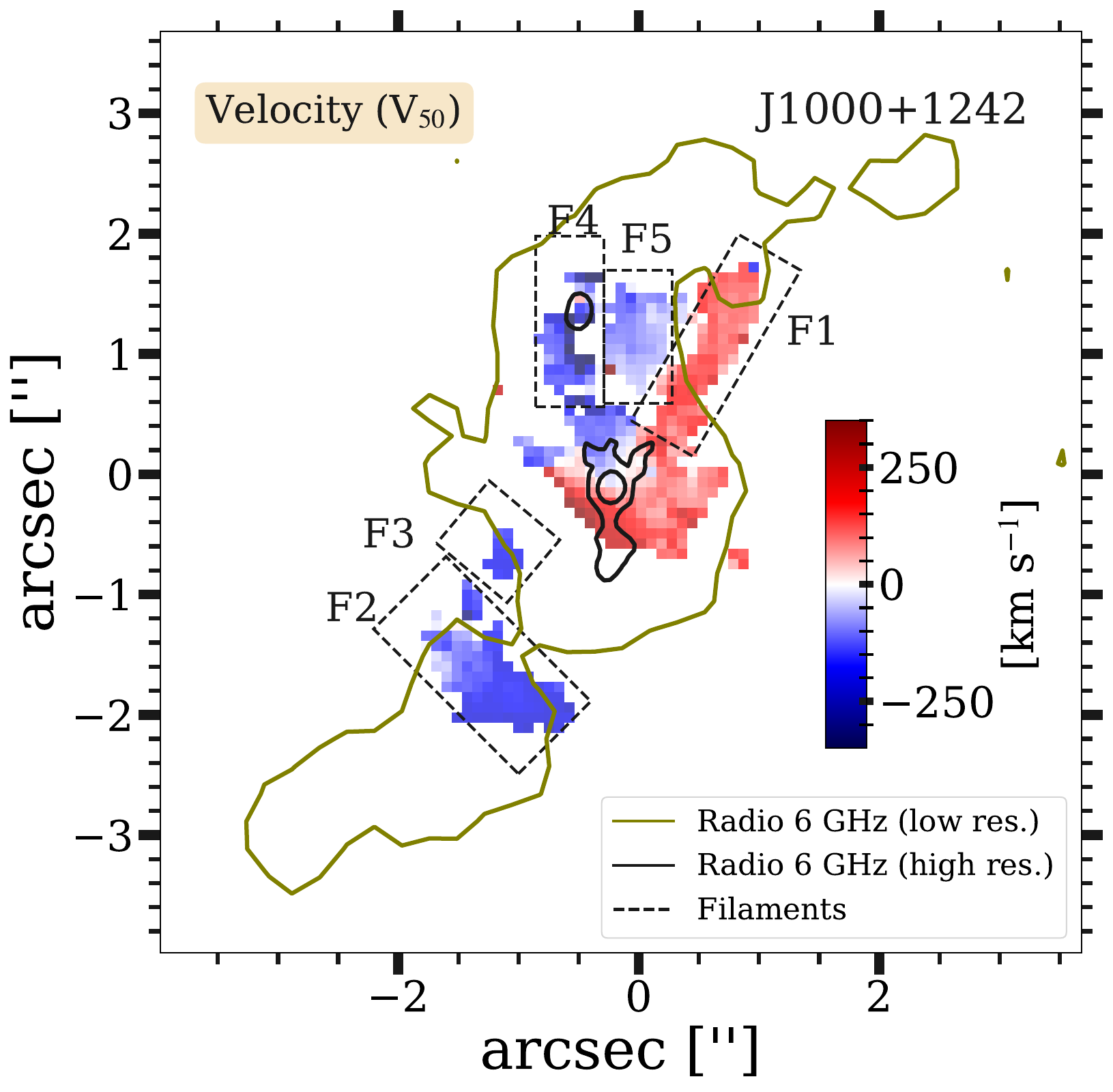}} & \subcaptionbox{Molecular Velocity Width (W$_{80}$; km\,s$^{-1}$)}{\includegraphics[width=.45\textwidth]{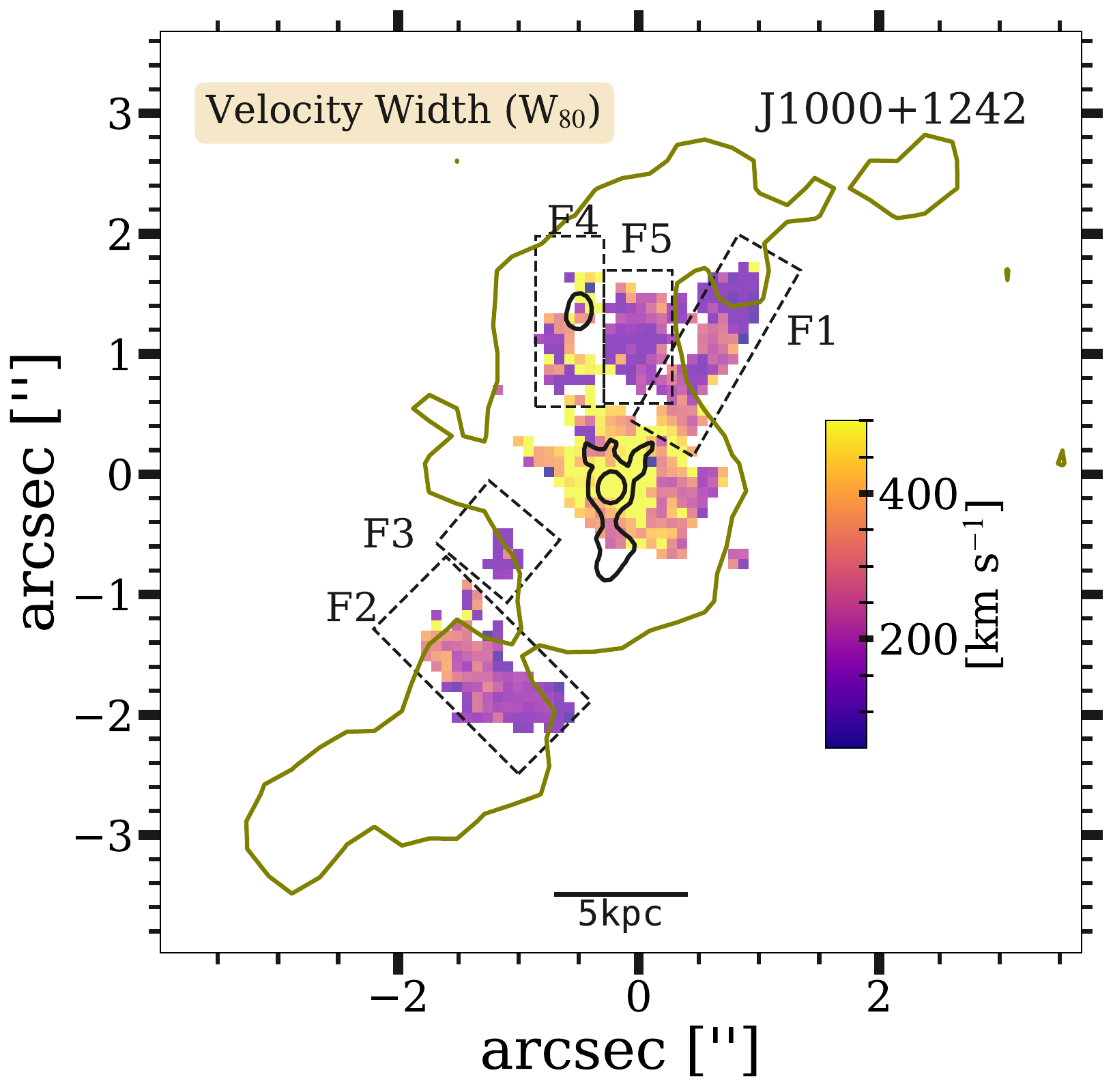}} \\
\end{tabular}
\caption{Kinematic maps of the CO(3--2) emission line for J1000+1242 (see Section \ref{sec: emLineProcedure}). The {\em left panel} is a velocity map (V$_{\mathrm{50}}$), the {\em right panel} is a map of velocity width (W$_{\mathrm{80}}$). Overlaid on each map, the black and olive-green contours correspond to the high- and low-resolution 6\,GHz radio images, same as in Figure \ref{fig2: all4targets}. The black, dashed boxes highlight the filamentary regions identified in Figure \ref{fig3: filaments_overview}, \textbf{and labeled respectively as F1\,--\,5}. A legend is shown in the left panel and a 5\,kpc scale bar in the right panel.}
\label{fig4: kinematics_J10001242}
\end{figure*}

\begin{figure*}
\begin{tabular}{cc}
\subcaptionbox{Molecular Gas Velocity (V$_{50}$; km\,s$^{-1}$)}{\includegraphics[width=.45\textwidth,height=8cm]{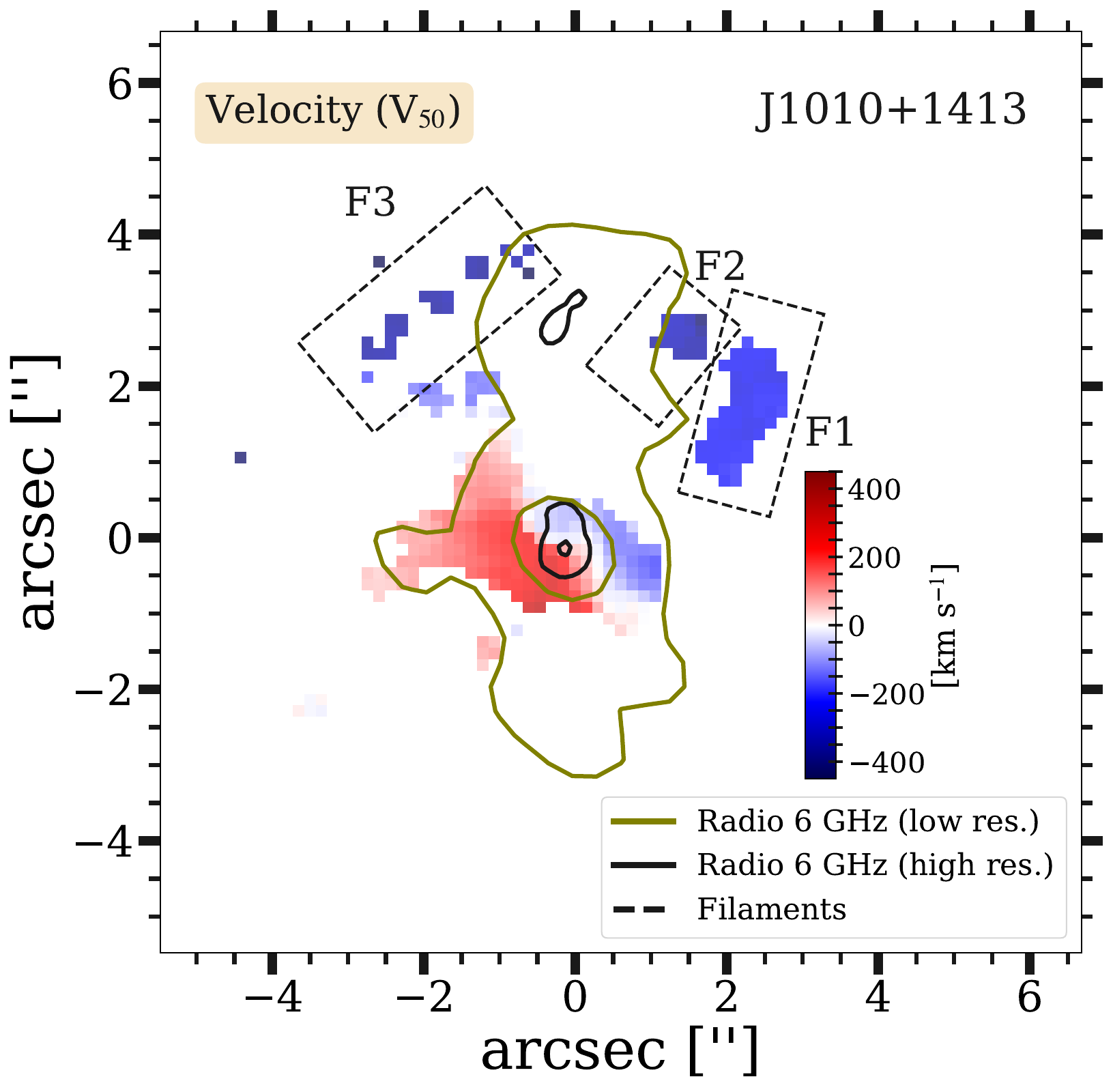}} & \subcaptionbox{Molecular Velocity Width (W$_{80}$; km\,s$^{-1}$)}{\includegraphics[width=.45\textwidth,height=8cm]{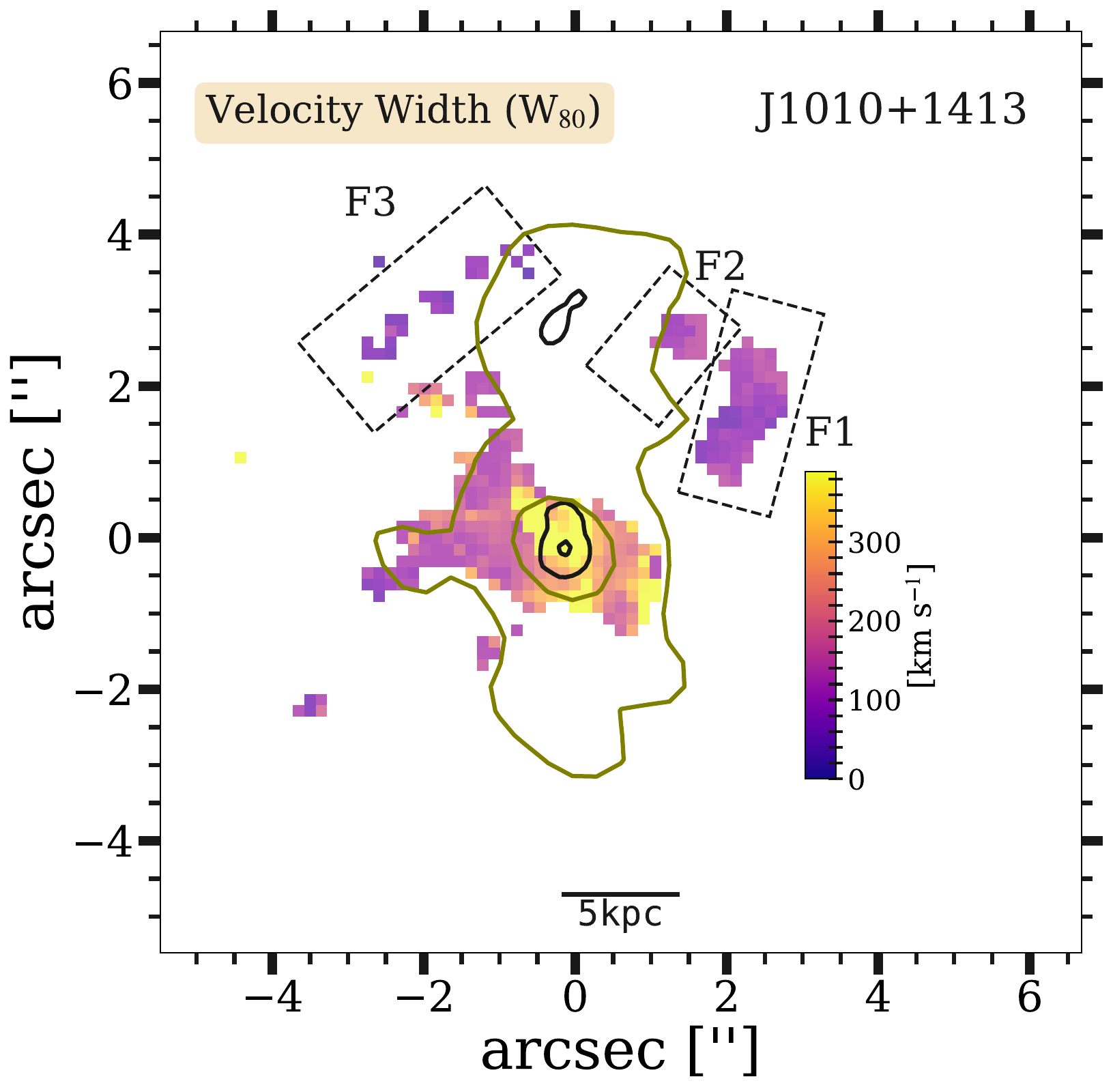}} \\ 
\end{tabular}
\caption{Kinematic maps of the CO(3--2) emission line for J1010+1413 (see Section \ref{sec: emLineProcedure}). The {\em left panel} is a velocity map (V$_{\mathrm{50}}$), the {\em right panel} is a map of velocity width (W$_{\mathrm{80}}$). The black, dashed boxes highlight the filamentary regions identified in Figure \ref{fig3: filaments_overview}, \textbf{and labeled respectively as F1\,--\,3}. The rest is the same as in Figure \ref{fig4: kinematics_J10001242}.}
\label{fig5: kinematics_J10101413}
\end{figure*}

\begin{itemize}
    \item \textbf{For the entire galaxy-scale}:\\
    For the total CO emission, we extract the spectrum over the region shown through a grey-dashed box in Figure \ref{fig2: all4targets} for all four targets. The obtained emission-line spectra along with the best fit following the fitting procedure above are shown for each galaxy in the panels \textbf{(a2), (b2), (c2), and (d2)}, respectively. \\ 

    \item \textbf{For individual spatial pixel fits}: \\
    To apply the emission-line fitting routine on individual spaxels, we first re-grided the ALMA cubes from an initial spatial resolution of 0.05\,$\times$\,0.05 arcsec to 0.15\,$\times$\,0.15 arcsec, to increase the SNR of each spatial-unit. We first checked that the SNR\,$\geq$\,3 for the emission line to be considered as detected. For a further conservative check, we then compared the single Gaussian fit, with a simple straight line fit using $\Delta$BIC. If the line fit had a lower BIC value,  the spaxel was discarded from further kinematic analysis. After these checks to confirm a detected emission line, the line fitting routine continued as above. For the two targets that show the filamentary molecular gas structures, the resulting kinematic maps are illustrated in Figure \ref{fig4: kinematics_J10001242} and Figure \ref{fig5: kinematics_J10101413} for J1000+1242 and J1010+1413, respectively. \\
        
    \item \textbf{For individual molecular gas structures}: \\
    To quantify the kinematic properties of the molecular gas structures (the `filaments'), we fit the integrated spectra from the region shown through dashed, gray boxes in Figure \ref{fig3: filaments_overview}. Kinematic maps for the individual structures, along with the extracted CO emission-line profiles and fits, are shown in the Appendix Figures \ref{appfig2: filamentsZoom1} and \ref{appfig3: filamentsZoom2}. 
\end{itemize}

\subsection{Molecular gas properties}\label{sec: filProps}
In this section, we measure the properties of the CO filaments seen in our sources. To compare the properties to similar structures seen in the BCGs at comparable redshifts (z$\lesssim$0.2); we follow methods motivated by the techniques used in \cite{russell19} and \cite{tamhane22}. These works have performed an extensive study of the morphology and kinematics of the filaments for 14 unique BCGs (across both samples). The measured CO properties for the whole galaxy measurements are listed in Table \ref{tab2: galprops}. For the individual filaments, the properties are provided in Tables \ref{tab3: filmprops} and \ref{tab4: fildprops}. 

\subsubsection{Molecular gas velocity}\label{sec: vflow}

To measure the bulk velocities in the molecular gas, we utilize the median velocity V$_{\mathrm{50}}$ (see Section \ref{sec: emLineProcedure}) derived from the emission-line profiles depending on the spatial level in consideration, i.e., at the galaxy-level, the velocity will be V$_{\mathrm{50, gal}}$ (or V$_{\mathrm{gal}}$ for simplicity). This median velocity is calculated from the fit to the galaxy-level spectra as shown in the top-right panels of Figure \ref{fig2: all4targets}. The velocity maps for J1000+1242 and J1010+1413 are shown in the left panels of Figure \ref{fig4: kinematics_J10001242} and \ref{fig5: kinematics_J10101413}, respectively. For the filamentary molecular gas structures, the velocity of the filaments (or V$_{\mathrm{fil}}$) is obtained from the spectra extracted from the filamentary regions (see appendix Figures \ref{appfig2: filamentsZoom1} and \ref{appfig3: filamentsZoom2}). To estimate the uncertainties for our velocity values, we performed Monte Carlo (MC) simulations. For this purpose, multiple simulated representations of the CO emission-line data were obtained by adding random Gaussian noise to our best-fit model (on the scale of the residual noise in the continuum). A fit was obtained for each of these simulated spectra with V$_{50}$ measured each time. The standard deviation of the distribution of the measured V$_{50}$ values was then used as the uncertainty. We also compared our method with that used by \cite{tamhane22}, where filament velocities are defined as `flow velocities' (V$_{\mathrm{flow}}$). They obtain their value from a flux-weighted average, over the filament regions, from a velocity map, and always use a single Gaussian component for their fits. Following the approach used by \citealt{tamhane22}, the resulting velocities are similar (i.e., within $\leq$\,5\%), as compared to the the previously described V$_{50}$ method. We also note that in the case of more than one filament for a galaxy, \cite{tamhane22} only provides an average value over all the filaments. 


\subsubsection{Molecular gas velocity dispersion}\label{sec: sigmaflow}

To obtain the velocity dispersion, we refer to the analysis of \cite{russell19}. They define the molecular velocity dispersion $\sigma_{\mathrm{gal}}$ ($\sigma_{\mathrm{mol}}$ in \citealt{russell19}) as the $\sigma$ width of a single Gaussian component, fitted to the CO line emission over the entire galaxy. They also separately estimate the CO line velocity dispersion of the filaments, $\sigma_{\mathrm{fil}}$, by fitting a single Gaussian for the emission-line profiles obtained only over the individual filamentary regions. We follow the same process to obtain the CO velocity dispersion for the total galaxy spectrum and for each of the extended filamentary structures in our sample. While we allow for multiple Gaussian fits, for consistency, we obtain an equivalent $\sigma$ value from the non-parametric values of velocity width, following \( \sigma\,=\,\mathrm{W}_{\mathrm{80}}/2.6 \). To obtain the uncertainty in the obtained values, we perform MC simulations as described for estimating the molecular gas velocity (see Section \ref{sec: vflow}).

\subsubsection{Radial extent of filaments}\label{sec: rflow} 

We quantify the maximum projected radial extent, R$_{\mathrm{fil}}$, of the filamentary structures as the maximum projected distance from the nucleus to the most distant part of each filament. For this, we used the images collapsed over the narrow velocity ranges in which the individual filaments were detected (see Figure \ref{appfig1: filNarrow}). We measured the spatial separation between the centre of the galaxy which was identified using the position of the radio core; and the farthest point of the filaments. This was measured for a few spaxels in the filaments and the maximum projected distance was used as a measure of the R$_{\mathrm{fil}}$ for each filament. This was done following the same method as \cite{tamhane22} (defined as R$_{\mathrm{flow}}$ in their work). The uncertainty on the R$_{\mathrm{fil}}$ values are estimated to be the equivalent size of the beam's major axis. 


\subsubsection{Molecular gas mass estimates}\label{sec: mass_calc}

To obtain the total molecular gas mass in the galaxy (M$_{\mathrm{tot}}$), we first measured the integrated line flux for the CO\,(3--2) emission from the galaxy spectra. Likewise, for the filament mass (M$_{\mathrm{fil}}$), we used the integrated flux from the spectra extracted from filamentary regions. The uncertainty in flux values is obtained from the emission line fitting routine as the square root of the diagonal of the covariance matrix of each free parameter used for the emission line fit. 

To estimate the molecular mass, we used the same conversion factors as \cite{tamhane22}, for a consistent comparison. We first converted CO\,(3--2) fluxes to CO\,(1--0), by using integrated line flux ratios of 7.2 (\citealt{vantyghem16}).  We then converted the integrated flux density of CO(1--0) line (S$_{\mathrm{CO}}\,\Delta\,$v) to molecular gas mass ($\mathrm{M}_{\mathrm{mol}}$) using the following relation (\citealt{solomonVandenBout05, bolatto13}): 
\begin{equation}
    \mathrm{M}_{\mathrm{mol}} = 1.05\times10^4\, \dfrac{X_{\mathrm{CO}}}{[X_{\mathrm{CO,Gal}}]}\,\dfrac{1}{1+z}\,\dfrac{S_{\mathrm{CO}}\,\Delta\,\mathrm{v}}{\mathrm{[Jy\,km\,s^{-1}]}}\,\dfrac{\mathrm{D}_{\mathrm{L}}\,^2}{\mathrm{[Mpc^2]}} \, M_{\odot}
\end{equation}

where \texttt{z} is the redshift of the galaxy, D$_\mathrm{L}$ is the luminosity distance, and X$_{\mathrm{CO}}$ is the CO-to-H$_2$ conversion factor, with X$_{\mathrm{CO,Gal}}$\,=\,2$\times$10$^{20}$\,cm$^{-2}$\,(K\,km\,s$^{-1}$)$^{-1}$ (\citealt{solomon87, solomonVandenBout05}). This relation is the corollary of the relation \(\mathrm{M}_{\mathrm{mol}}\,=\,\alpha\,\mathrm{L}'_{\mathrm{CO(1-0)}}\), where X$_{\mathrm{CO}}$ and $\alpha_{\mathrm{CO}}$ are both referred to as ``CO-to-H$_2$'' conversion factor. For X$_{\mathrm{CO}}$=\,2$\times$10$^{20}$\,cm$^{-2}$\,(K\,km\,s$^{-1}$)$^{-1}$, the corresponding $\alpha_{\mathrm{CO}}$ = 4.3\,M$_{\odot}$\,(K\,km\,s$^{-1}$\,pc$^2$)$^{-1}$. It is a caveat that this exact X$_{\mathrm{CO,Gal}}$ factor may not apply to our galaxies (and also neither to BCGs), and there may be significant variation, pertaining to environmental variations (reviewed by \citealt{bolatto13}). However, we use these factors for a consistent comparison between these different studies, and we assume a systematic uncertainty of $\sim$\,0.5\,dex on any derived quantities related to molecular gas masses, throughout \cite[following][]{tamhane22}. 

\begin{figure}
\includegraphics[width=\columnwidth]{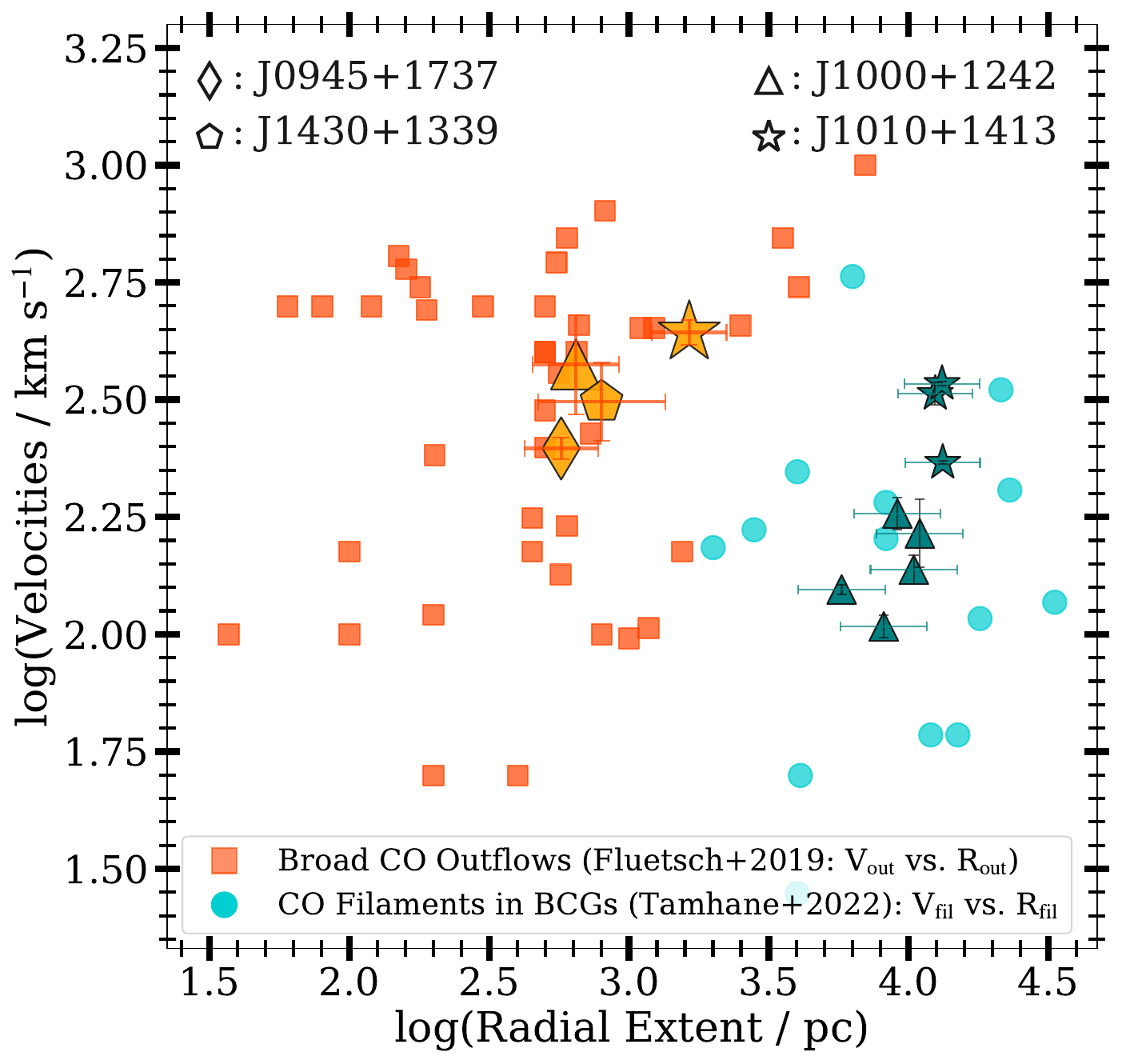}
\caption{Velocity (V$_{\mathrm{fil}}$) versus radial extent (R$_{\mathrm{fil}}$) of the filamentary structures identified in this work. Small teal triangles and stars represent the filaments in J1000+1242 and J1010+1413, respectively, and the measurements for BCGs are represented with light-blue circles (\citealt{tamhane22}). The larger orange symbols, now including J1430+1339 and J0945+1737 plotted as a pentagon and diamond, respectively, represent V$_{\mathrm{out}}$ and R$_{\mathrm{out}}$ of the central broad outflows identified using CO\,(3--2) emission-line wings (see Section \ref{sec: nucOutMap}). A comparison is shown with the AGN and starburst galaxies sample as orange squares (from \citealt{fluetsch19}).}
\label{fig6: rflow_vflow}
\end{figure}
\begin{figure*}
\includegraphics[width=\textwidth]{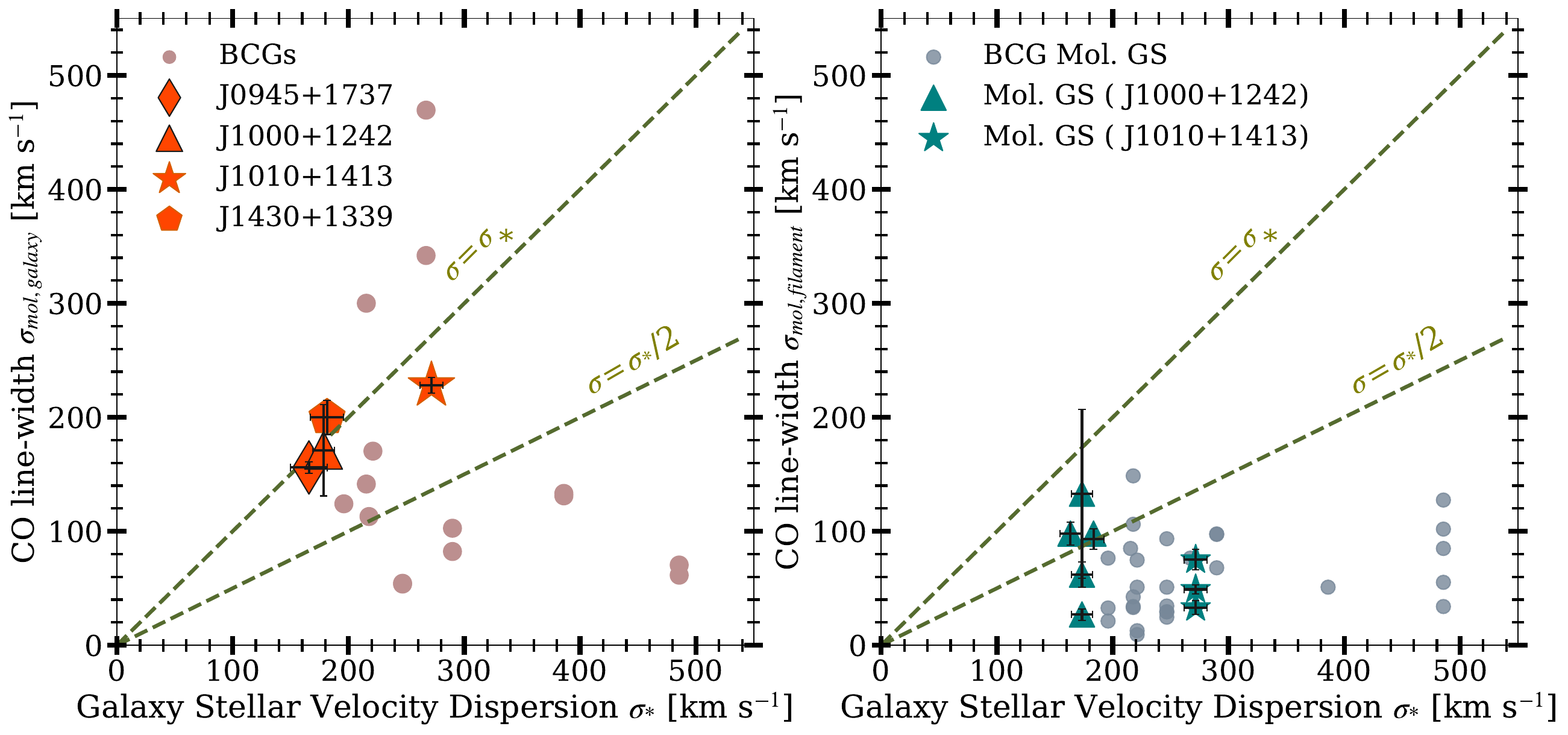}
\caption{{\em Left:} CO velocity dispersion from the galaxy-wide emission-line profile ($\sigma_{\mathrm{mol,galaxy}}$; as in Figure \ref{fig2: all4targets}) and {\em Right:} from the individual filaments ($\sigma_{\mathrm{mol,filament}}$; see spectra in Figures \ref{appfig2: filamentsZoom1} and \ref{appfig3: filamentsZoom2}) versus stellar velocity dispersion ($\sigma_{\mathrm{*}}$), for our targets (different symbols, as in legend) compared to the 10/12 BCGs sample with the appropriate values (from \citealt{russell19}; shown as small circles). The two dashed lines in each panel show a 1:1 and 1:0.5 relationship of $\sigma_{\mathrm{mol}}$:$\sigma_{\mathrm{*}}$. For stellar velocities, we obtain values from the stellar template fits for the whole galaxy (see Section \ref{sec: stelKin}). The $\sigma$ values are obtained following the method summarised in Section \ref{sec: sigmaflow}.}
\label{fig8: sigmaComp}
\end{figure*}

\subsection{Spatially mapping central outflows in molecular gas phase}\label{sec: nucOutMap}

We also aim to characterise any central molecular outflows, which are often traced with underlying wings in the CO emission-line components. Although a detailed kinematic analysis of molecular outflows is beyond the scope of this work (following e.g., \citealt{ramosAlmeida22}), we compare the CO properties to previous works that investigate such CO components, under the assumption that they are tracing outflows. We focus our comparison to \cite{fluetsch19}, who study the CO outflow kinematics for 45 active galaxies (starburst and AGN) at z$<$0.2, with L$_{\mathrm{AGN}}\,\sim\,$10$^{40-46}$\,erg\,s$^{-1}$. Therefore, we are motivated by the methods of \cite{fluetsch19} and consequently, we measure the spatial extent of the region over which CO emission-line wings are identified.

Following Section~\ref{sec: emLineProcedure}, we mapped the CO emission in the central regions, identifying pixels where two emission-line components were required. The velocity width maps within the central regions reveal broad velocity widths across all four galaxies (i.e., $\gtrsim$\,400\,km\,s$^{-1}$). This motivated us to undertake a more comprehensive analysis to identify any disturbed gas in the central regions of the four targets. The BIC-based selection was effective in selecting the required number of Gaussian components for obtaining the fits. However, acknowledging the complexities of emission-line kinematics, we also visually inspected the fits to identify the regions that show clear signs of a secondary, underlying high-velocity wing component (as opposed to two narrow components). The regions over which broad CO wings are clearly identified are shown as dashed purple boxes in the panels \textbf{a1, b1, c1, and d1} of Figure \ref{fig2: all4targets}. Using these regions, we extracted the spectral profile for studying the properties of the central outflows (shown in the respective \textbf{a3, b3, c3, and d3} panels of Figure \ref{fig2: all4targets}). We measured R$_{\mathrm{out}}$ as the projected distance between the farthest spaxel from the central spaxel over these regions. The uncertainty in the projected distance was taken as the major axis of the respective beams. 

We use the CO emission-line profiles from these central regions to obtain the velocities of the outflowing gas as: \(\mathrm{V}_{\mathrm{out}}=\mathrm{FWHM_{broad}}/2 + \mathrm{V}_{\mathrm{broad}}\), i.e., the same definition as \cite{fluetsch19}. For the uncertainty in the velocity values, we combined the uncertainties in FWHM and V$_{\mathrm{broad}}$ (see Section \ref{sec: filProps}). The central outflow properties are listed for all four targets in Table \ref{tab5: nucoutprops} and plotted in Figure \ref{fig6: rflow_vflow}. We discuss these properties later in Section \ref{sec: central_res}. We note that when we employ the same methods used by \cite{fluetsch19} (which involves a simplified approach of producing CO images over the high-velocity wings of the profiles), we obtain very similar values, and our derived values are also close to the previous studies of the CO emission for the case of J1430+1339 (see \citealt{audibert23}). In summary, our values are sufficient for the simple parameter-space comparison of CO emission-line profile properties presented in Figure~\ref{fig6: rflow_vflow}.


\section{Results and Discussion}\label{sec: results}

In this section, we present our results, and discuss their interpretation, from our analysis of the molecular gas (traced via CO\,(3--2) emission) of four quasars from the QFeedS (Figure~\ref{fig1: paramplot}). Specifically, in Section \ref{sec: res_filProps}, we summarise the properties of the identified extended molecular gas structures in terms of their morphology, radial extent, and kinematics. In Section~\ref{sec: compFilProps} we make a comparison to similar structures found in BCGs. Hence, in the first two sections we discuss the interaction of the radio lobes with the molecular gas at larger scales. This is followed by Section \ref{sec: central_res}, where we present the observations of the central molecular gas outflows. Finally, in Section \ref{sec: dualMode}, we discuss the evidence for two feedback mechanisms acting on the molecular gas, in the same targets, and discuss possible implications for an evolutionary sequence of feedback via low- and moderate-power radio jets in `radio quiet' quasars. 

\begin{figure*}
\centering
\begin{tabular}{cc}
\subcaptionbox{}{\includegraphics[width=.47\textwidth,height=9cm]{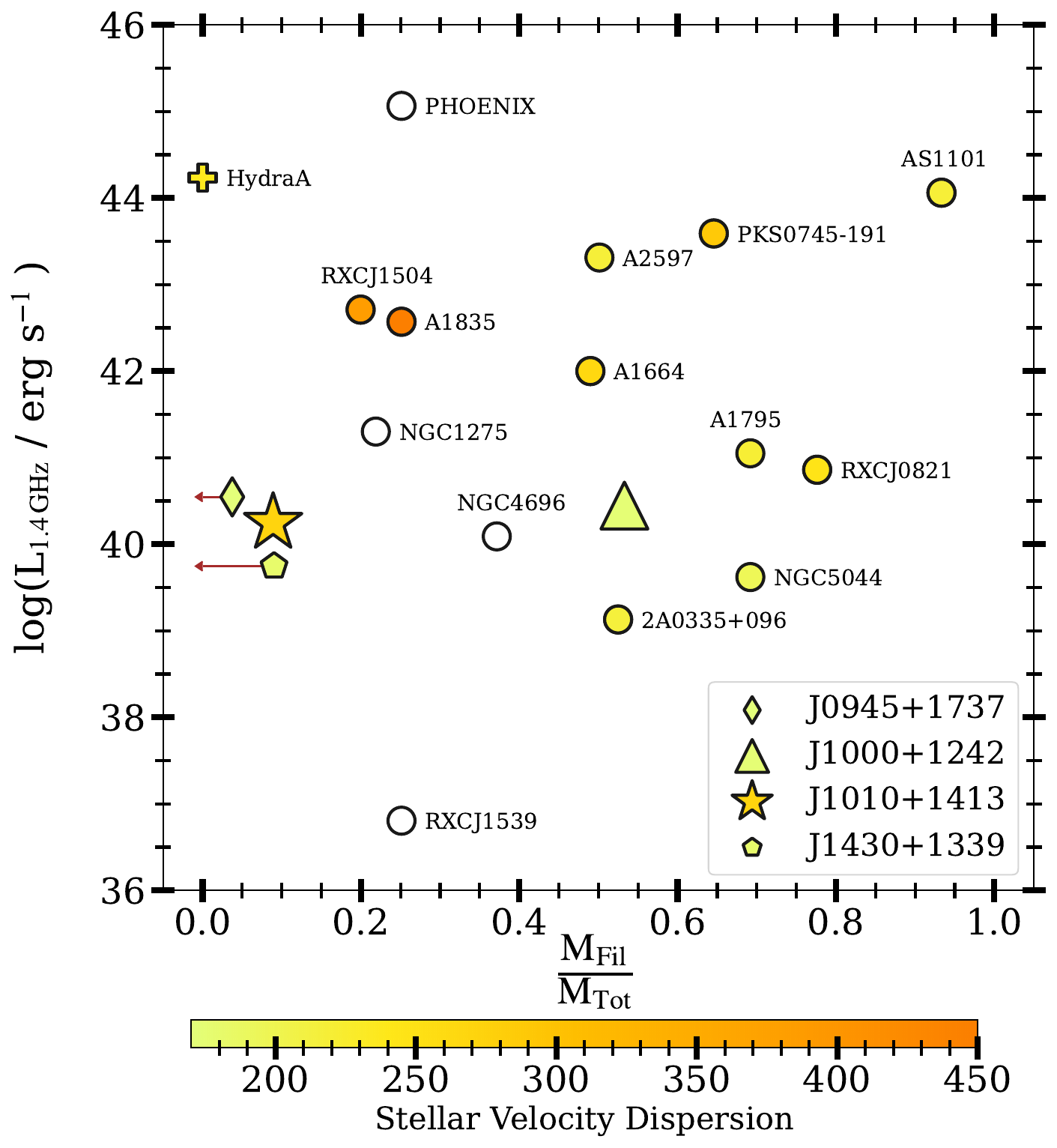}} & \subcaptionbox{}{\includegraphics[width=.47\textwidth,height=9cm]{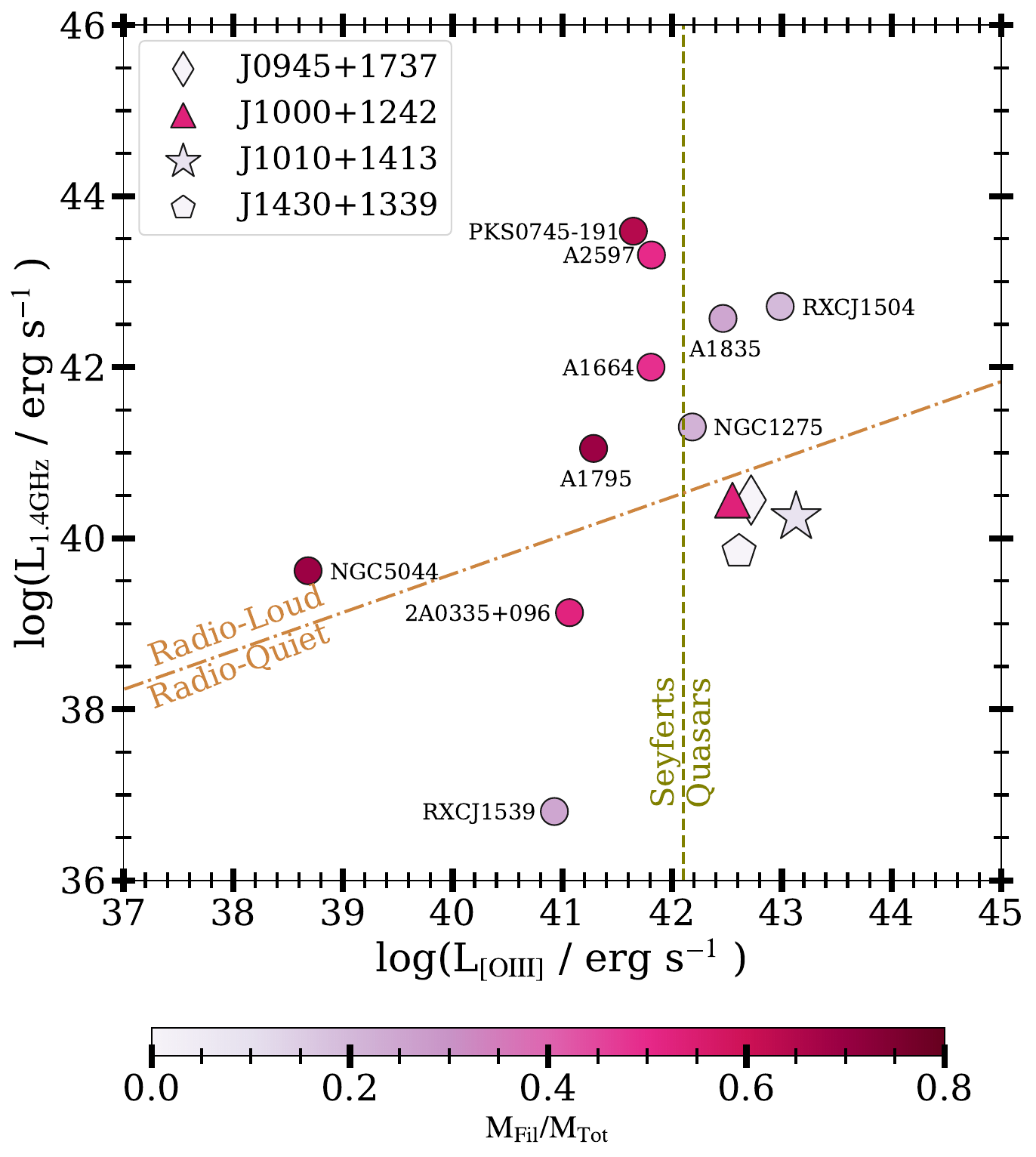}} \\
\end{tabular}
\caption{ \emph{Left}: Ratio of molecular gas in filaments to the total molecular gas as a function of 1.4 GHz radio luminosity of the respective target for both the samples: symbols as in the legend, and the 14 BCGs from \citealt{tamhane22}, represented as circles. Hydra A is also shown which has no identified filaments, through a plus symbol (\citealt{rose19}). The data points are colour-coded by stellar velocity dispersion of the galaxies, except where these data were not found (shown as empty circles). \emph{Right}: 1.4\,GHz radio luminosity versus [O\,\textsc{iii}] luminosity for the targets as in the left panel, but for the entire galaxy (excluding 4 BCGs for which L$_{\mathrm{[O III]}}$ was not available), and colour-coded by the ratio of molecular mass found in filaments to the total. The dot-dashed line separates `radio loud' and `radio quiet' sources (following \citealt{xu99}) and the vertical dashed line is the `quasar' luminosity threshold used to select our sample (i.e., L$_{\mathrm{[OIII]}}$\,$\geq$\,10$^{42.1}$\,erg\,s$^{-1}$; see Section \ref{sec: samplSelect}). }
\label{fig7: massFracs}
\end{figure*}


\subsection{Properties of the extended molecular gas structures}\label{sec: res_filProps}

Figure \ref{fig2: all4targets} reveals molecular gas in the form of extended filamentary structures for two of the four galaxies.  As presented in Section \ref{sec: filDefine}, these structures have morphologies that are typically elongated (with a median axis ratio of 2.7; see Table \ref{tab3: filmprops}). Following the terminology used for similar morphological structures seen in BCGs, we refer to these gas structures as filaments. As shown in Figure \ref{fig3: filaments_overview}, we identify five filaments for J1000+1242 and three for J1010+1413. Figures \ref{fig4: kinematics_J10001242} and \ref{fig5: kinematics_J10101413} show the velocity and velocity-width maps (in terms of W$_{80}$) over the entire CO emitting regions. A zoomed-in version of these maps, and corresponding CO\,(3--2) emission-line profiles extracted from the regions of the filaments, are provided for the individual filaments in Appendix~\ref{app: filamentMaps}. The observed filament properties are listed in Table \ref{tab3: filmprops}. 

We present values of filament velocity and radial extent in Figure \ref{fig6: rflow_vflow}, as teal-coloured triangles for J1000+1242 and stars for J1010+1413. For J1000+1242, across the 5 filaments, there is a radial extent range of R$_{\mathrm{fil}}$\,=\,5\,--\,11\,kpc and velocities of V$_{\mathrm{fil}}$\,=\,|\,100\,--\,190\,|\,km\,s$^{-1}$. In case of the 3 filaments in J1010+1413, we see a radial extent range of 12\,--\,13\,kpc with comparatively higher velocities of 220\,--\,340\,km\,s$^{-1}$. This gives us an average radial extent of 8\,kpc and 12\,kpc; and an average velocity of $\sim$\,150\,km\,s$^{-1}$ and $\sim$\,280\,km\,s$^{-1}$ for J1000+1242, and J1010+1413, respectively. 

In Figure \ref{fig8: sigmaComp}, we compare the molecular velocity dispersion of the filaments ($\sigma_{\mathrm{fil}}$) with the stellar velocity dispersion ($\sigma_{\mathrm{*}}$) of the host galaxies. The filaments show velocity dispersion values in the range of 30\,--\,130\,km\,s$^{-1}$ for J1000+1242 and 47\,--\,90\,km\,s$^{-1}$ for J1010+1413. In general, the velocity dispersion of the filaments is much lower than the stellar velocity dispersion values, with a median ratio of $\sigma_{\mathrm{fil}}$/$\sigma_{\mathrm{*}}$\,=\,0.32 across all 8 filaments. 

In the left panel of Figure \ref{fig7: massFracs}, we present the fraction of total molecular gas mass located in the filaments, with respect to the radio luminosity (L$_{\mathrm{1.4 GHz}}$), where the colour-scaling corresponds to the galaxy's stellar velocity dispersion. These ratios are simply the ratio of CO\,(3--2) flux across all filaments, divided by the total CO\,(3--2) flux for each galaxy. That is, we are assuming the same CO flux to mass conversion factor for both the filaments and total gas mass. J1000+1242 is observed to have the highest value, with 53\% of the gas located within these structures. J1010+1413 has only 9\% of the CO(3--2) emitting gas located in these structures. For J0945+1737 and J1430+1339, where no filaments were detected, we estimated upper limits for the molecular gas mass of the filaments of 4\%, and 9\% respectively. For estimating the upper limits, we used the flux ratio of the faintest detected filament (i.e., filament 3 of J1010+1413) to total galaxy flux and scaled it by the noise in the respective cubes of J0945+1737 and J1430+1339. We note that these mass ratio measurements can be affected by the sensitivity to structures on different scales, depending on the distribution of the CO\,(3--2) emitting gas. For example, we may be missing low surface brightness CO\,(3--2) emission (either contained in filaments or the central galaxy). Towards this, we compared our total CO flux measurements from the 12\,m ALMA observations with single-dish observations of the three targets detected in CO\,(3--2) in APEX data (i.e., all but J0945+1737; \citealt{molyneux23}). We found that ALMA/APEX flux ratios range from 0.53--1.3. Although this adds some additional uncertainty (at the $\sim$0.3\,dex level on these mass ratios), our measurements are sufficient for a broad comparison to the values for similar structures, obtained using similar datasets, seen in BCGs (Section~\ref{sec: compFilProps}).  

The CO\,(3--2) emitting molecular gas seen as elongated structures in the two galaxies, appear to be entrained along or around the radio bubbles seen in these targets (see Figure~\ref{fig3: filaments_overview}; Figure~\ref{fig4: kinematics_J10001242}; and Figure~\ref{fig5: kinematics_J10101413}). The surface brightness and velocity maps of the filaments reveal some clumpy sub-structures, but the velocity gradients across the filaments are relatively smooth, and are typically small (30\,--\,100\,km\,s$^{-1}$), following the major axes of the structures. Further, the more clumpy molecular gas is seen to be coincident with bends in the radio bubbles. {\bf We note that for J1000+1242, the velocity structures of the northern filaments seen in Figure~\ref{fig4: kinematics_J10001242}, could be consistent with seeing both the blueshifted (filaments 4 and 5) and redshifted parts (filament 1) of an expanding bubble.} Qualitatively, all of these morphological and kinematic structures, and their spatial connection to expanding bubbles, are similar to those we see associated with BCGs, hosted in cool core clusters that are rich in molecular gas (e.g., see \citealt{tremblay18, balmaverde18, russell19, tamhane22, capetti22}). Therefore, it is warranted to make a more quantitative comparison between the structures observed in our new observations of quasars, with those seen in BCGs, which are not classed as `radio quiet' quasars (Figure~\ref{fig7: massFracs}).


 \subsection{Comparison with BCGs}\label{sec: compFilProps}

We compare our observations of molecular gas structures with a sample of 14 BCGs (at z\,$\leq$\,0.2) compiled from \cite{tamhane22} in terms of their radial extent, velocity, and mass. For comparing in terms of velocity dispersion, we use the velocity dispersion data for only the 10 BCGs from \cite{russell19}, that are in common over both the samples. We also note that the filament properties from \cite{russell19} are provided as individual filaments, whilst for \cite{tamhane22}, they are an average over all the filaments. When referred to, we make this clear for each comparison. These works make use of ALMA to measure the properties of gas structures observed via low CO transitions (i.e., CO\,(3--2), CO\,(2--1), and CO\,(1--0)), similar to our observations and approaches. The stellar velocity dispersion, the radio fluxes, and the [O\,\textsc{iii}] luminosities for the BCGs are taken from \cite{hogan15, hogan17} and \cite{pulido18}\footnote{When radio flux and [O\,\textsc{iii}] luminosities were not available in these work, we obtained these values using NASA/IPAC Extragalactic Database (\citealt{ned}). However, for 5/15 BCGs we didn't obtain either the radio flux or the [O\,\textsc{iii}] luminosities, and hence these are excluded from the Figure \ref{fig7: massFracs}\,(right panel). For 4/15 BCGs, we could not recover the stellar dispersion values and hence these are represented as empty symbols in Figure \ref{fig7: massFracs}\,(left panel), and discussed later.}. 

The BCGs are typically massive galaxies (with $\sigma_{\star}~\approx$200\,--\,500\,km\,s$^{-1}$; see Figure~\ref{fig8: sigmaComp}; and M$_{\star}$\,=\,10$^{10.6\,-\,12.5}$\,M$_{\odot}$) and `radio loud' galaxies (see Figure \ref{fig7: massFracs}; right panel). In contrast, our sample uniquely consists of `radio quiet' quasars (see Section \ref{sec: samplSelect}; Figure~\ref{fig7: massFracs}) and has comparatively lower stellar masses (i.e., M$_{\star}$\,=\,10$^{9.9\,-\,11}$\,M$_{\odot}$; \citealt{jarvis20}). The BCGs, with known strong `radio-mode' feedback and large reservoirs of molecular gas (\(10^9-10^{11} \mathrm{M}_\odot\)) serve as an interesting comparison for our targets. This helps to explore the feedback processes across different populations, and over an extended parameter space in terms of radiative and radio luminosities (Figure~\ref{fig7: massFracs}). 

In comparison to the BCGs, the CO filaments observed in our targets have comparable properties in the V$_{\mathrm{fil}}$ vs. R$_{\mathrm{fil}}$ parameter space (Figure~\ref{fig6: rflow_vflow}). Further, we see similar velocity dispersion values of $\sim$\,10\,--\,160\,km\,s$^{-1}$, shown in the right panel of Figure \ref{fig8: sigmaComp}, across both samples\footnote{ For consistency, only the 10/12 galaxies common between \cite{russell19} and \cite{tamhane22} have been shown in this plot. The two remaining BCGs from \cite{russell19} that are not shown are A262 and A2052, which do not affect the overall scientific interpretation in this context.}. Furthermore, the vast majority of BCG filaments fall significantly lower than half of the stellar velocity dispersion, as is also seen for those in our sample.

In the left panel of Figure \ref{fig7: massFracs},\, a comparison sample of 15 BCGs is used to observe the spread in filament mass fractions. Along with the 14 BCGs comparison sample, we also add Hydra-A to this comparison list (studied in \citealt{rose19}), only for this plot, for an overall representation of the parameter space covered by the BCGs. They show a significant spread in the filament mass fraction from 0\% in Hydra-A (disk-dominated) to $\sim$90\% in AS1101 (filament-dominated). Three of the four galaxies from our sample lie towards the lower end of this filament mass-fraction, with J1000+1242 lying towards the middle at 53\%. Due to the archival and inhomogeneous nature of the BCG sample, and the small sample of `radio quiet' quasars investigated here, it is not yet possible to rigorously assess if the distribution of filament-to-total molecular mass fractions of the two populations is consistent. A more complete, systematic survey of the two populations is required. 

Overall, in Figure \ref{fig7: massFracs}, we look for trends in the observed filament-to-total molecular gas fractions, with respect to the galaxy properties, such as, radio luminosity (L$_{\mathrm{1.4GHz}}$), [O\,\textsc{iii}] luminosity (L$_{\mathrm{[OIII]}}$), and stellar velocity dispersion ($\sigma_*$) for the combined sample of BCGs plus our sample. In the left panel of Figure \ref{fig7: massFracs}, we see no clear trends with radio luminosity, nor with stellar velocity dispersion (represented by the colour-scaling). In the right panel of Figure \ref{fig7: massFracs}, we compare our targets to the BCG sample in the L$_{\mathrm{1.4 GHz}}$ vs. L$_{\mathrm{[OIII]}}$ plane; with the colour-scaling corresponding to the filament mass fraction. Our sources uniquely lie in the radio-quiet but radiative quasar regime. However, in terms of radio-luminosity, we see no obvious trend combining this sample and the BCG sample. In contrast, in terms of the radiative luminosity, the BCG sources with a higher [O {\sc iii}] luminosity (L$_{\mathrm{[OIII]}}$\,$\geq$\,10$^{42.1}$\,erg\,s$^{-1}$; in the quasar regime) have a typically {\em lower} mass-fraction of gas in the filaments ($\sim$\,22\%), compared to the higher average mass-fraction ($\sim$\,52\%) seen for those with {\em lower} [O {\sc iii}] luminosities. The average filament mass-fraction for luminous [O\,\textsc{iii}] sources reduces further when our targets are also included ($\leq$\,16\%). This appears to indicate that higher radiative power does not result in higher fractions of mass involved in these filaments. Nonetheless, we reiterate that a more homogeneous and complete investigation across both samples is required to confirm any such trend (or lack thereof) between radiative power and the fraction of molecular gas located in filamentary structures. 

\subsection{Central multi-phase outflows}\label{sec: central_res}
All four quasars are already known to contain central ionized gas outflows (traced by broad emission-line components) from previous work (see \citealt{harrison14, harrison15, ramosAlmeida17, speranza22, venturi23}), with velocities reaching 1000\,km\,s$^{-1}$, and extending to spatial extents of $\gtrsim$1--10\,kpc. In all cases, the jet-ISM interactions have been proposed as an important driving mechanism of outflows and turbulence, with possible additional contributions from quasar-driven winds.

As reported in Section \ref{sec: nucOutMap}, we have found evidence of high-velocity wings in the CO\,(3--2) emission line profiles in the central regions of the galaxies. Such signatures are attributed to outflowing molecular gas in \cite{fluetsch19}. Following their definition (also see Section~\ref{sec: nucOutMap}), we observe the velocities of the central disturbed gas to be V$_{\rm out}$\,=\,249, 376, 441, and 331 km\,s$^{-1}$ for J0945+1737, J1000+1242, J1010+1413, and J1430+1339, respectively. We measure the projected radial extent of this outflowing phase to be R$_{\rm out}$\,=\,0.60, 0.65, 1.64, 0.80\,kpc, for these same targets. In Table \ref{tab5: nucoutprops}, we summarise all the properties, along with the measured uncertainties of the central outflows. We note that for J1430+1339, similar properties of outflowing molecular phase using the CO\,(3--2) and CO\,(2--1) emitting gas were presented in  \cite{ramosAlmeida17} and \cite{audibert23}, which is attributed to the inner radio jet seen in this source \citep[][]{harrison15,jarvis19}. These molecular outflow components are less extreme in both velocity and spatial extent than seen in the corresponding ionized gas, in agreement with the multi-phase study of \cite{girdhar22} for a different QFeedS target. 

The sample studied by \cite{fluetsch19} covers an AGN luminosity range of 10$^{40-46}$\,erg\,s$^{-1}$; in comparison to our sample that lies in the higher quasar luminosity regime, i.e., 10$^{45-46}$\,erg\,s$^{-1}$. Further, both the samples lie in the same redshift range, z$\,\sim\,$0.2. In Figure~\ref{fig6: rflow_vflow} we show that the velocities and radial extents of molecular outflow properties estimated by \cite{fluetsch19} (shown as orange squares) lie closely in the parameter space to our observed central outflow properties (yellow symbols), despite the differences in AGN luminosities.

\begin{figure*}
\centering
\begin{tabular}{cc}
\subcaptionbox{}{\includegraphics[width=.47\textwidth,height=8cm]{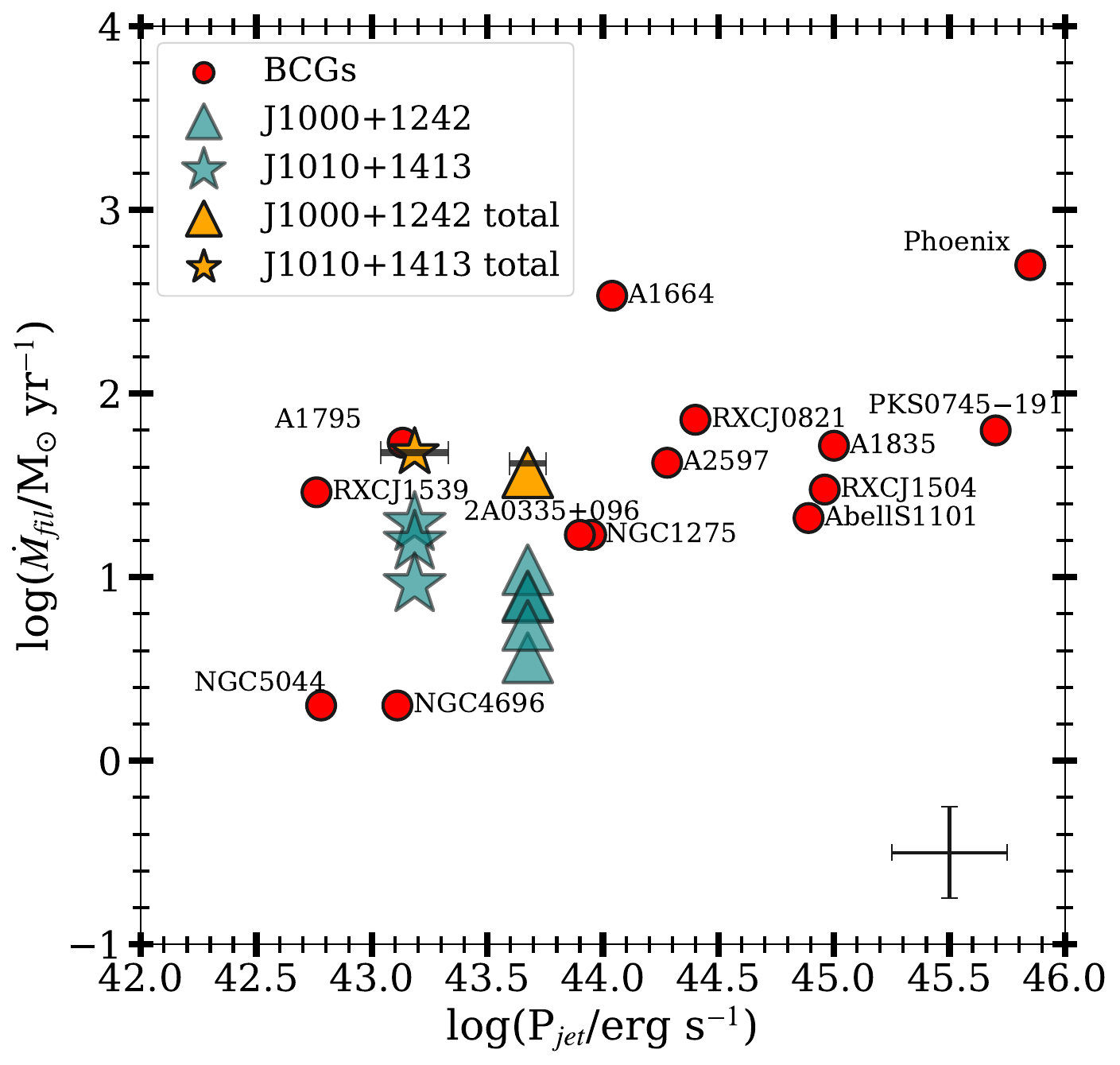}} & \subcaptionbox{}{\includegraphics[width=.47\textwidth,height=8cm]{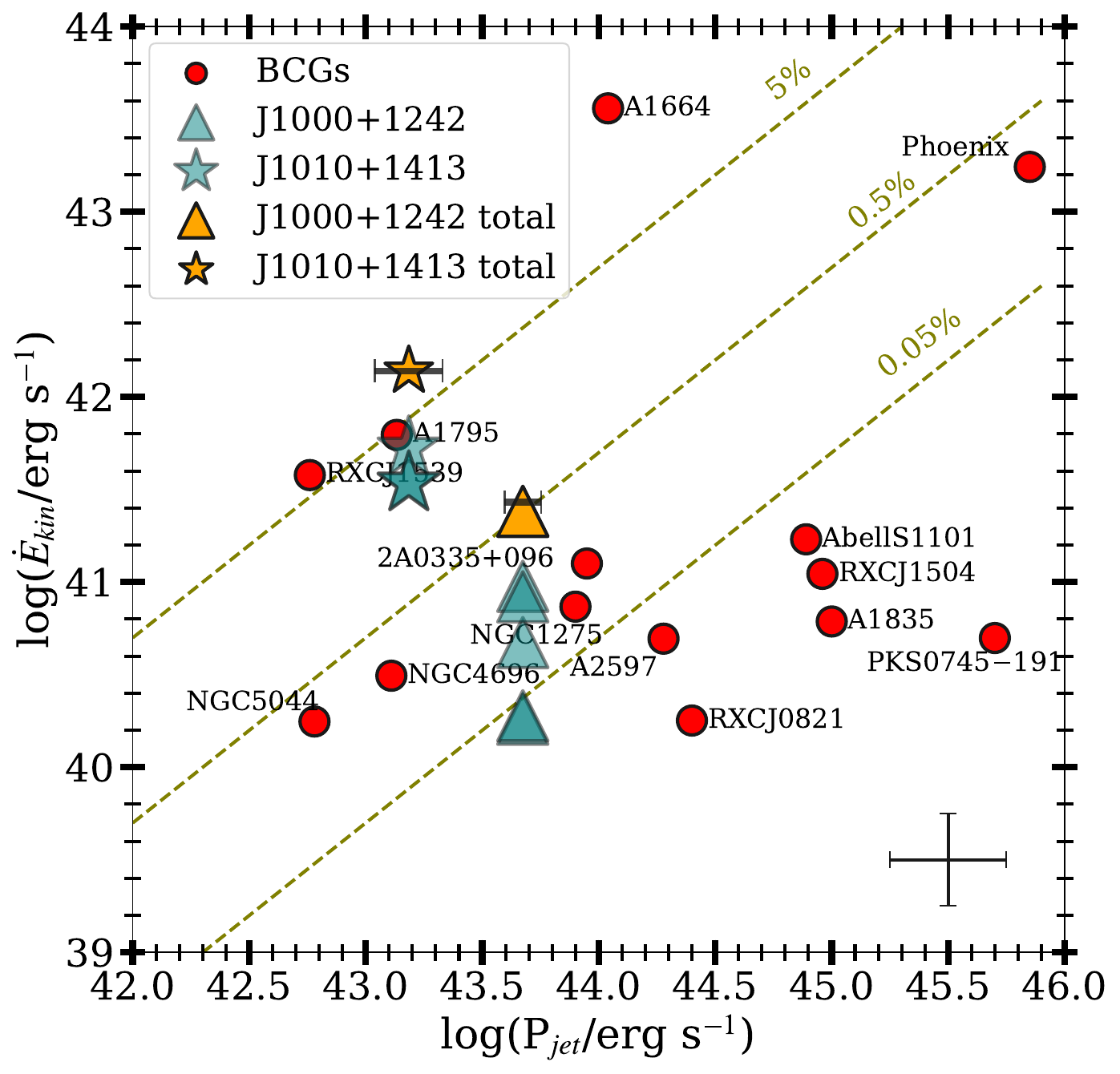}} \\ 
\end{tabular}
\caption{\emph{Left panel:} Rate of Molecular Mass Outflow vs. the Jet Kinetic Power (P$_{\mathrm{jet}}$) for our filaments and the 14 comparison BCG. \emph{Right panel:} The rate of kinetic energy vs. the jet kinetic power (P$_{\mathrm{jet}}$). The teal symbols show the individual filaments for our targets. The yellow symbols show the total rates over all the filaments for each galaxy and the red circles show the values for the BCG sample (which are averaged over multiple filaments). The black error bar shows the variation in the jet power depending on whether the core or total radio luminosity was used. The dashed olive lines on the right correspond to coupling efficiencies of 5\%, 0.5\%, and 0.05\%. In the bottom-right the average systemic error assumed for each derived value is shown.}
\label{fig9: energetics}
\end{figure*}


\subsection{Feedback on two spatial scales}\label{sec: dualMode}

With a goal to understand the relative importance of feedback processes in different populations, a recent study by \cite{tamhane22} compared the properties of molecular filaments (located around radio bubbles) in $z\le0.2$ BCGs with the properties of high-velocity wings observed in CO emission lines (as a tracer of central molecular outflows) in the sample of $z\le0.2$ AGN and starburst galaxies in \cite{fluetsch19}. They conclude that radio feedback is generally more effective at lifting the gas in galaxies compared to the AGN and starburst winds. However, as acknowledged by \cite{tamhane22}, there is not a systematic investigation of possible `radio feedback' in the \cite{fluetsch19} sample. Nor is the same CO broad wing analyses, as performed by \cite{fluetsch19}, applied to the BCG sample. In our study we have searched for {\em both} types of molecular gas features (extended filaments and central outflows) in our sample of four `radio quiet' quasars. In Section~\ref{sec: res_filProps} and Section~\ref{sec: compFilProps} we showed that two of the four targets show molecular gas filaments located around radio lobes, with similar properties to those seen in BCGs. Further, in Section \ref{sec: central_res} we show the presence of central outflows in all 4 of the sample. In this section, we discuss the implications for the observed feedback effects on multiple scales.

\subsubsection{Radio lobes impact on $\sim$10\,kpc scale molecular gas}\label{sec: impactLargeScale}

Possible explanations of the molecular gas structures seen around the radio bubbles (which contain radio jets; see Figure \ref{fig3: filaments_overview}), is either a thin cover of clumpy molecular gas, expanding along with the expanding radio bubbles, or molecular gas that is in-situ condensed in the updrafts \citep[e.g.,][]{mcnamara14,mcnamara16,russell19, zanchettin23}. The gas is then expected to appear the brightest around the edges of the bubbles, aligned with the line of sight, thus giving a filamentary appearance. In general, the filaments are observed to have slow velocities and narrow velocity widths (Figures~\ref{fig6: rflow_vflow}, \ref{fig8: sigmaComp}; and also in \citealt{russell19,tamhane22}). This could be because they retain the velocity structure of the rising bubbles which themselves may be relatively cool, and not shock-heated (\citealt{mcnamara00,fabian00}) as opposed to the typically energetic jet-ISM interactions. 

Whilst we can not be conclusive about the origin of the molecular gas structures observed, following \cite{tamhane22}, we assume the off-nuclear molecular gas structures as a flow, noting that some of this gas may be flowing towards the central galaxies as opposed to a pure outflow (see \citealt{russell16,balmaverde22}). \cite{tamhane22} noted that these molecular flows in BCGs are 1--3 orders of magnitude larger than the central outflows found in \cite{fluetsch19}. In two of our sample, we find evidence of both types of molecular gas structures in the same sources. We find that those molecular gas structures identified around the radio lobes are roughly an order of magnitude larger in size. However, it is important to note the approaches taken to search for these types of flows are very different and are somewhat biased by the requirement to have large radio bubbles outwith the central molecular gas disks. 

Again, following \cite{tamhane22}, we compute the mass flow rate for the filaments ($\dot{\mathrm{M}}_{\mathrm{fil}}$) as the molecular mass in the filament (M$_{\mathrm{fil}}$) divided by the time (t$_{\mathrm{fil}}$) it would have taken the filament to reach the projected radial extent (R$_{\mathrm{fil}}$) at the velocity of the filament (V$_{\mathrm{fil}}$); which is computed as t$_{\mathrm{fil}}$\,=\,R$_{\mathrm{fil}}$/V$_{\mathrm{fil}}$. We compute the kinetic power of the filament as: $\dot{\mathrm{E}}_{\mathrm{kin}}$\,=\,1/2\,$\dot{\mathrm{M}}_{\mathrm{fil}}$\,V$_{\mathrm{fil}}^2$. For all these derived quantities, we assume a systemic error of 0.5 dex on our computed values (following \citealt{tamhane22}). These derived values for the filaments are listed in Table \ref{tab4: fildprops}.

In Figure \ref{fig9: energetics}, we compare the filaments in our targets to BCGs in terms of their mass outflow rates (left panel) and kinetic powers (in the right panel) in relation to estimated jet powers. We obtained the jet power P$_{\rm jet}$ using the \cite{merloniHeinz07} relation. For each target, a range of jet powers were calculated using the 5.2\,GHz radio luminosity corresponding to the radio core (component HR:\,A, in \citealt{jarvis19}) and across all the radio structures combined (LR:\,Total, in \citealt{jarvis19}). The variation in jet powers depending on whether the core luminosity was used or the total radio luminosity was used, varies by 0.08\,dex and 0.15\,dex for J1000+1242 and J1010+1413, respectively. We use the median of these two values for our primary calculations and data points and the range for an error bar in the figures (also quoted in Table~\ref{tab2: galprops}). In Figure \ref{fig9: energetics}, individual filaments are shown as triangles for J1000+1242 and as stars for J1010+1413, and the total for all filaments is represented using yellow symbols. 

It can be seen that the mass outflow rates in the filaments for our targets (4\,--\,20\,M$_{\odot}$\,yr$^{-1}$) are comparable to the BCGs. At the same time, the kinetic power in the filaments (10$^{40-42}$\,erg\,s$^{-1}$) also seems to be comparable to the BCGs. However, similar to the case of the BCGs,  the kinetic energy transferred in the filaments is typically a small fraction of the energy available in the radio jet. The rate of energy transferred to the filaments, under these assumptions, is observed to be lower than 5\,\% percent of the jet kinetic power (consistent with all but \textbf{2} of the BCGs). Summing over all filaments, for our targets, this suggests a jet coupling efficiency of \textbf{0.0005\,--\,0.04, i.e., 0.05\,--\,4\,\%} for our sample, which is significantly higher than if we were to assume the AGN bolometric luminosity was responsible for driving these flows, i.e., $\dot{E}_{\rm kin}/L_{\rm bol}\sim$\textbf{\,10$^{-6}$ i.e., \,10$^{-4}$\,\%} (see Table \ref{tab4: fildprops}).

This brings us to the conclusion that Figure \ref{fig6: rflow_vflow} does not necessarily always correspond to two different galaxy populations for different feedback mechanisms on the molecular gas, i.e., where radiative energy (`quasar') drives central CO outflows in luminous, high accretion rate AGN, and radio jets drive molecular flows in typically `radio loud', low accretion rate sources.  In our sample, we see both central outflows and turbulence (in all four galaxies) and large-scale filamentary structures (in two). Furthermore, despite the high radiative output, it appears that radio jets and lobes also have a significant role in the impact on the molecular gas on multiple scales. 


\subsubsection{Dual feedback effects and evolution of moderate power radio jets}\label{sec: evolving_feedback}

We have found two different types of impact on molecular gas in `radio quiet' quasars, acting on different scales. Indeed, theoretical studies do show that specific AGN physical mechanisms are expected to result in distinct concomitant forms of AGN feedback, operating on different spatial and temporal scales for example: a) ultra-fast outflows/small-scale winds (\citealt{costa14,costa20}), (b) jets (\citealt{talbot22,talbotSijacki22}), (c) radiation pressure (\citealt{costa18}) and even the more phenomenological AGN 
feedback models used in state-of-the-art cosmological boxes (e.g. \citealt{zinger20}).

In the case of radio jets, even low- and moderate-power jets, have gained recognition as potentially causing significant disturbance through direct jet-ISM interactions from observations (\citealt{alatalo11, tadhunter14, morganti15, venturi21, girdhar22, morganti23,nandi23}). Hydrodynamic simulations of jets (see \citealt{sutherlandbicknell07,wagner12,mukherjee16,mukherjee18,mandal21,talbot22,talbot23,tannerWeaver22,talbotSijacki22}) have studied the progression of a jet through a clumpy interstellar medium to understand its impact on the ISM. 

Through these works comes a possible evolutionary sequence of jet-ISM interactions as motivated by both simulations and observations (see \citealt{morganti23}). On small spatial scales (and shorter timescales), the jet directly interacts with the ISM causing turbulence, and as the jet propagates, it grows along with the cocoon of shocked and heated gas and ISM plasma, that takes over at larger scales. This may cause the molecular gas to couple and rise in the wake of this growing radio bubble to greater radial extents. On larger scales ($\sim$\,5\,--\,10\,kpc), the feedback may hence be moderated by the jet-cocoon which heats or disperses the molecular gas causing cavities and pushing it aside, which is then observed as filaments. However, we cannot be certain about the origin of the filaments and it could also be possible they were cooled in situ around the radio bubbles (see \citealt{russell19}). Furthermore, a single radio event can continue to drive gas outwards for a long time after a quasar of similar power has shut down (see discussion in \citealt{tamhane22}). We suggest the presence of two ongoing effects in our targets, i.e., (i) effects on central scales due to ongoing accretion activity; and (ii) effects due to the impact of the larger-scale radio bubbles. We suspect the latter may take over as the dominant mechanism on longer timescales, as is also observed in the case of more evolved BCGs as the dominant feedback mechanism.

Our observations stem from a small number of objects and need to be confirmed by a larger sample. One significant step towards this would be to quantify the feedback from jets depending on their properties of inclination, power, and evolutionary stage. Additionally, it will be important to establish the relative importance of jets, quasar-driven winds, and radiation pressure for driving multi-phase outflows and turbulence across a homogenous sample. Further, obtaining observations of other resolved CO transitions could be crucial for deriving their excitation and physical properties and will also help unveil the full ``population'' of filaments. 



\section{Conclusions}\label{sec: conclusions}

We present the study of the molecular gas properties in four z$<$0.2, `radio quiet' type 2 quasars from QFeedS (L$_{\mathrm{bol}}\sim10^{45.3-46.2}$\,erg\,s$^{-1}$; L$_{\mathrm{1.4\,GHz}}\sim10^{23.7-24.3}$\,W\,Hz$^{-1}$), namely, J0945+1737, J1000+1242, J1010+1413, J1430+1339. These targets were selected based on the availability of high spatial-resolution ALMA data, to trace the CO(3--2) emission, and their projected radio linear sizes of LLS$_{\rm radio}$\,$\geq$\,10\,kpc (see Figure \ref{fig1: paramplot}, and Table \ref{tab1: litprops} for an overview of properties). We explored the kinematics and spatial distribution of the CO\,(3--2) emission with $\sim$\,0.33\,-\,1.09\,kpc spatial resolution. This was compared to the morphology seen in 6\,GHz radio images, previously obtained from the VLA (see Figure \ref{fig2: all4targets} for a data overview). Our main findings are summarised below: \\

\textbf{1. We identify filamentary molecular gas structures in and around $\sim$10\,kpc radio lobes in two out of the four `radio quiet' quasars (Figure \ref{fig3: filaments_overview}, \ref{fig4: kinematics_J10001242}, and \ref{fig5: kinematics_J10101413}}). Both J1000+1242 and J1010+1413 show filamentary molecular gas structures that appear to wrap around the radio lobes. They have maximal radial extents of 5\,--\,13\,kpc, velocities of V$_{\mathrm{fil}}$\,=\,|\,100\,--\,340\,| km\,s$^{-1}$ and velocity dispersion values of $\sigma_{\mathrm{mol,filament}}$\,=\,30\,--\,130\,km\,s$^{-1}$. We observe that $\sim$\,53\% and $\sim$\,9\% of the total molecular gas mass is contained within these structures for J1000+1242 and J1010+1413, respectively. For J0945+1737 and J1430+1339, we do not see any such structures, but estimate a maximum of 4\% and 9\% of the total CO\,(3--2) emitting gas could be contained in such structures (see Figure \ref{fig7: massFracs}; left panel). Our observations of radial extents and mass fractions are in close agreement to simulations of low- and moderate-power radio jets predicting the relation between the radial extent and the mass of molecular gas being driven (see Figure 20 in \citealt{mukherjee16}). \\

 \textbf{2. The molecular gas filaments in these `radio quiet' quasars have properties comparable to those seen driven by radio jets in, predominantly radio-loud, BCGs (Figures \ref{fig6: rflow_vflow}, \ref{fig7: massFracs}, \ref{fig8: sigmaComp}, and \ref{fig9: energetics}}). The velocities, velocity dispersion and maximal radial extent of the molecular filaments from our `radio quiet' quasars show very similar values to those seen in BCGs, for which similar analyses have been performed (see Figures \ref{fig6: rflow_vflow} and \ref{fig8: sigmaComp}). 
 
 The inferred mass outflow rates (4\,--\,20\,M$_{\odot}$\,yr$^{-1}$) and kinetic powers (10$^{40.3-41.7}$\,erg\,s$^{-1}$) are also comparable to those seen in the BCGs (Figure \ref{fig9: energetics}). Combining our sample with the BCGs, we observe no obvious trends with stellar velocity dispersion or radio luminosity and the fraction of mass found in filaments. Although limited by source statistics, there is tentative evidence that less [O\,\textsc{iii}] luminous sources (i.e., L$_{\rm [O III]}\leq10^{42}$\,erg\,s$^{-1}$) tend to show higher fractions of the molecular mass in filaments, with an average filament mass-fraction of 55\,\%, while the higher-[O\,\textsc{iii}] luminous sources, including our targets, have an average filament mass-fraction of $\leq$16\,\% (see Figure \ref{fig7: massFracs}; right panel). This may tentatively suggest kinetic power to be a more compelling driver of these molecular filaments than radiative power at larger scales ($\sim$\,10\,kpc). \\

\textbf{3. Evidence for both central molecular outflows and large-scale radio feedback on the molecular gas in `radio quiet' quasars (Figure \ref{fig4: kinematics_J10001242}, \ref{fig5: kinematics_J10101413}, and \ref{fig6: rflow_vflow})}. In all four quasars, we see evidence for central (0.6\,--\,1.6\,kpc), outflows traced by high-velocity wings (V$_{\mathrm{out}}$\,=\,250\,--\,440\,km\,s$^{-1}$) of the CO(3--2) emission-line profiles (see Figure \ref{fig4: kinematics_J10001242} and \ref{fig5: kinematics_J10101413}). These have properties comparable to those seen for the archival AGN and starburst galaxies (of typically lower L$_{\mathrm{bol}}$; \citealt{fluetsch19}). This adds to the evidence for central multi-phase outflows in these systems, likely caused by an interaction between moderate-power radio jets and the ISM. 

We have shown that both central molecular outflows, typically associated with luminous AGN, and molecular gas filaments around radio lobes on $\sim$10\,kpc-scales (analogous to those found in BCGs), can also be found in `radio quiet' quasars. This implies that both feedback mechanisms can act within the same systems. Our observations are consistent with recent simulations and observations (e.g., \citealt{talbot22, morganti23}) that suggest that two feedback effects can take place due to low- and moderate-power radio jets (P$_{\mathrm{jet}}\lesssim10^{44}$\,erg\,s$^{-1}$). On the smaller scales, jet-ISM interactions can drive turbulence and central outflows. On larger scales, radio lobes have penetrated beyond galaxy disks and cause a more gentle pushing aside of molecular gas. Unlike high-power jets that escape swiftly, low- and moderate-power jets are seen to be trapped for longer in simulations. Their effect is amplified due to the development of an energy bubble which is basically a cocoon of a shocked ISM and plasma. This interaction leads to the constant stirring of the ISM with the energy bubble, thus inhibiting the star formation (see \citealt{mukherjee16}).

Our results underscore that the availability of higher radiative energy as we see in quasars, does not necessarily imply that it would also be the dominant feedback mechanism on all spatial scales. We should therefore take caution in assuming different dominant mechanisms, based on if a quasar is `radio quiet' or `radio loud'. A radio jet, even if low in power, has considerable potential to couple with the galaxy ISM and lead to a significant impact on the host galaxy gas (\citealt{mukherjee16,mukherjee18,meenakshi22}). We acknowledge that all four studied targets are different to some extent and the two targets showing the presence of filaments are themselves quite unalike. To draw firm conclusions about the population as a whole, calls for a larger study, across a wider sample, to understand: how common each of these different mechanisms are; the relative importance of radio jets compared to radiative processes; and what AGN or galaxy properties determine the amount of molecular mass associated with both central outflows and larger-scale radio lobe interactions.


\section*{Acknowledgements}
We thank the referee for their valuable comments. CMH \& AN acknowledge funding from the United Kingdom Research and Innovation grant (code: MR/V022830/1). ACE acknowledges support from Science and Technology Facilities Council (STFC) grant ST/P00541/1. EPF is supported by the international Gemini Observatory, a program of NSF’s NOIRLab, which is managed by the Association of Universities for Research in Astronomy (AURA) under a cooperative agreement with the National Science Foundation, on behalf of the Gemini partnership of Argentina, Brazil, Canada, Chile, the Republic of Korea, and the United States of America. MB acknowledges funding support from program JWST-GO-01717, which was provided by NASA through a grant from the Space Telescope Science Institute, which is operated by the Association of Universities for Research in Astronomy, Inc., under NASA contract NAS 5-03127. PK \& SS acknowledges the support of the Department of Atomic Energy, Government of India, under the project 12-R\&D-TFR-5.02-0700. SS acknowledges financial support from Millenium Nucleus NCN19\_058 (TITANs).

We thank Helen Russell for helping us obtain the values for the CO velocity widths.     This paper makes use of the following ALMA data: ADS/JAO.ALMA\#2016.1.01535.S and ADS/JAO.ALMA\#2018.1.01767.S. ALMA is a partnership of ESO (representing its member states), NSF (USA) and NINS (Japan), together with NRC (Canada), MOST and ASIAA (Taiwan), and KASI (Republic of Korea), in cooperation with the Republic of Chile. The Joint ALMA Observatory is operated by ESO, AUI/NRAO and NAOJ.  This research has made use of the NASA/IPAC Extragalactic Database (NED), which is funded by the National Aeronautics and Space Administration and operated by the California Institute of Technology. This research mainly uses the Python packages: Astropy,\footnote{http://www.astropy.org} a community-developed core Python package for Astronomy \citep{astropy2013, astropy2018}; SciPy (\citealt{virtanen20}); NumPy (\citealt{harris20}); Matplotlib (\citealt{hunter07}). 

\section*{Data Availability}
The MUSE and ALMA data presented in this analysis were accessed from the ESO and ALMA archives under the proposal ids: 0103.B-0071, 0102.B-107, and 0104.B-0476 for the MUSE data and ADS/JAO.ALMA\#2016.1.01535.S and ADS/JAO.ALMA\#2018.1.01767.S for the ALMA data. The VLA images used in this work are available at Newcastle University's data repository (\url{https://data.ncl.ac.uk}) and can also be accessed through our \href{https://blogs.ncl.ac.uk/quasarfeedbacksurvey/data/}{Quasar Feedback Survey website}. 


\bibliographystyle{mnras}
\bibliography{citations_list} 

\bsp	
\label{lastpage}

\appendix

\section{Individual Filament Narrow-band Images}\label{app: filSlices}

\begin{figure*}
\centering
\begin{tabular}{cc}
{\includegraphics[width=.32\textwidth]
{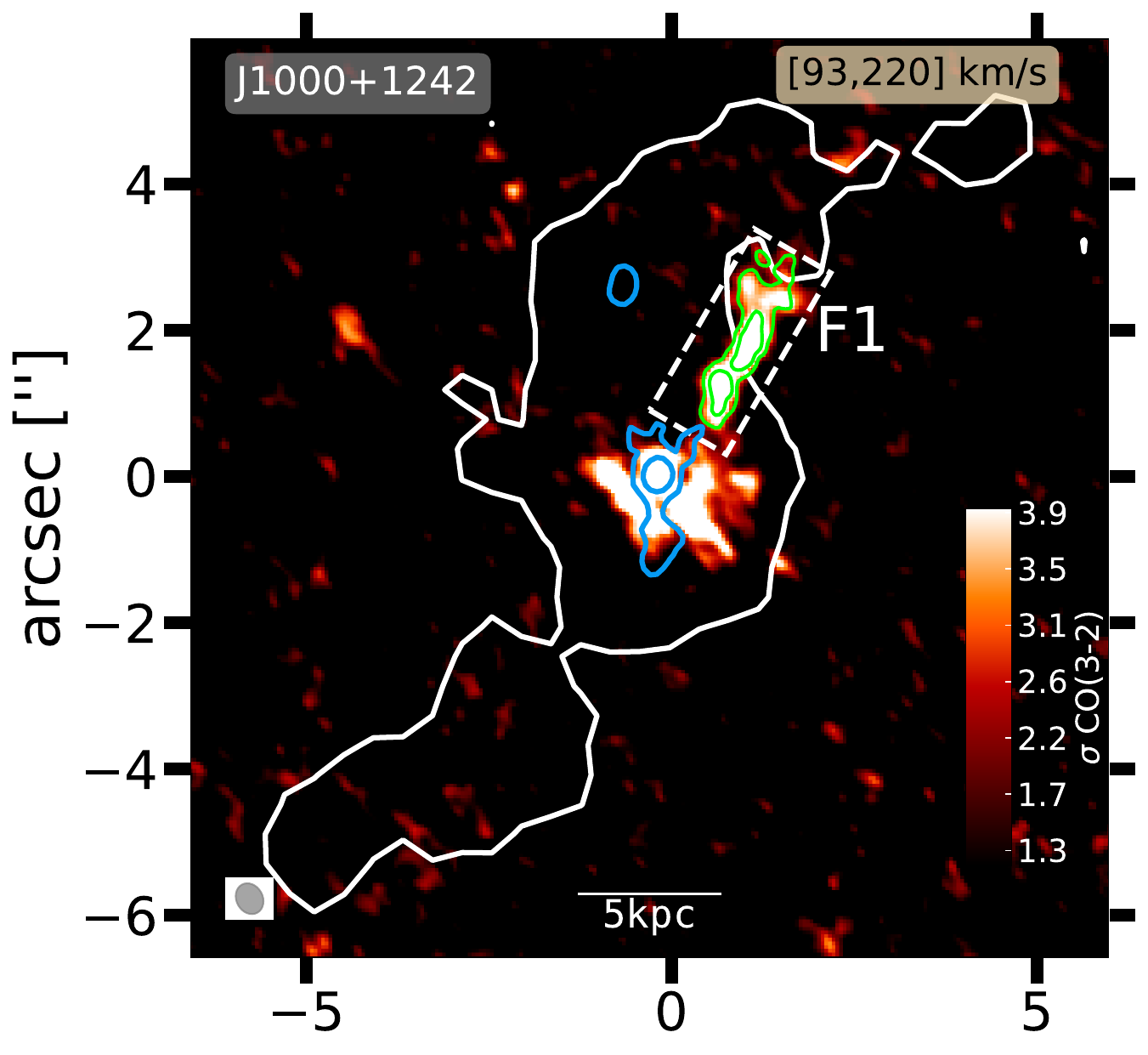}} & {\includegraphics[width=.32\textwidth]
{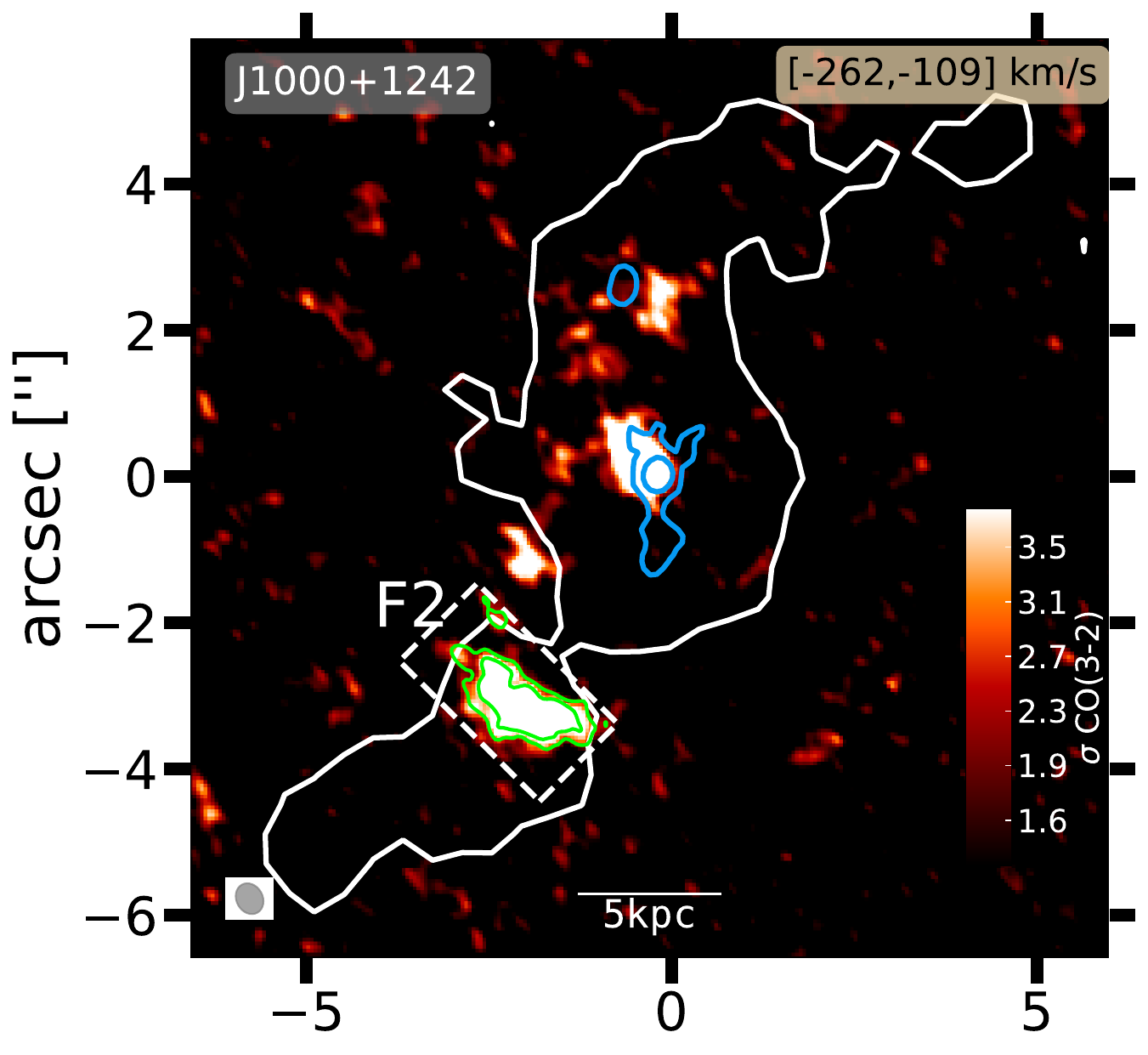}} \\
{\includegraphics[width=.32\textwidth]
{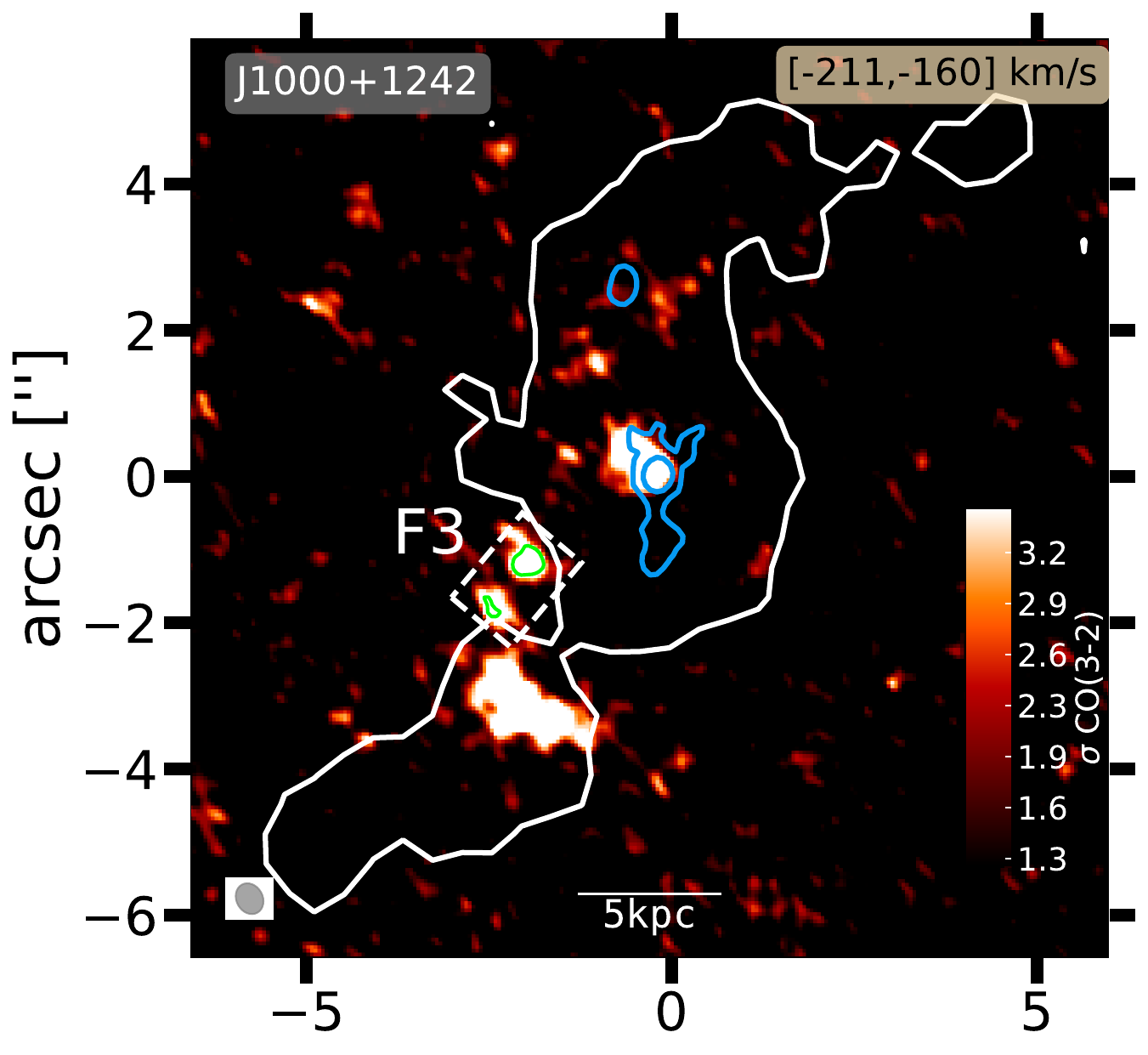}} & {\includegraphics[width=.32\textwidth]
{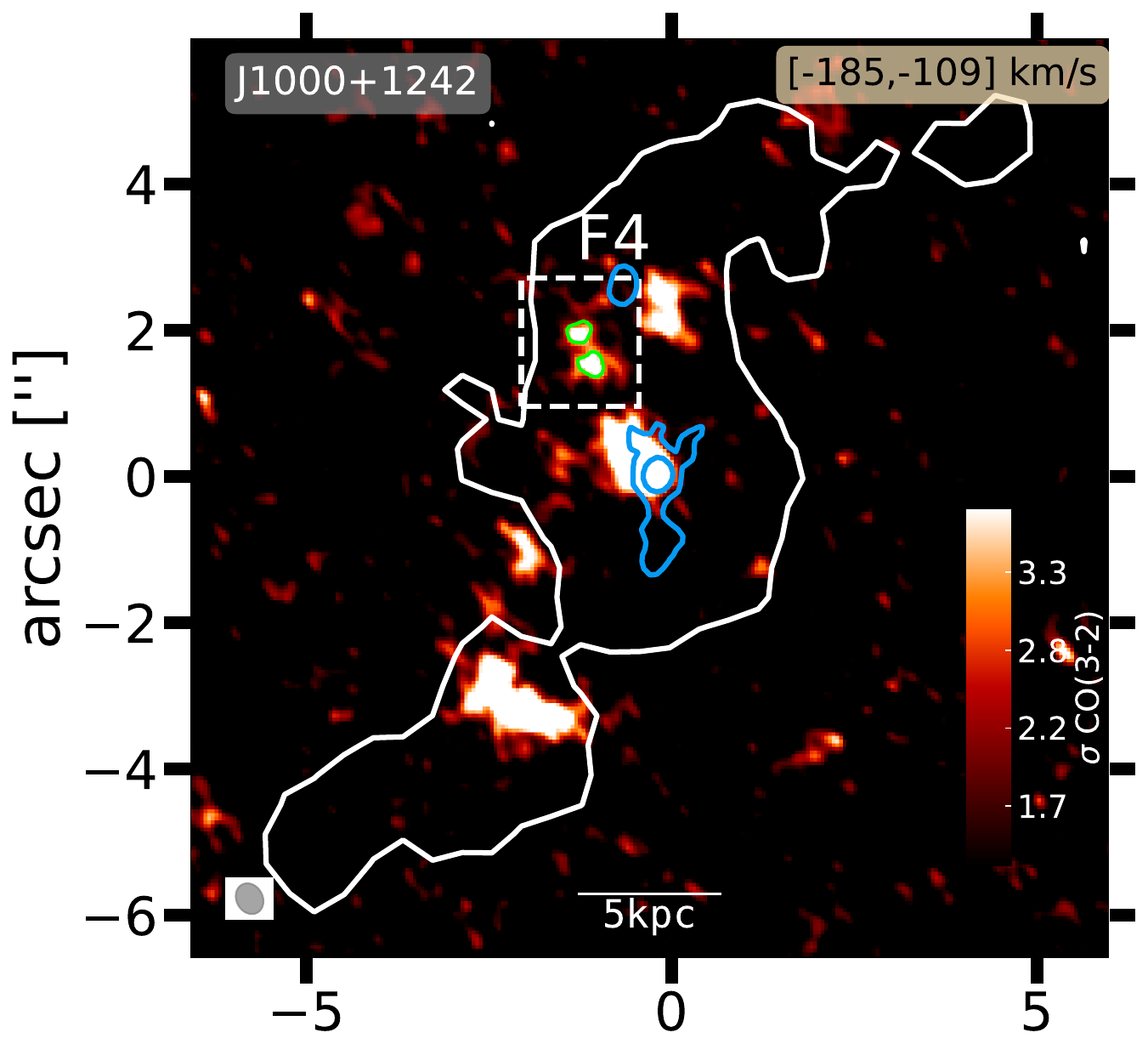}} \\
{\includegraphics[width=.32\textwidth]
{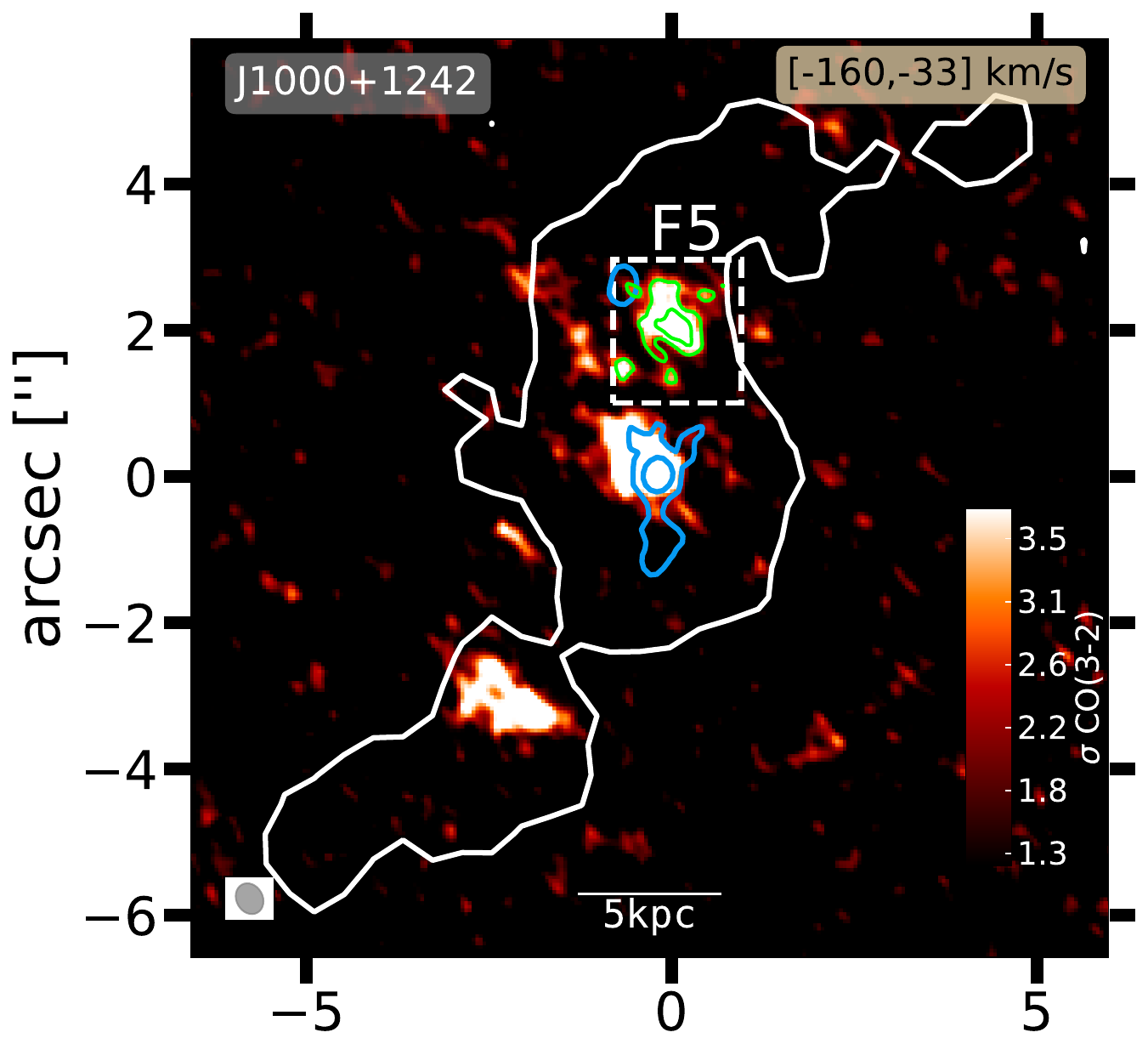}} & {\includegraphics[width=.32\textwidth]
{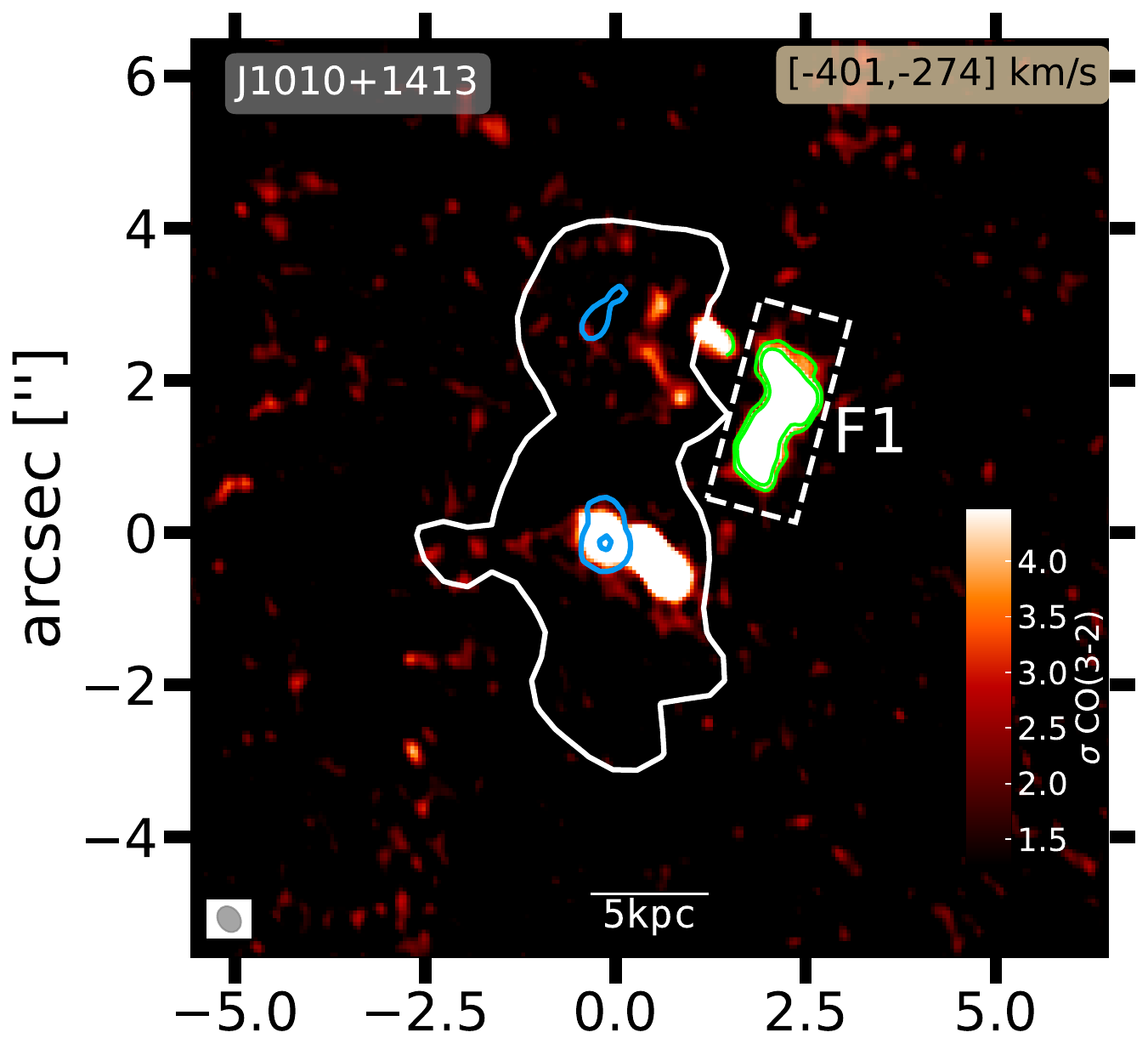}} \\
{\includegraphics[width=.32\textwidth]
{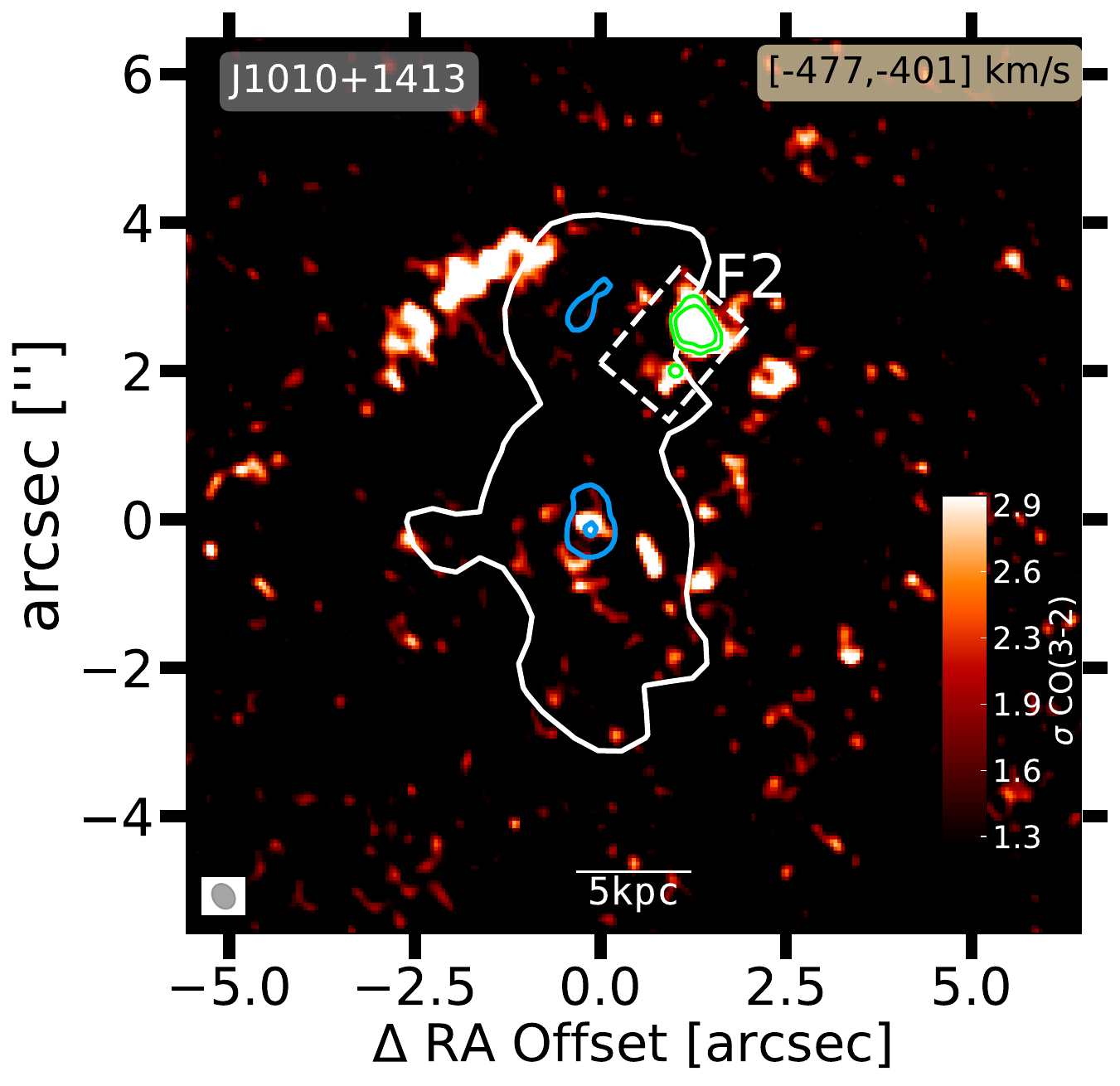}} & {\includegraphics[width=.32\textwidth]
{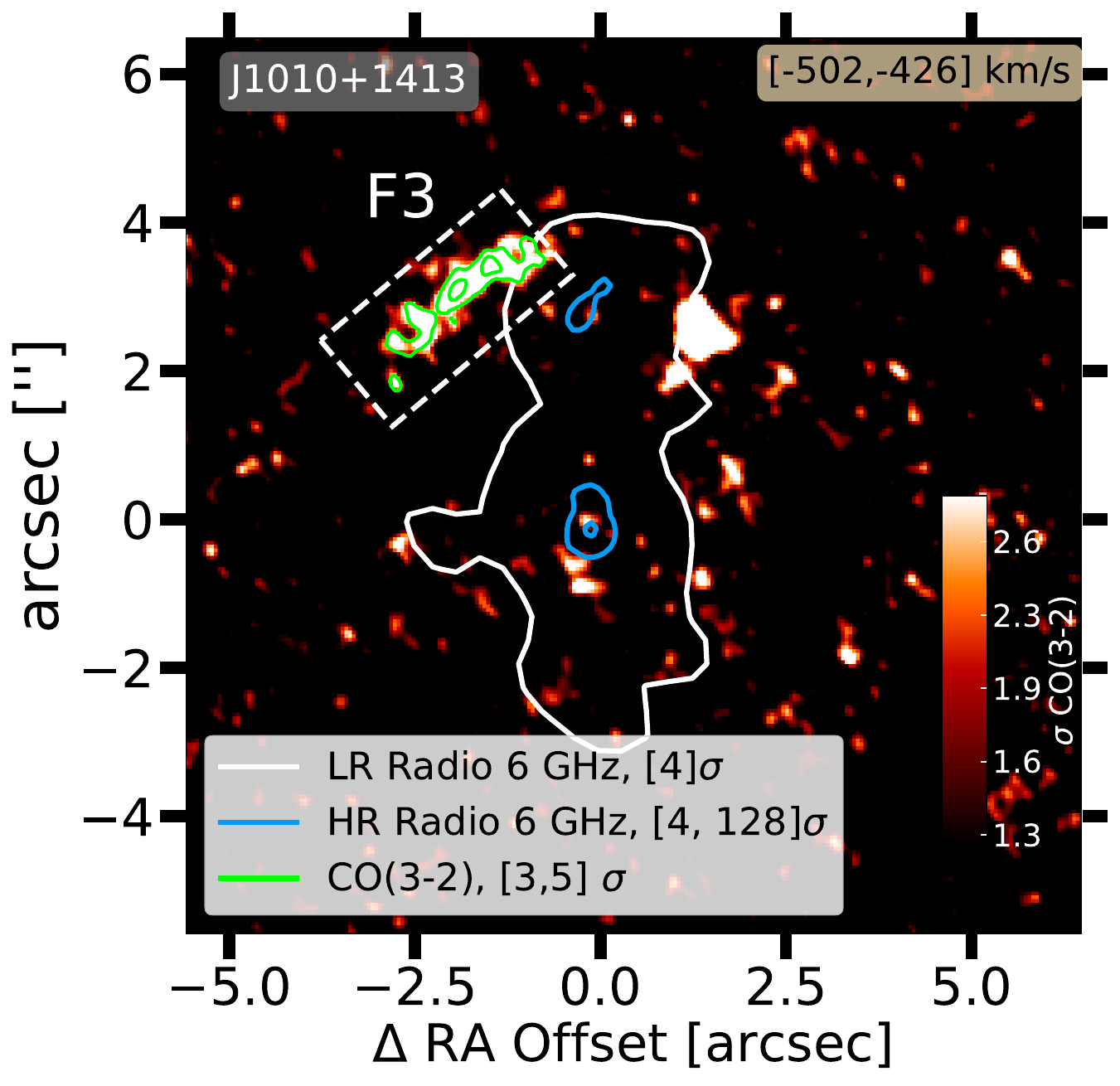}} \\
\end{tabular}
\caption{CO(3--2) emission-line maps, integrated over specific velocity ranges (quoted in the top-right of each panel), to highlight the identified filamentary structures (see Section \ref{sec: samplSelect}). The filament regions are highlighted with dashed-white boxes, \textbf{and labeled respectively as F1\,--\,5 for J1000+1242 and as F1\,--\,3, for J1010+1413}. The green contours are the CO(3--2) emission inside these boxes, at the 3 and 5$\sigma$ level. The radio contours are the same as in Figure \ref{fig2: all4targets} just shown here in blue and white colours for high- and low-resolution at 6 GHz respectively. A legend is shown on the bottom-right panel.}
\label{appfig1: filNarrow}
\end{figure*}

To identify the filaments following our approach in Section \ref{sec: filDefine}, we created narrow velocity slice CO images by collapsing over the consecutive velocity channels where any emission $\geq\,5\,\sigma$ was seen associated with these structures. These are shown in Figure~\ref{appfig1: filNarrow}. For J1000+1242, we identified 5 filaments and for J1010+1413, we identified 3 filaments, as shown in each of the panels below.  The velocity windows used for collapsing each image are labeled on the top-right of each panel, and each filament is highlighted with a surrounding dashed white box. These boxes cover the full observed extent (at $\geq3\,\sigma$) of the structures, where the $3\,\sigma$ contours are also shown. A combined overview figure for each of the targets can be seen in Figure \ref{fig3: filaments_overview}. 

\section{Figures of the Filamentary Molecular Gas Structures}\label{app: filamentMaps}

\begin{figure*}
\centering
\begin{tabular}{cccc}
\includegraphics[width=.2\textwidth,height=5cm]{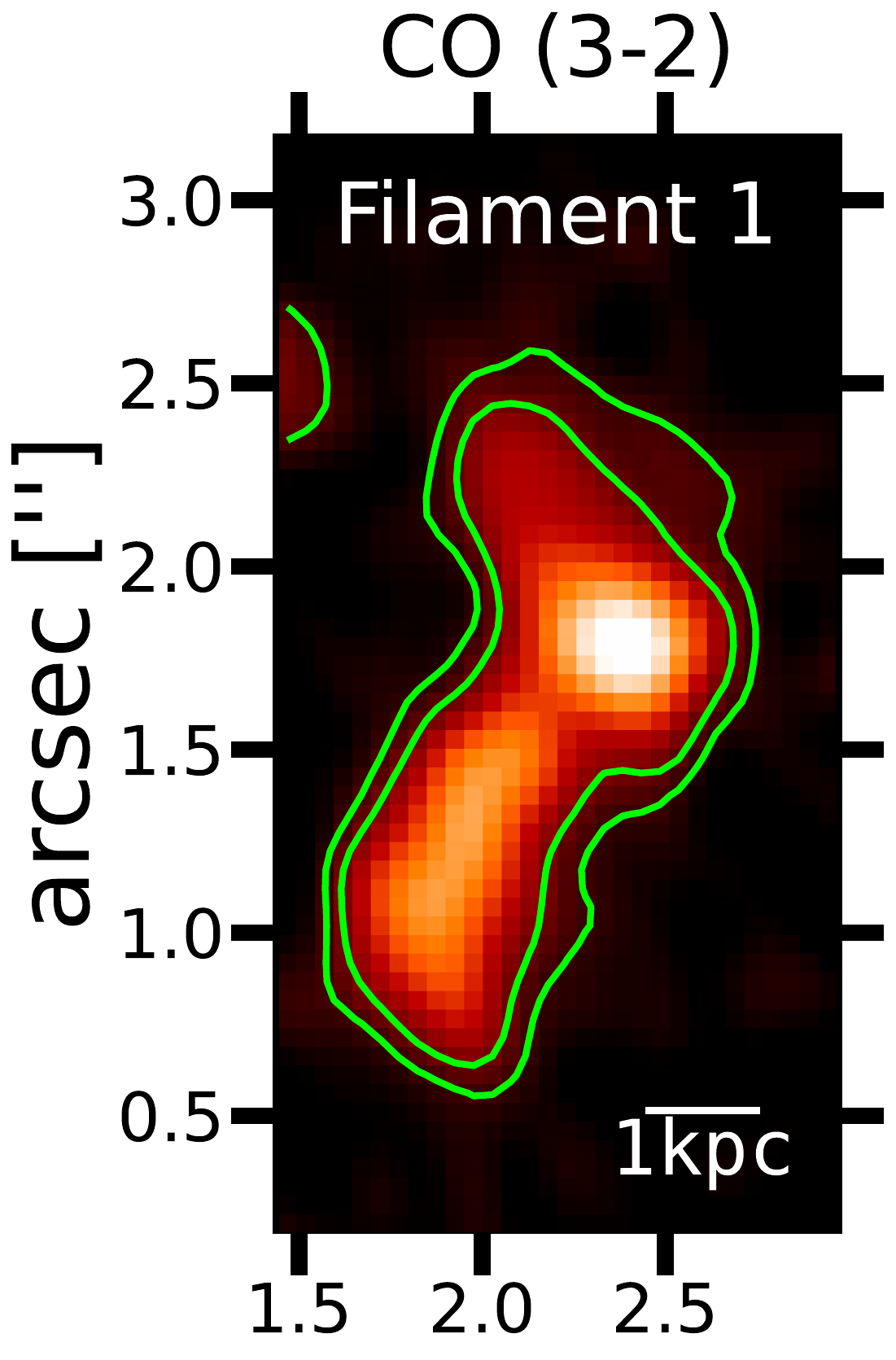 }&\includegraphics[width=.2\textwidth,height=5cm]{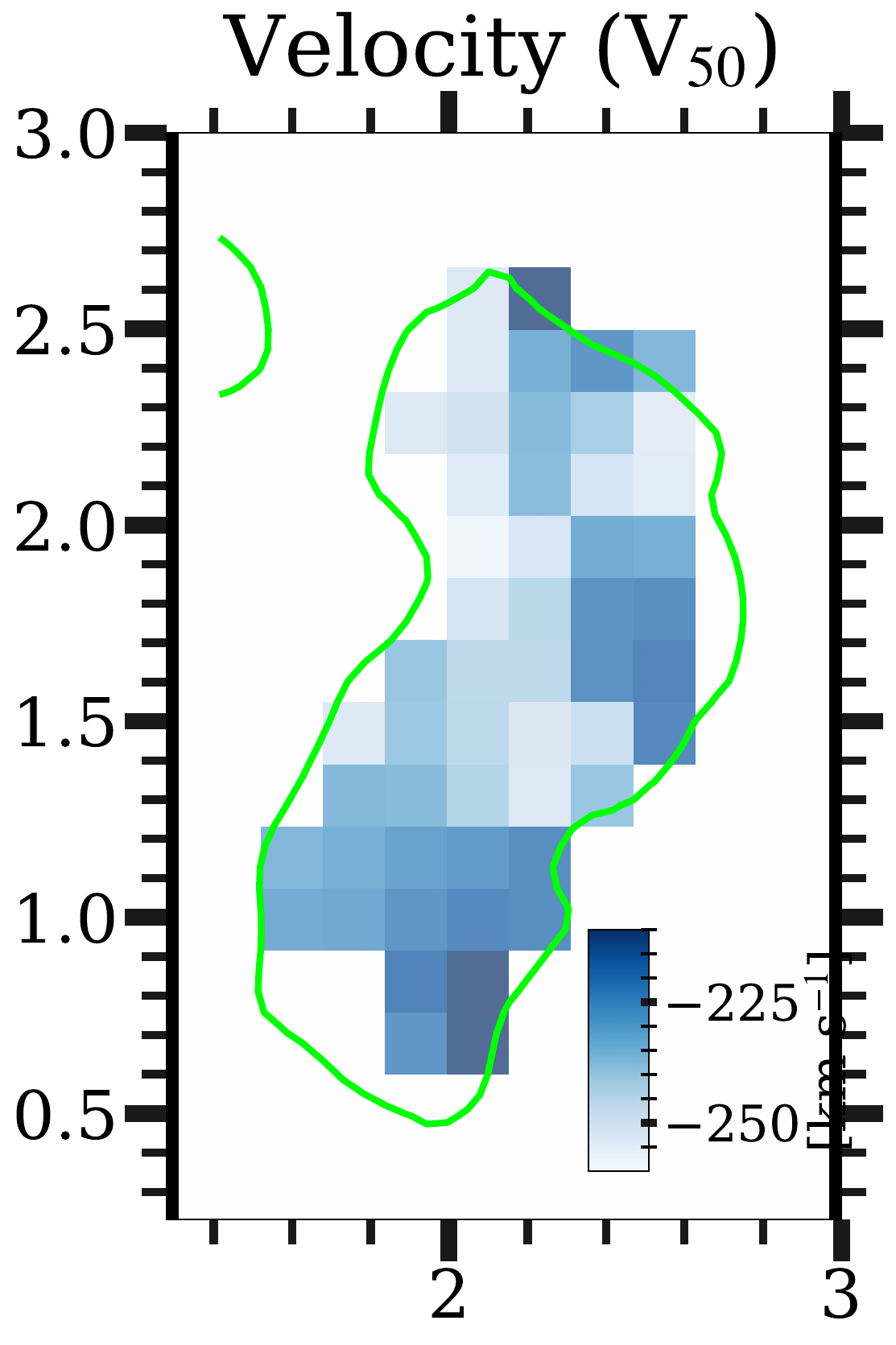 }&\includegraphics[width=.2\textwidth,height=5cm]{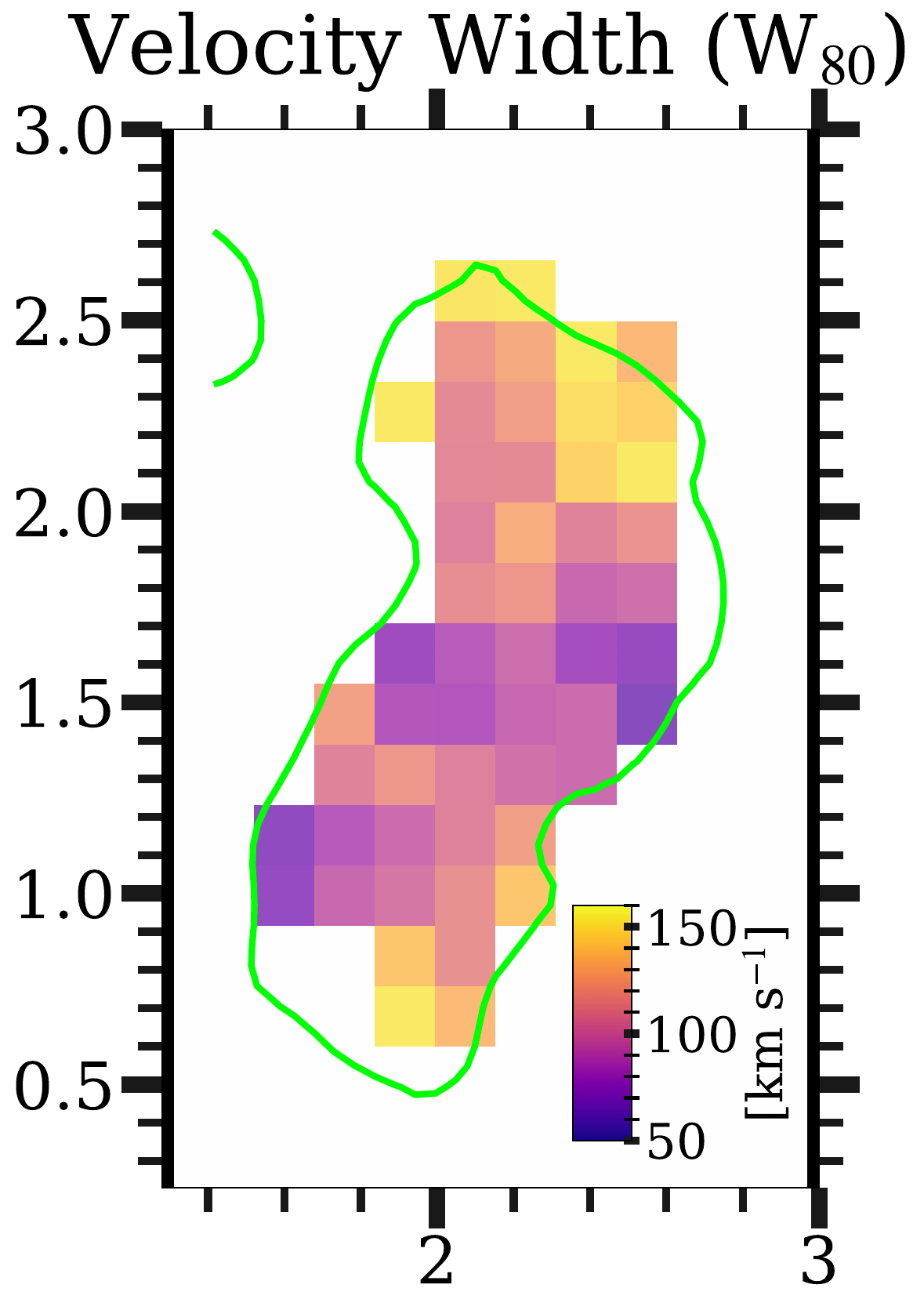 }&\includegraphics[width=.3\textwidth,height=3cm]{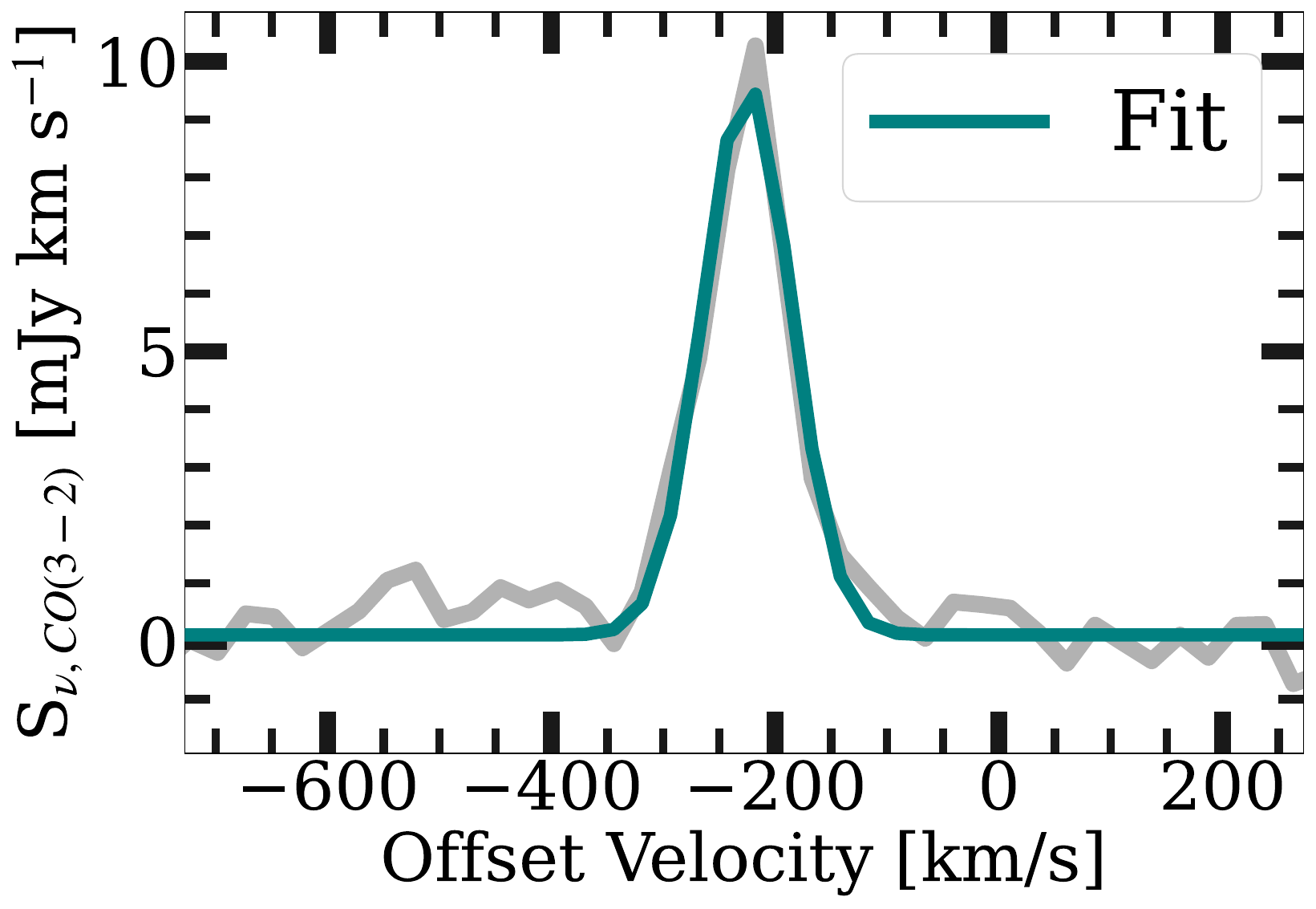 } \\

\includegraphics[width=.2\textwidth,height=3.8cm]{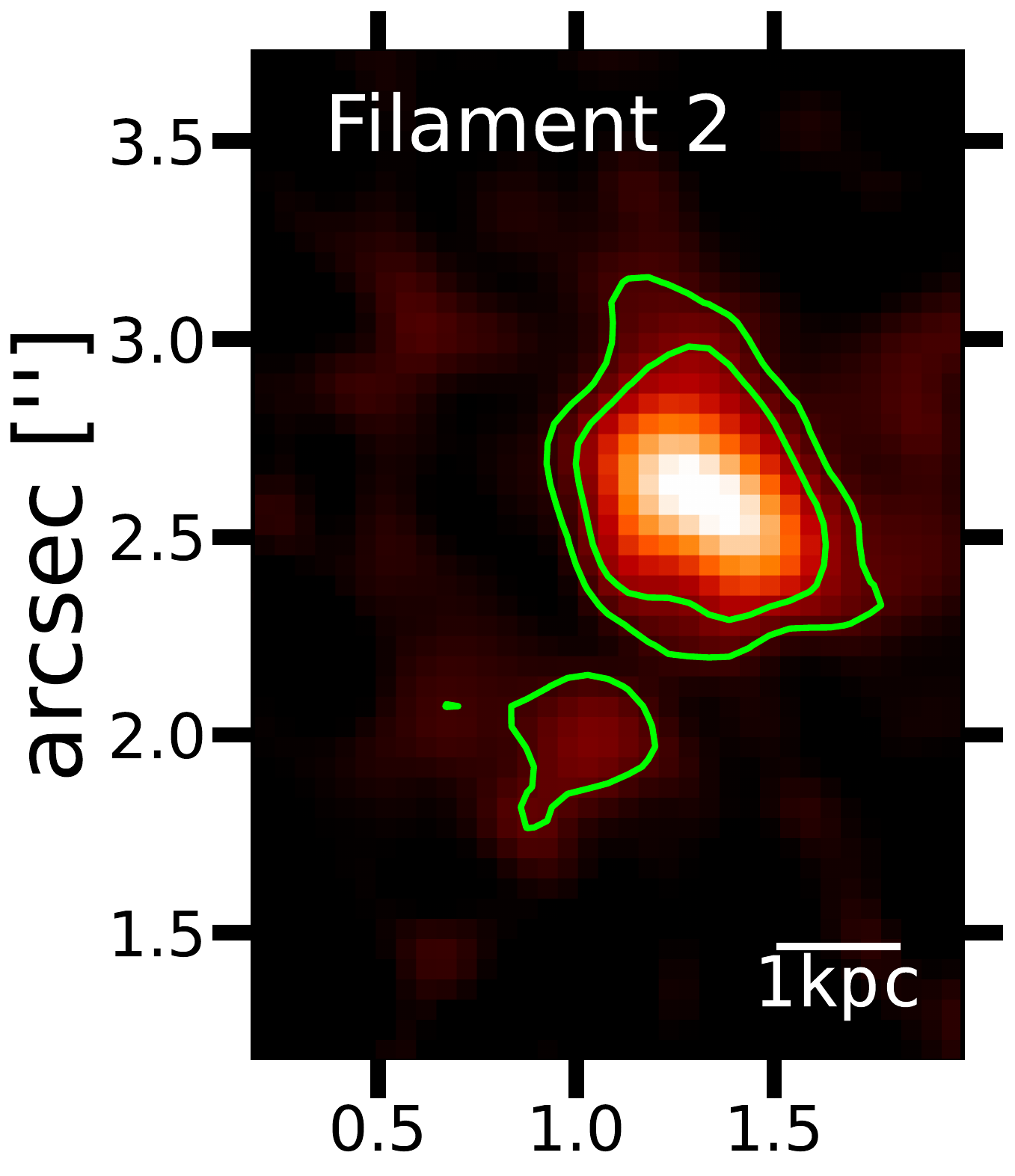 }&\includegraphics[width=.2\textwidth,height=3.8cm]{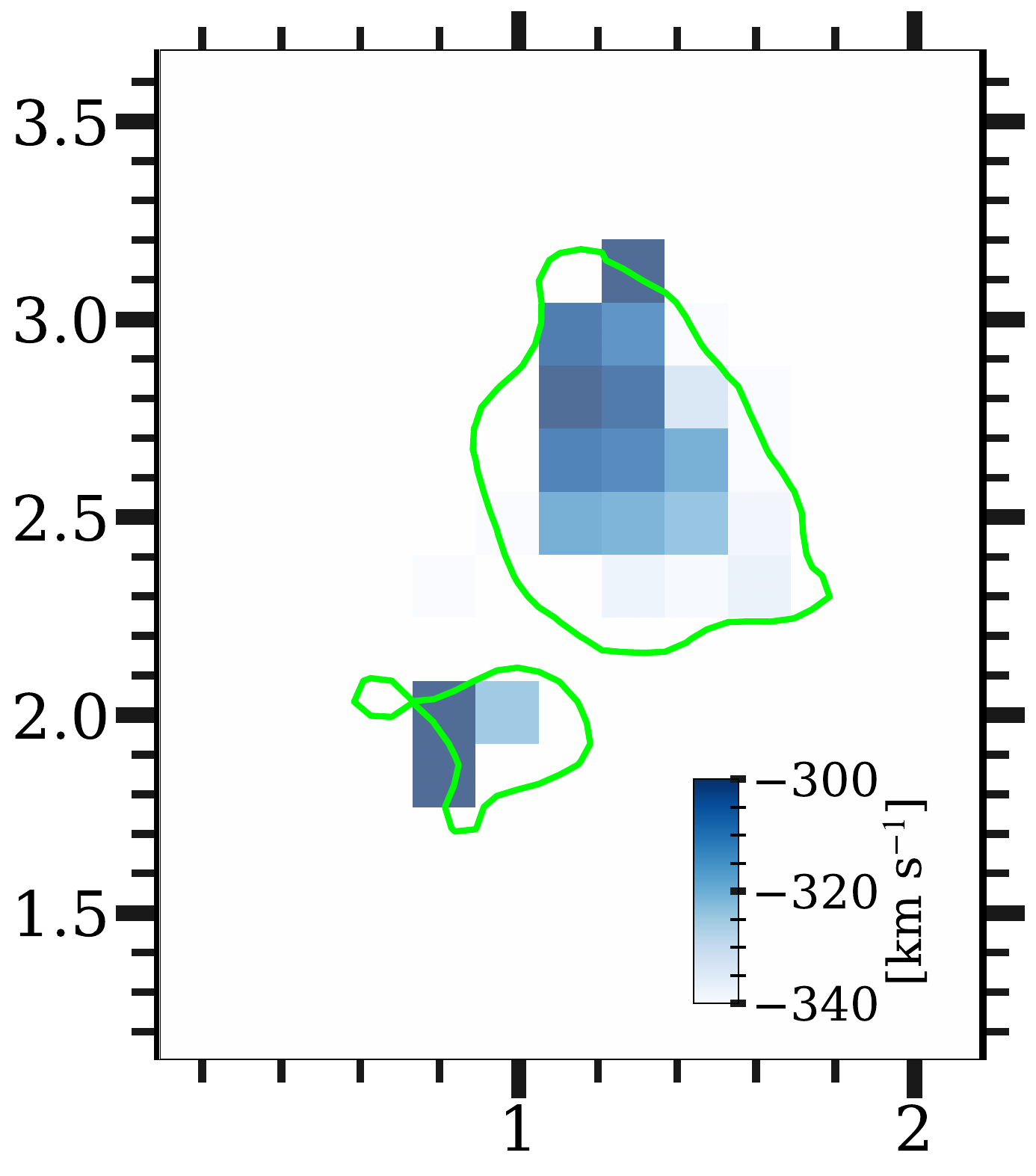 }&\includegraphics[width=.2\textwidth,height=3.8cm]{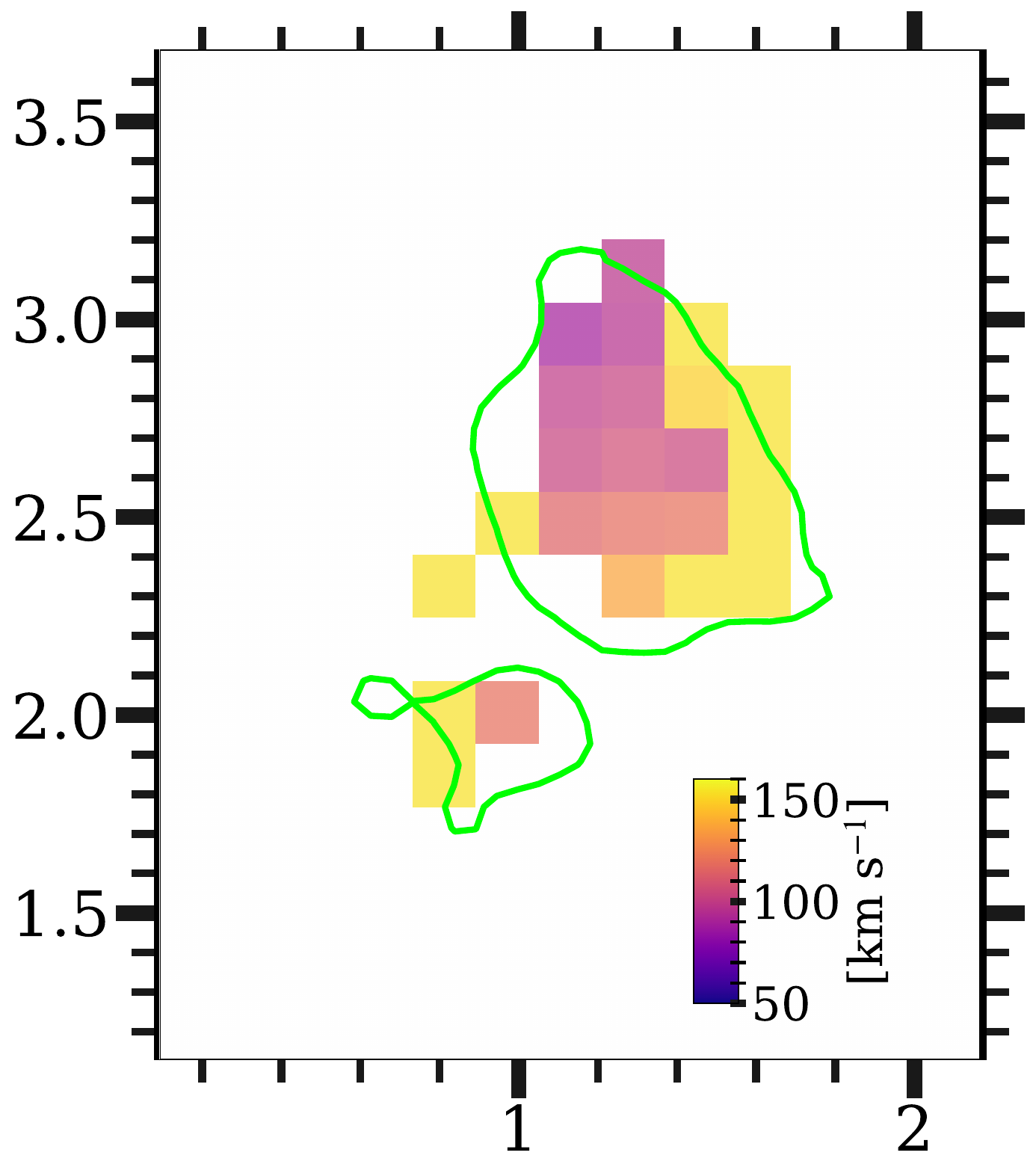 }&\includegraphics[width=.3\textwidth,height=3cm]{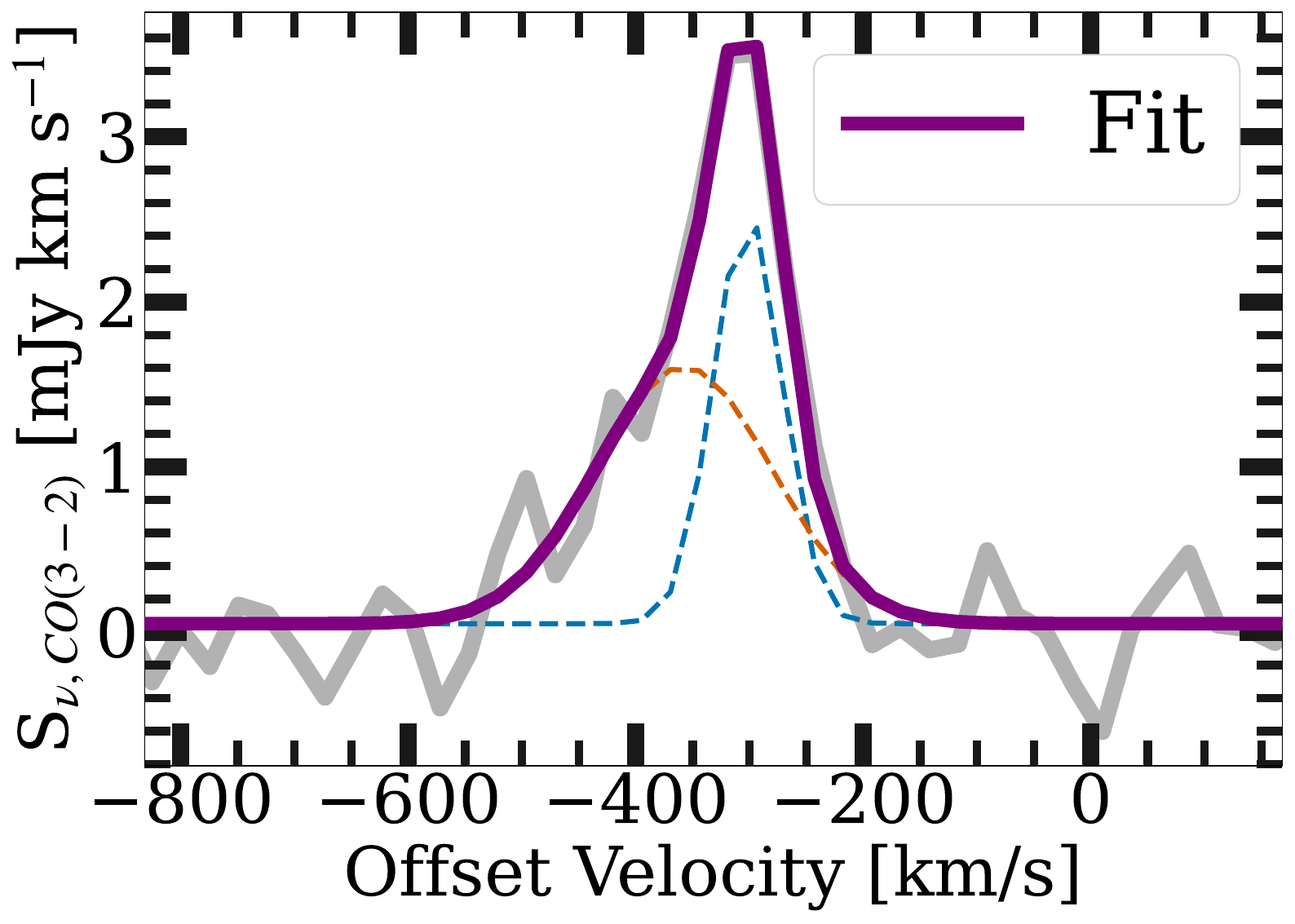 } \\

& & & \\

\includegraphics[width=.2\textwidth,height=3cm]{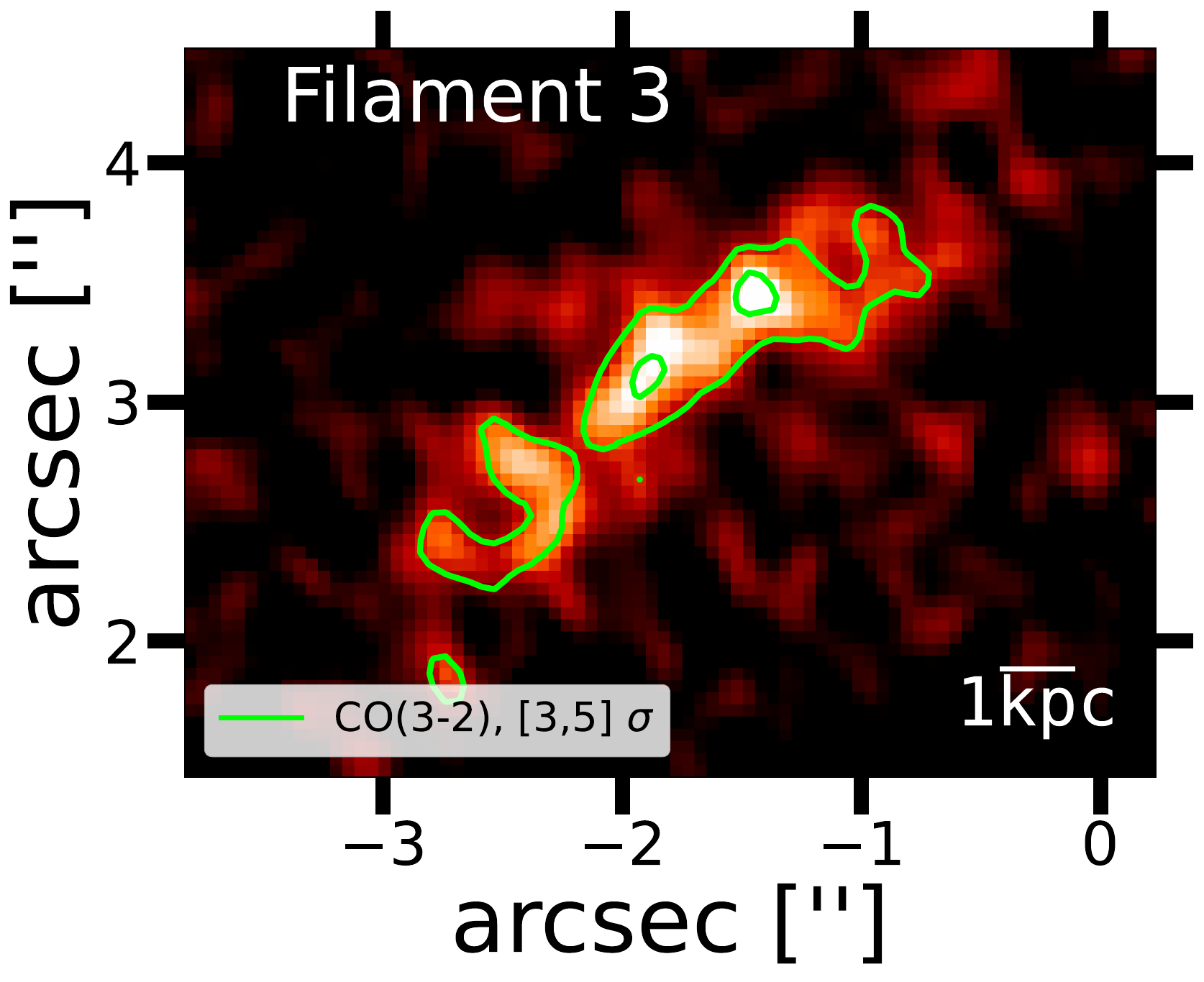 }&\includegraphics[width=.2\textwidth,height=3cm]{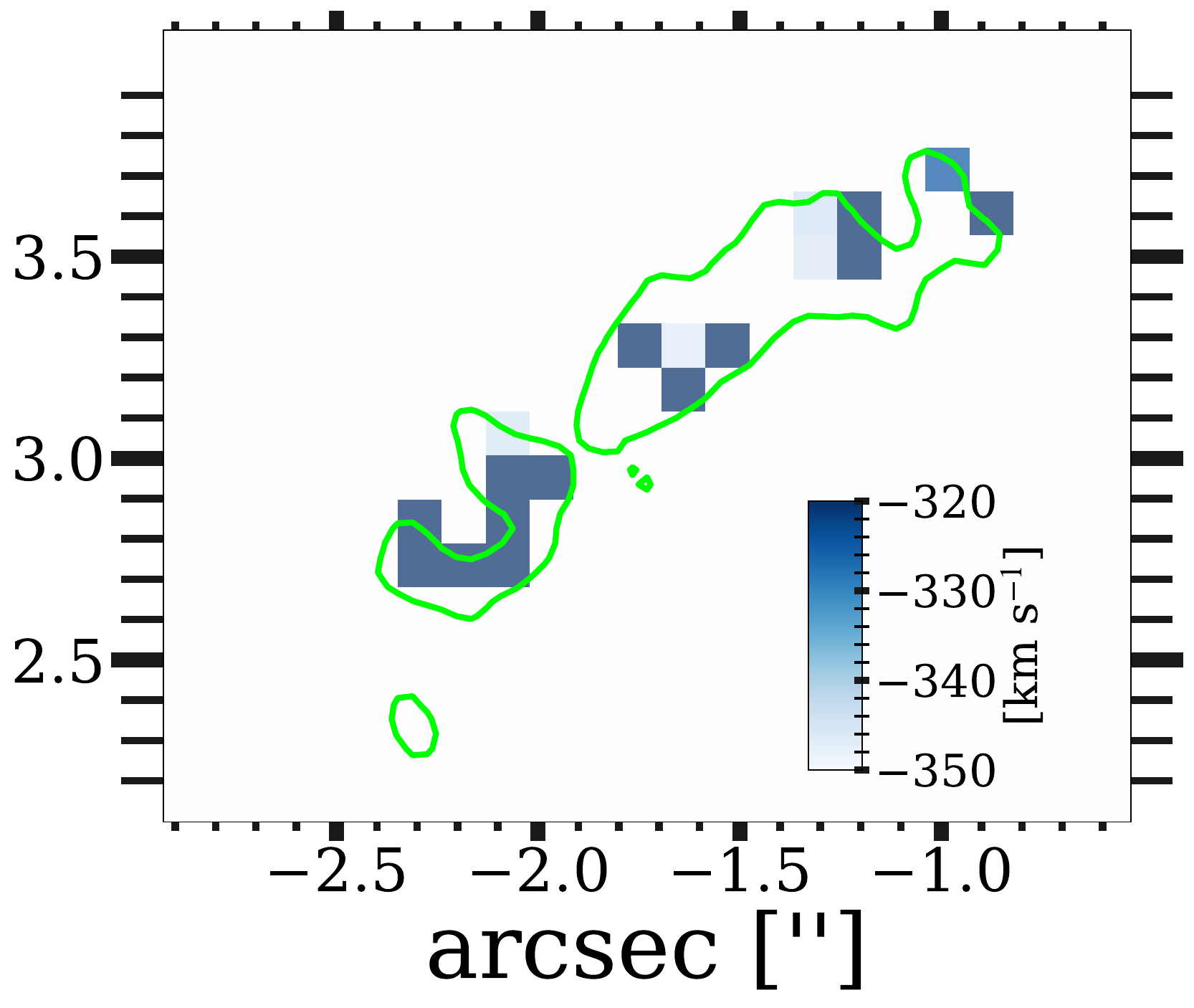 }&\includegraphics[width=.2\textwidth,height=3cm]{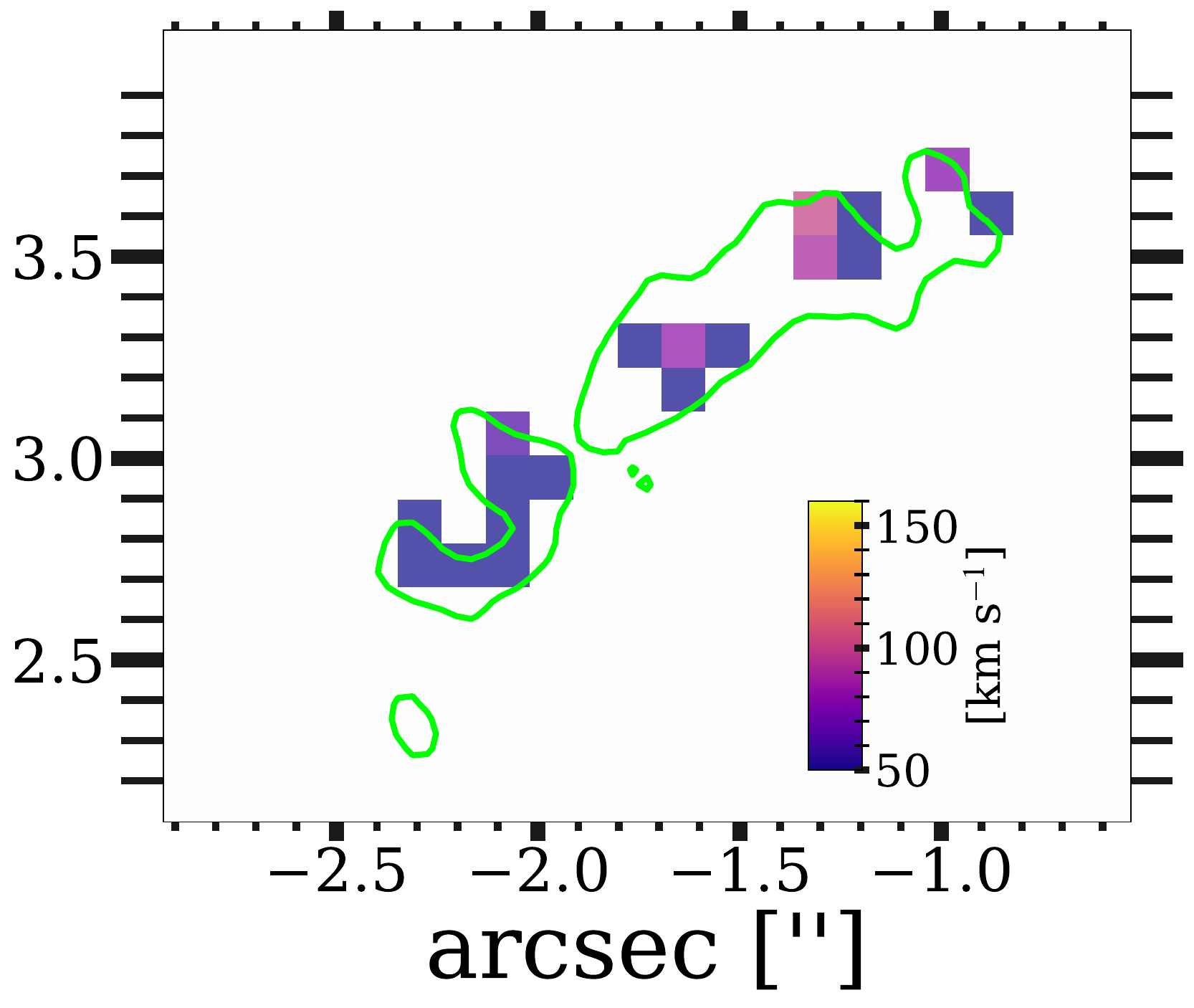 }&\includegraphics[width=.3\textwidth,height=3cm]{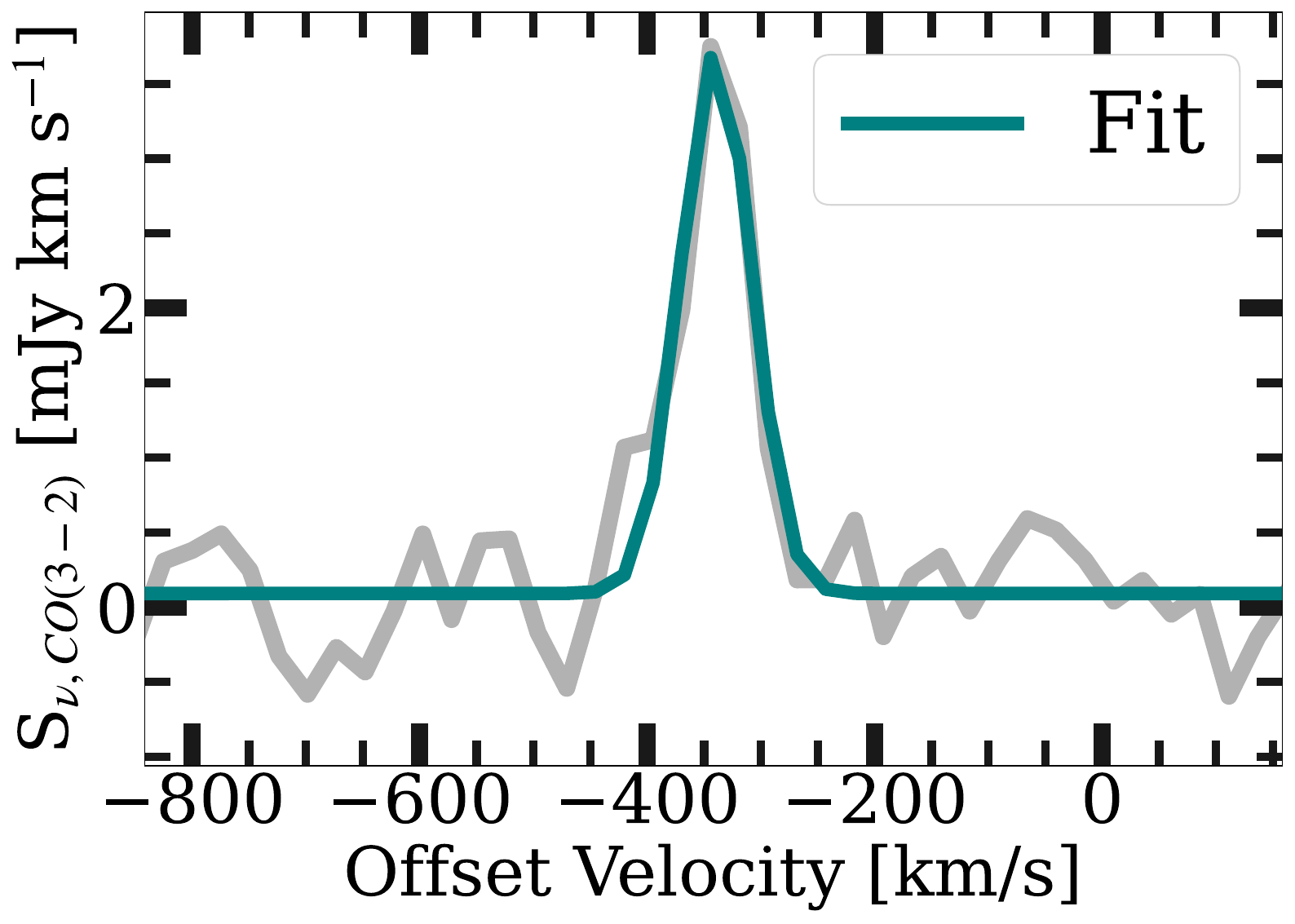 } 
\end{tabular}
\caption{Kinematic maps of the filamentary molecular gas structures in J1010+1413. Each row corresponds to a single filament, numbered 1\,--\,3, as shown in Figure 3. \textit{(From Left to Right:)} (1) CO(3--2) emission-line map, collapsed over the velocity range of the filament (see Section 3.1); (2) velocity map (V$_{50}$); (3) velocity width map (W$_{80}$); (4) CO(3--2) emission-line profile obtained over the entire filament region. The CO(3--2) emission is highlighted with 3$\sigma$ and 5$\sigma$ contours in the first column and overlaid on the kinematic maps the same 3$\sigma$ contour is shown. In the final row, the coloured solid curves show the emission-line fits, and the dashed curves show the individual Gaussian components (when more than one).}
\label{appfig2: filamentsZoom1}
\end{figure*}

\begin{figure*}
\centering
\begin{tabular}{cccc}
\includegraphics[width=.23\textwidth,height=4.5cm]{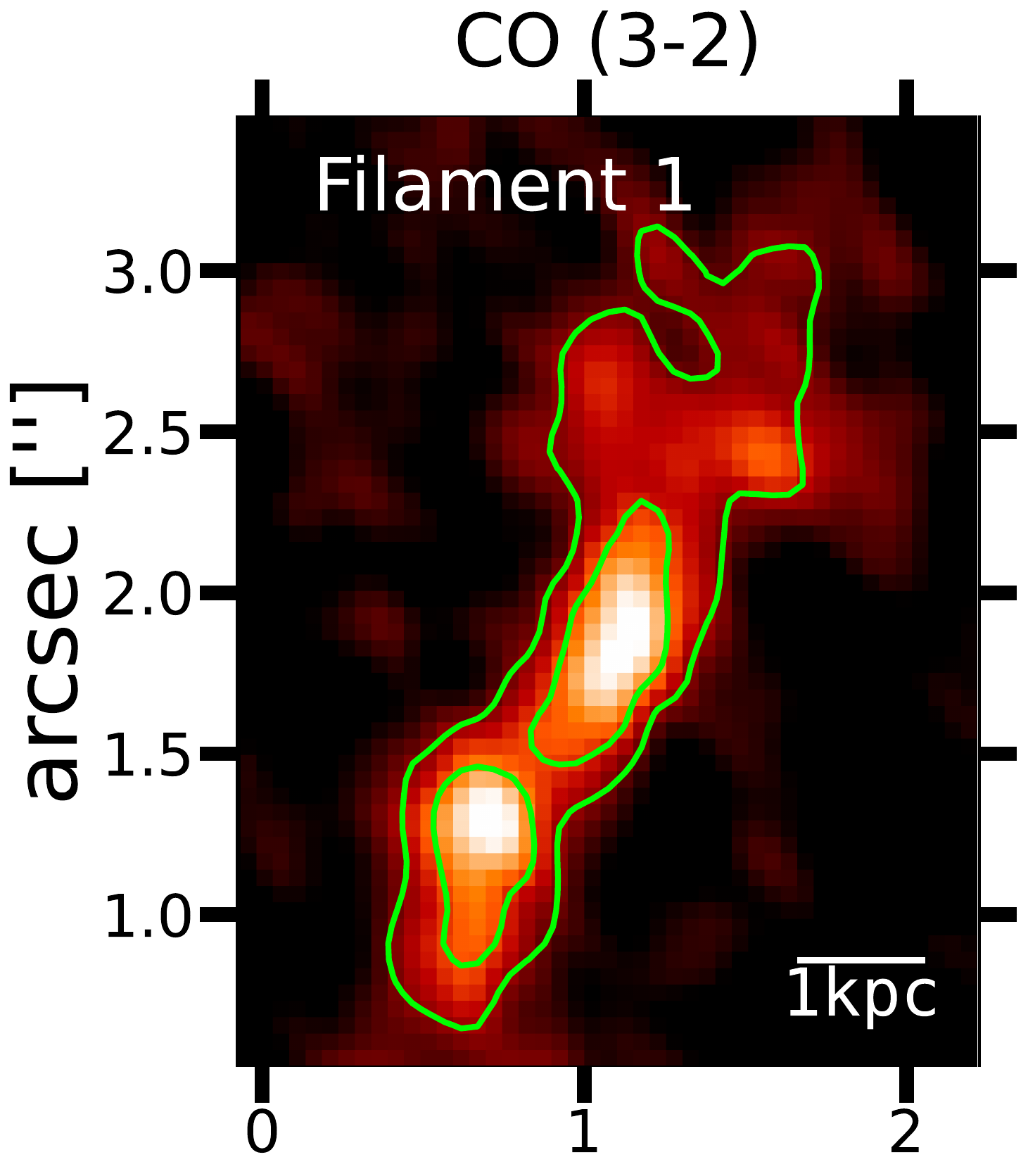} & 
\includegraphics[width=.22\textwidth,height=4.5cm]{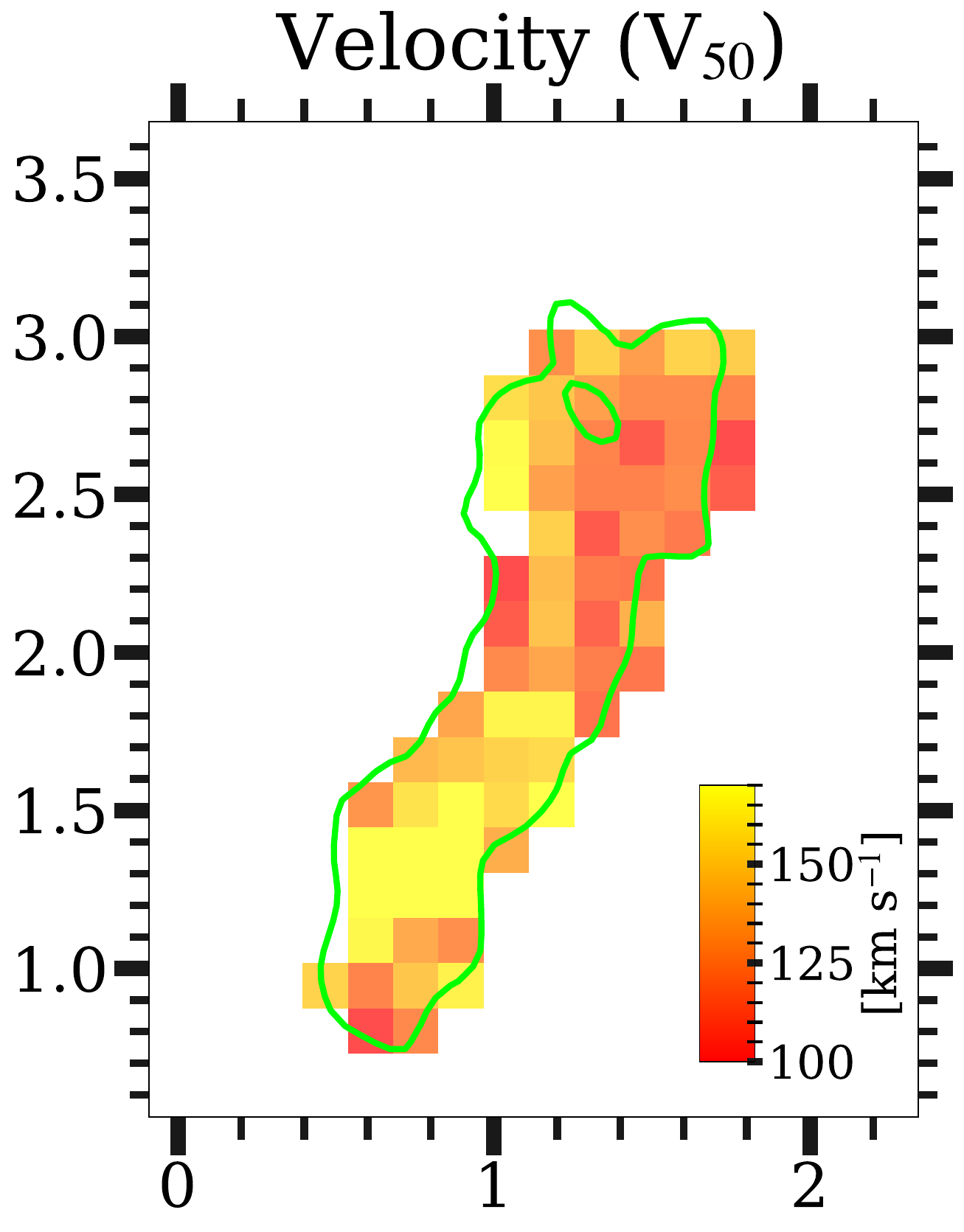} & 
 \includegraphics[width=.22\textwidth,height=4.5cm]{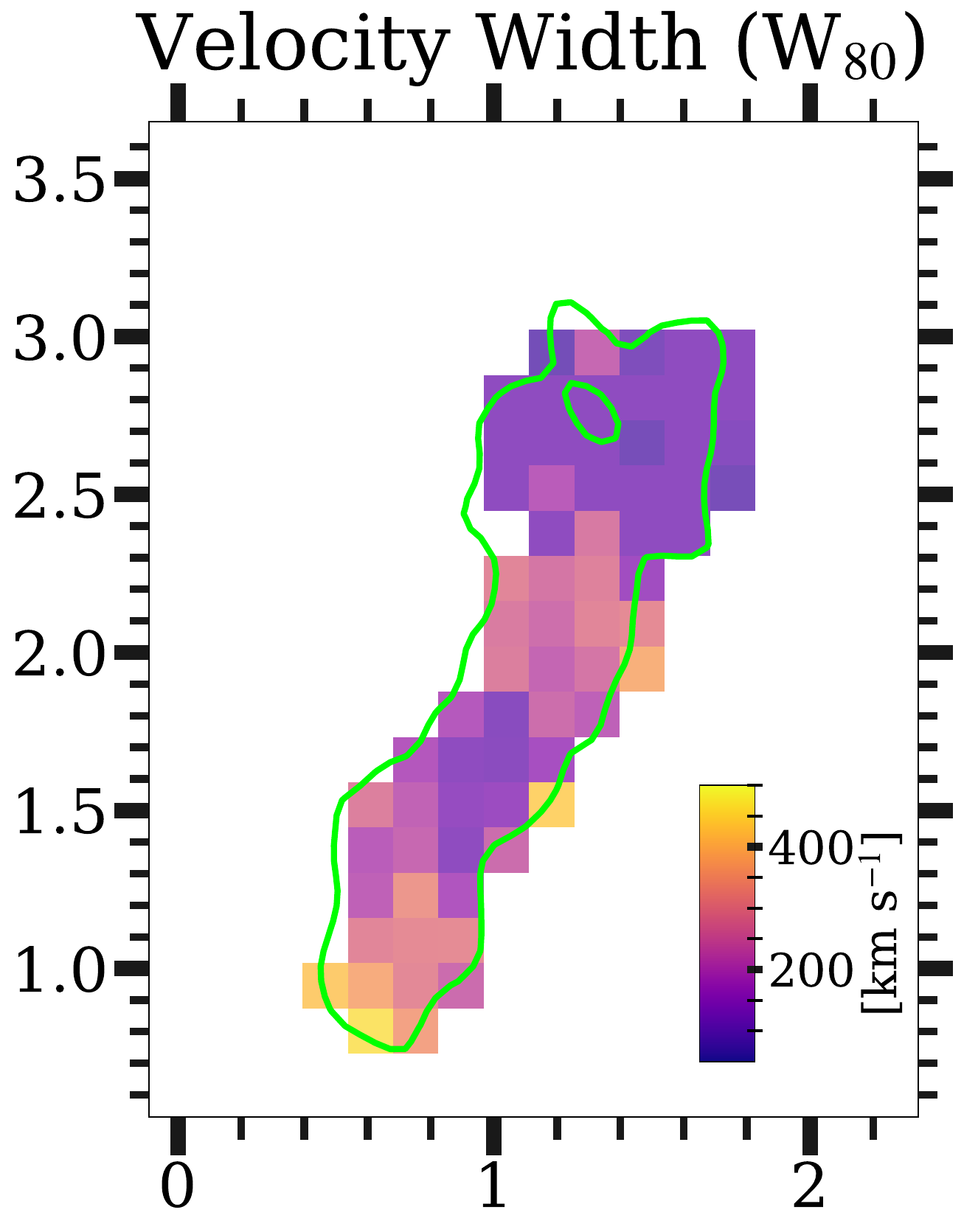} &
 \includegraphics[width=.3\textwidth,height=3cm]{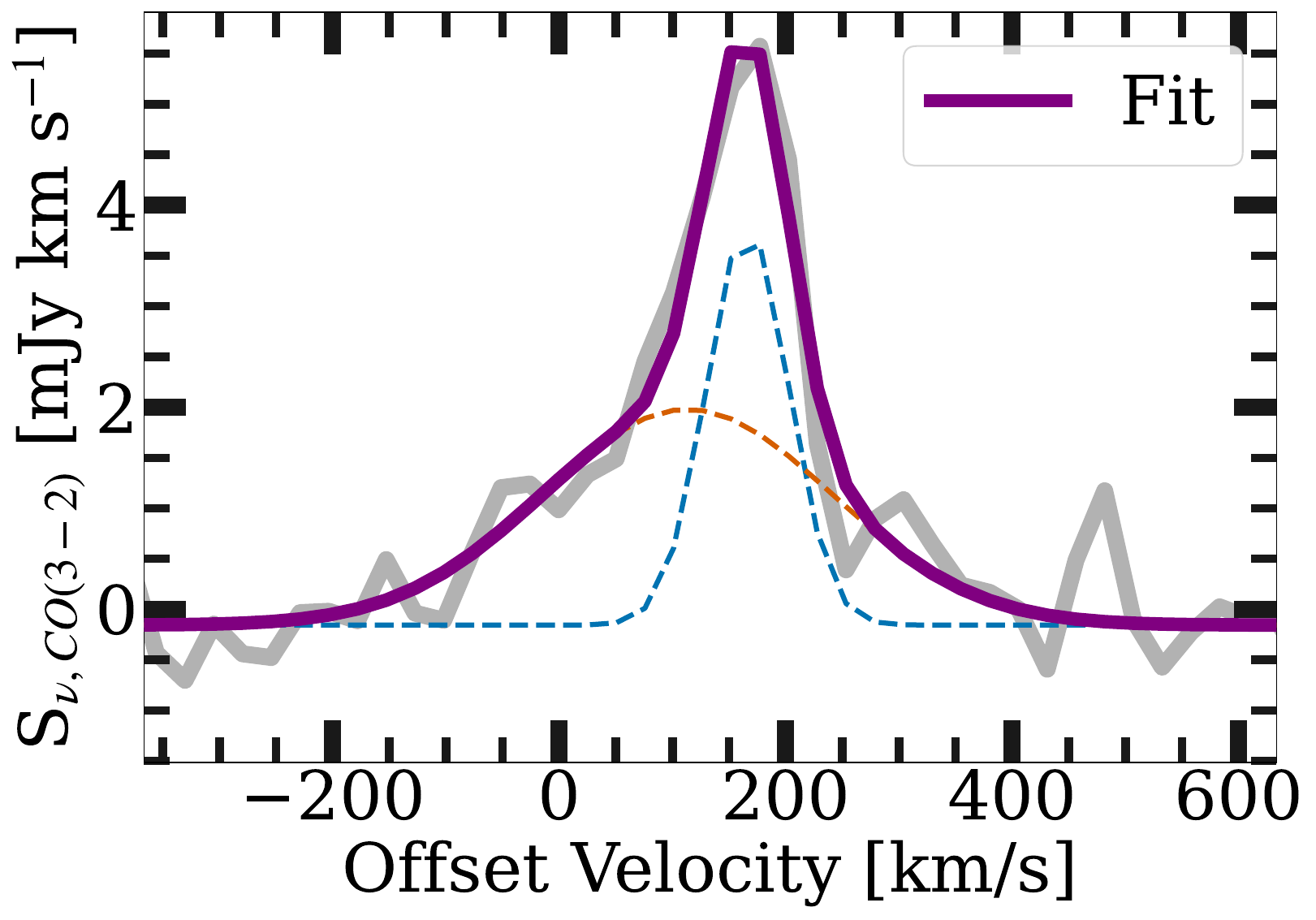} \\

\includegraphics[width=.23\textwidth,height=3.5cm]{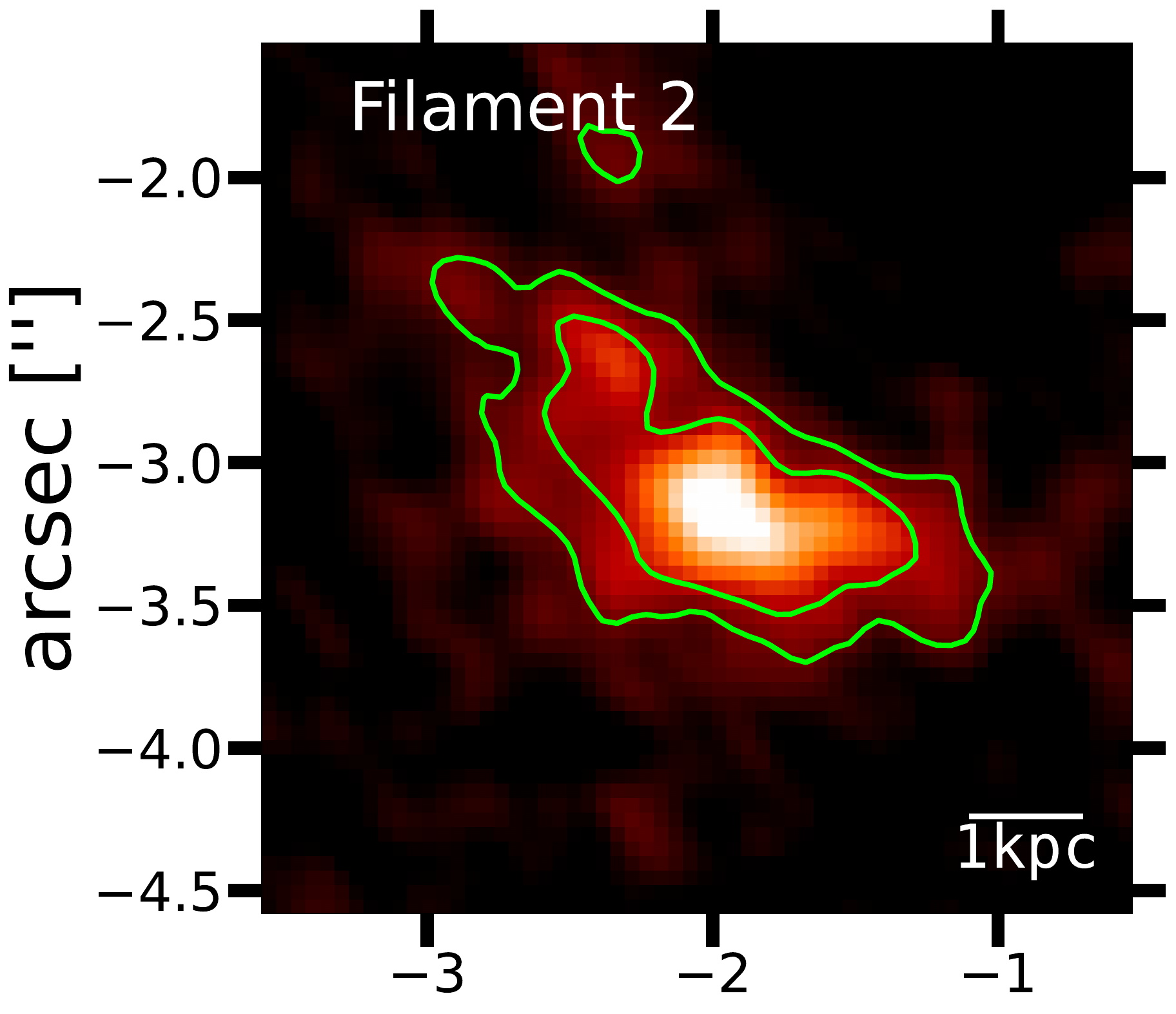} & 
\includegraphics[width=.22\textwidth,height=3.5cm]{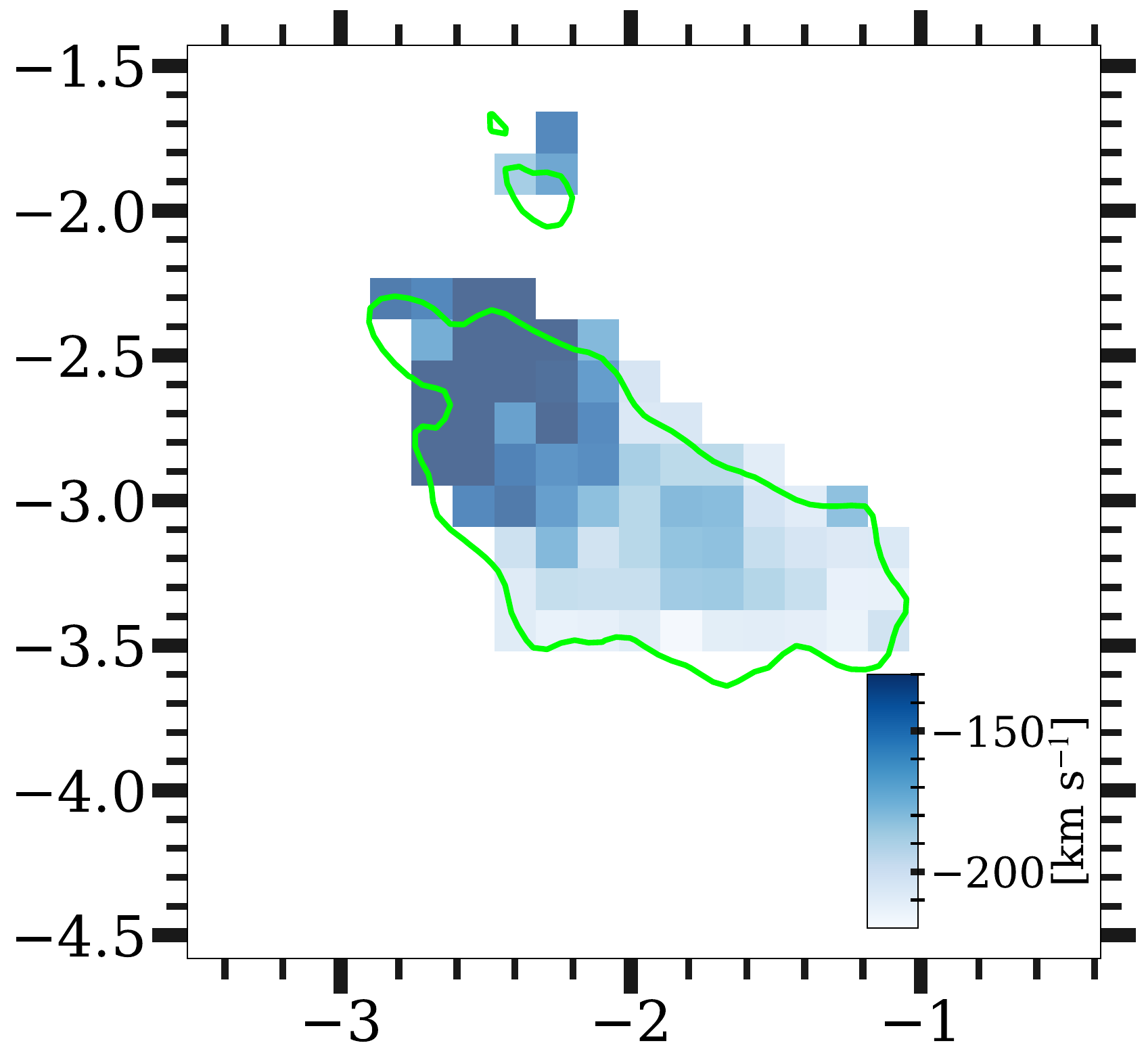} & 
 \includegraphics[width=.22\textwidth,height=3.5cm]{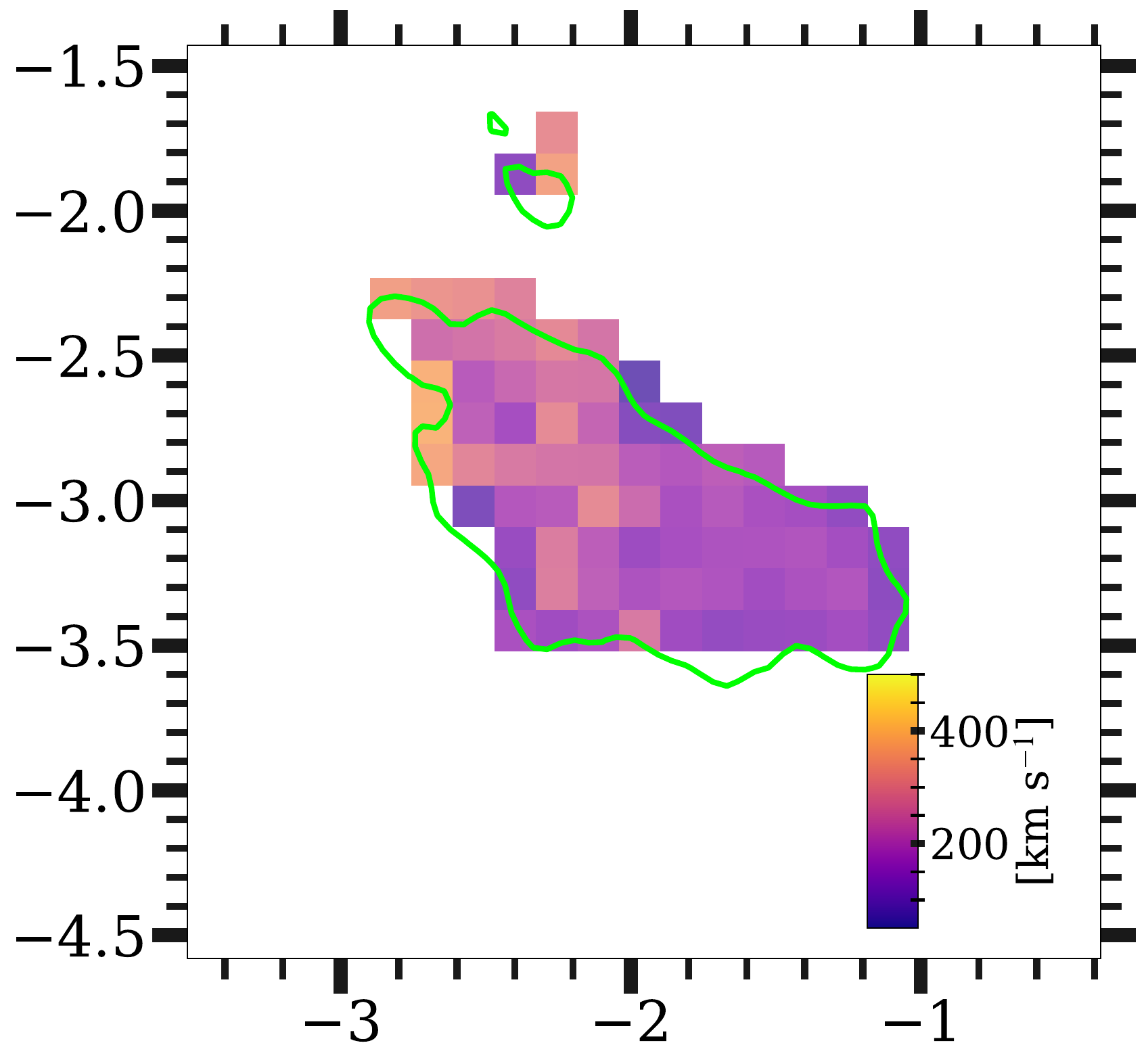} &
 \includegraphics[width=.3\textwidth,height=3cm]{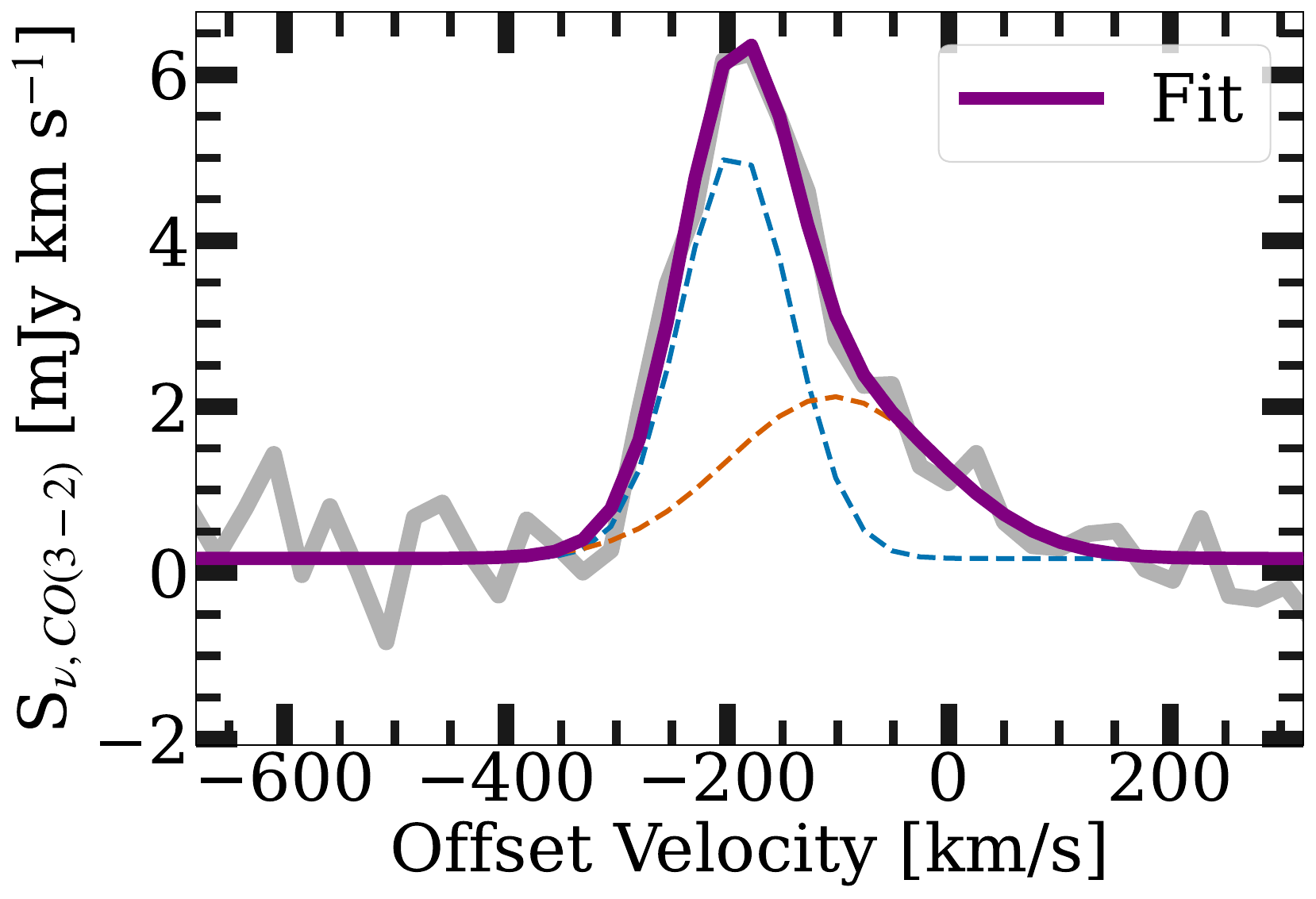} \\

\includegraphics[width=.23\textwidth,height=3.5cm]{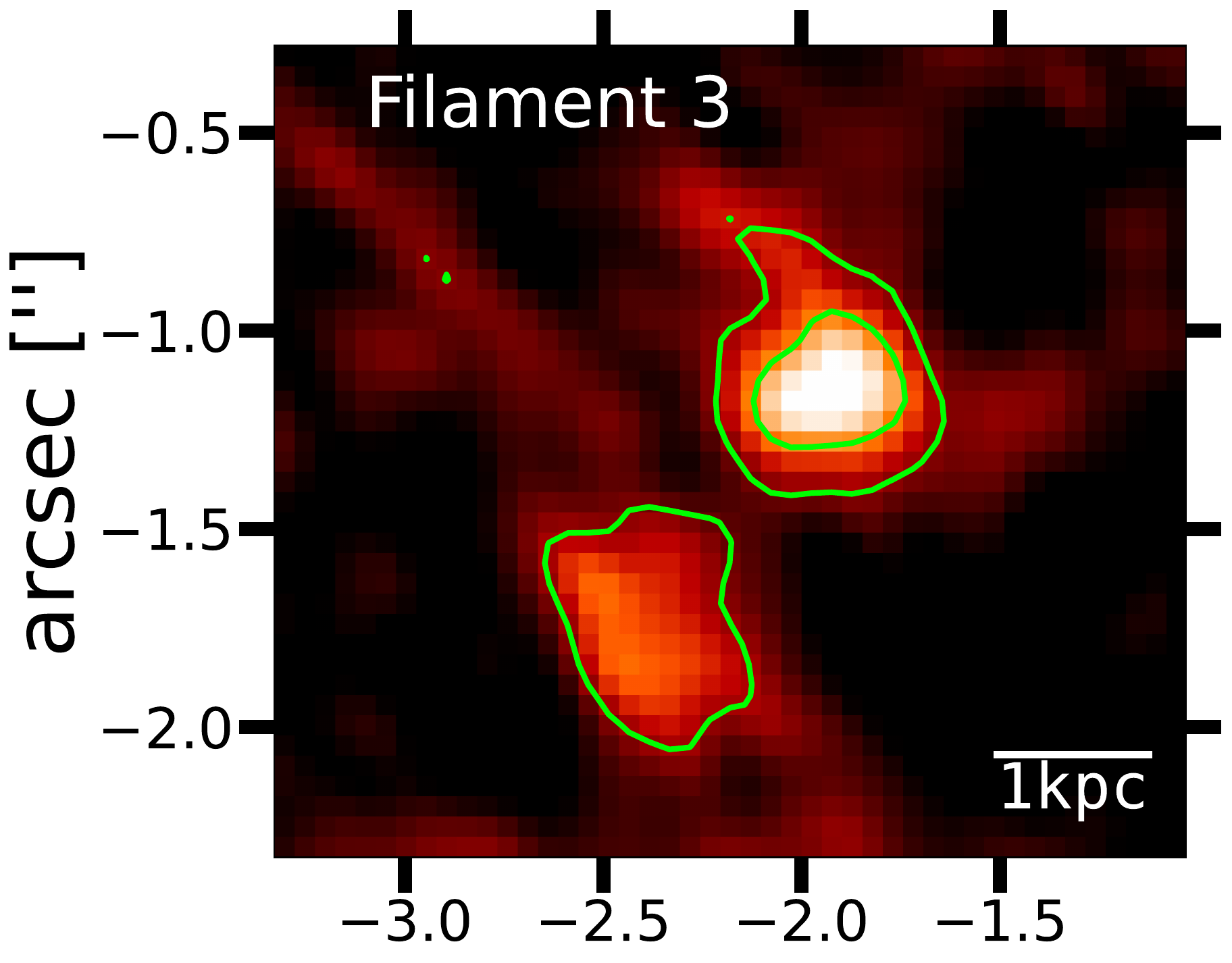} & 
\includegraphics[width=.22\textwidth,height=3.5cm]{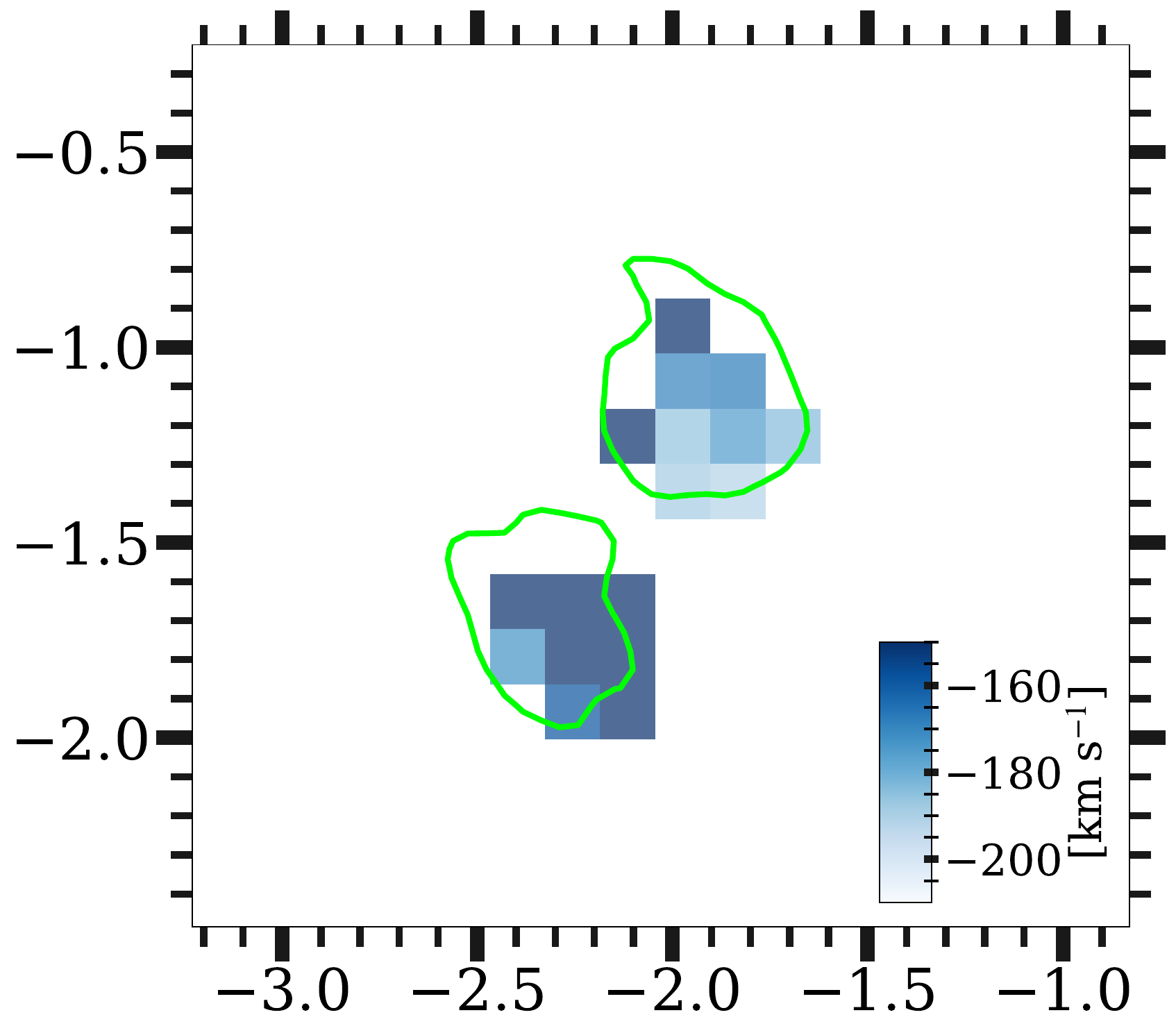} & 
 \includegraphics[width=.22\textwidth,height=3.5cm]{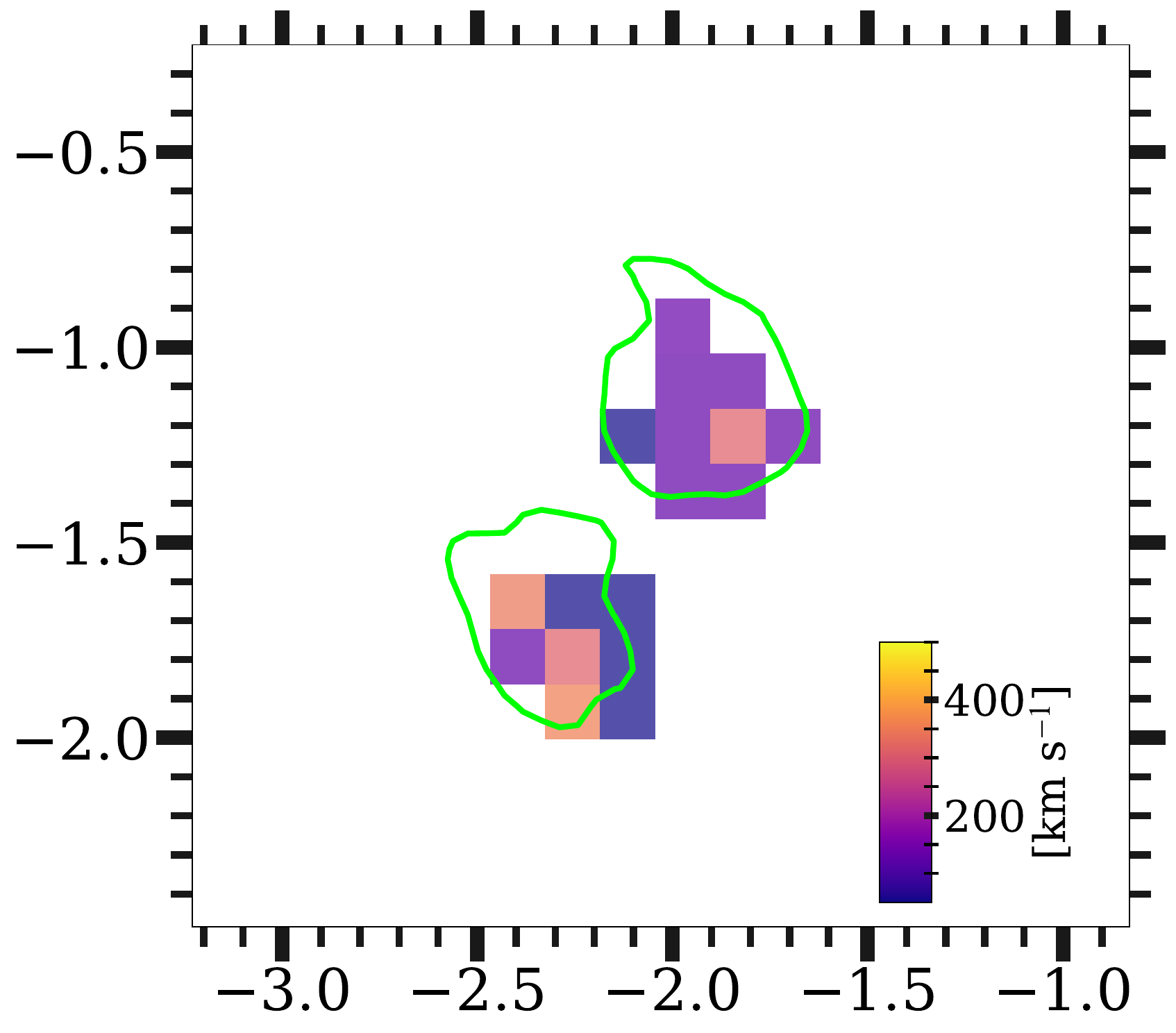} &
 \includegraphics[width=.3\textwidth,height=3cm]{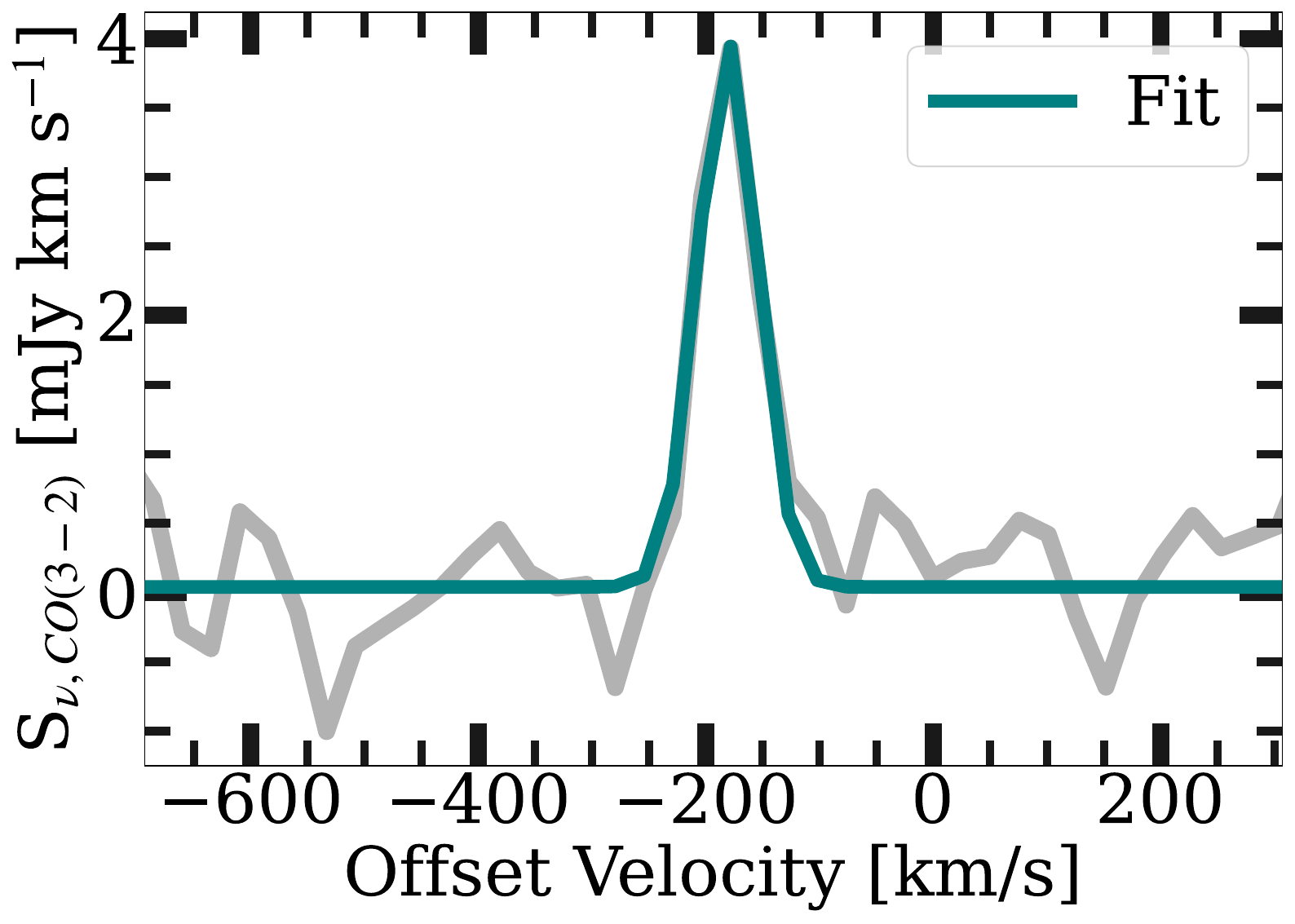} \\

\includegraphics[width=.23\textwidth,height=3.5cm]{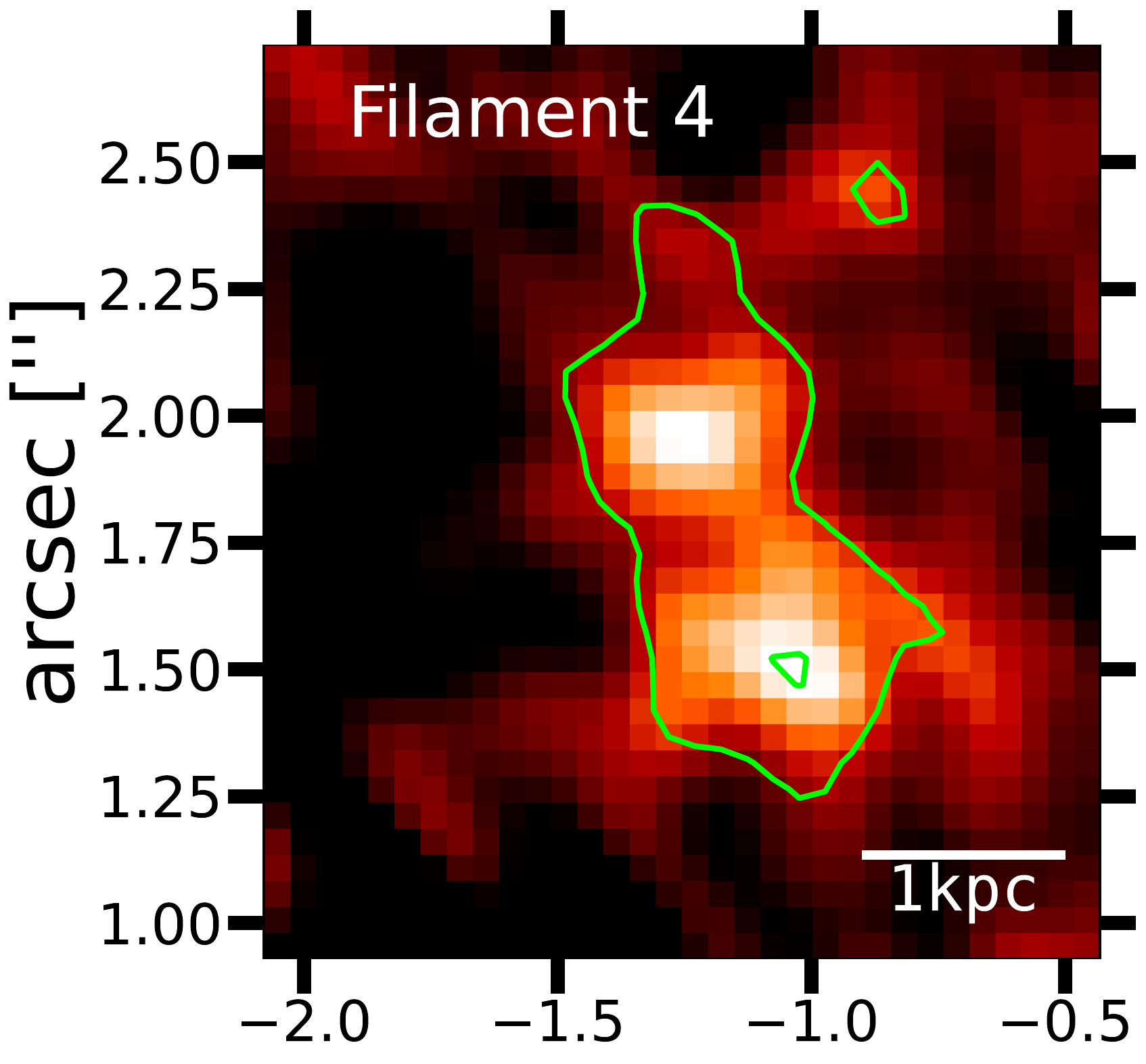} & 
\includegraphics[width=.22\textwidth,height=3.5cm]{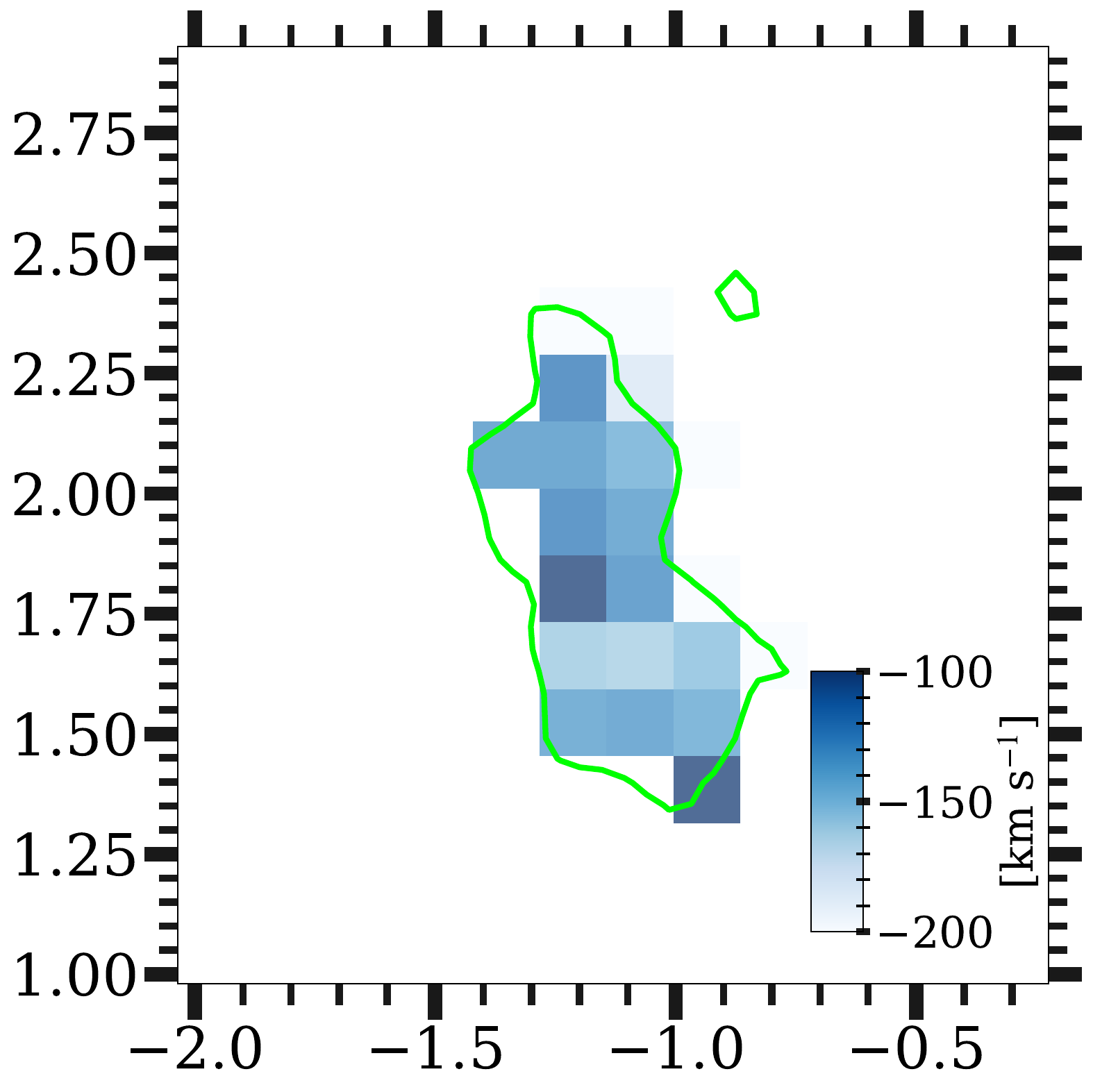} & 
 \includegraphics[width=.22\textwidth,height=3.5cm]{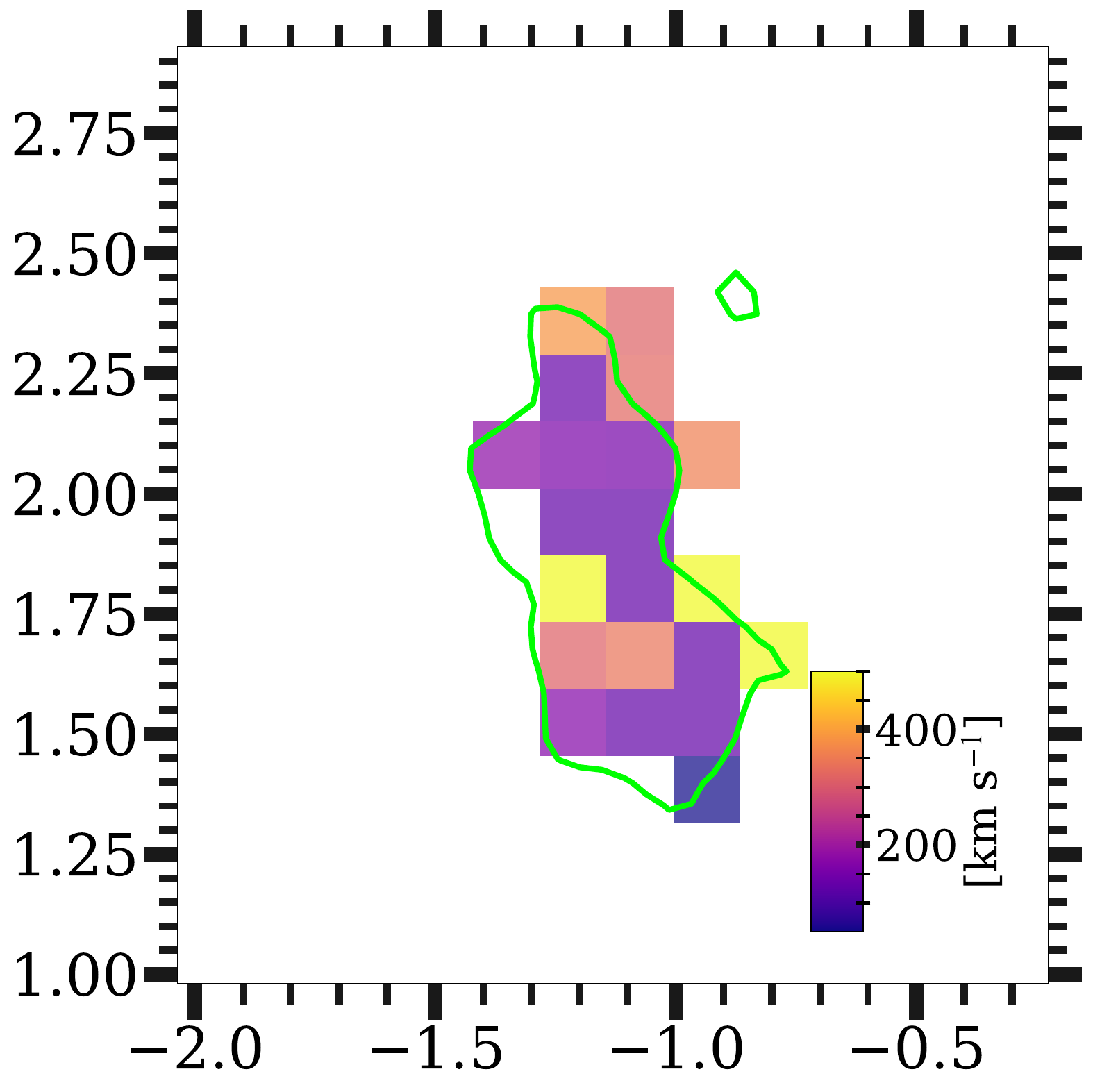} &
 \includegraphics[width=.3\textwidth,height=3cm]{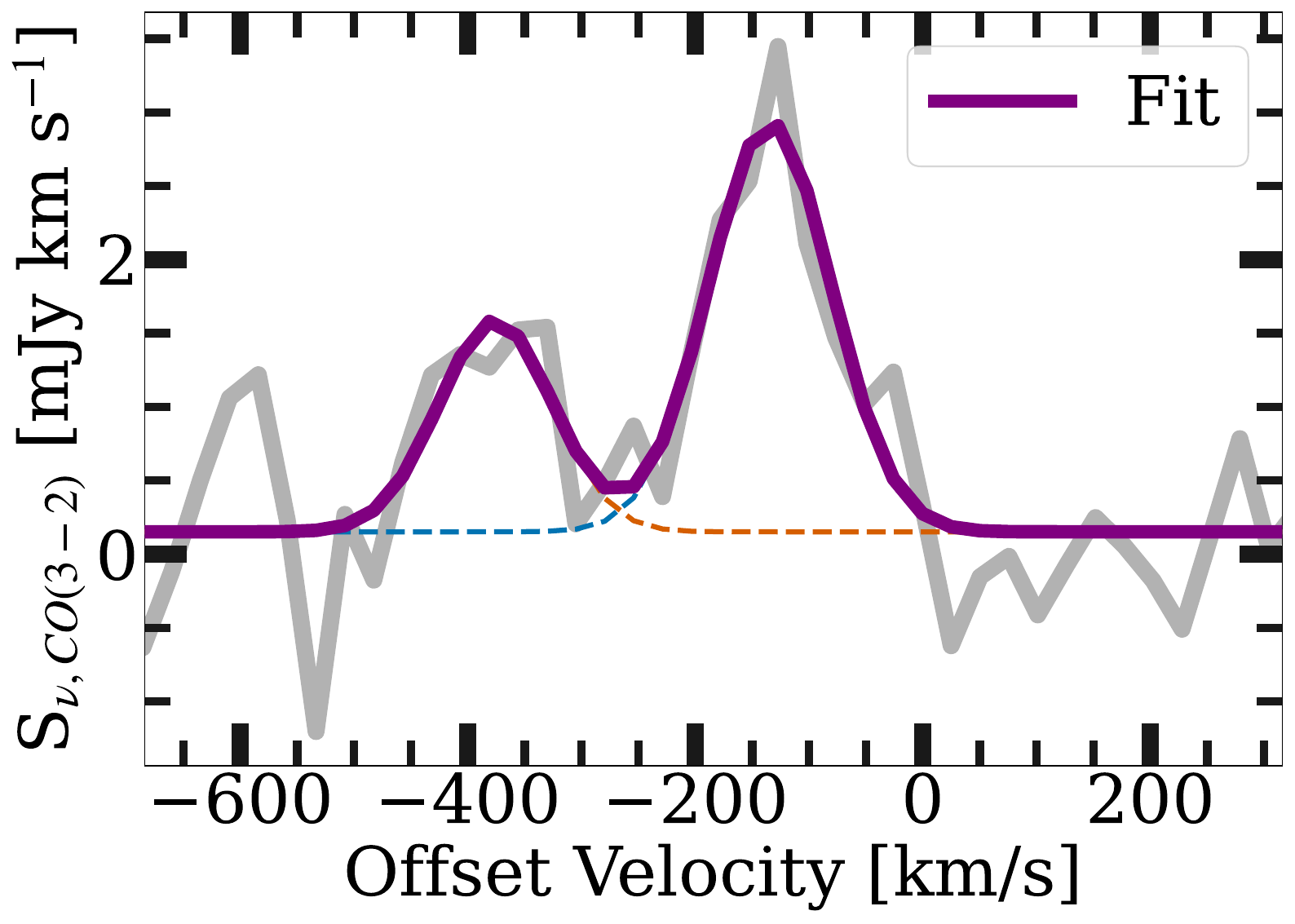} \\

\includegraphics[width=.23\textwidth,height=3.5cm]{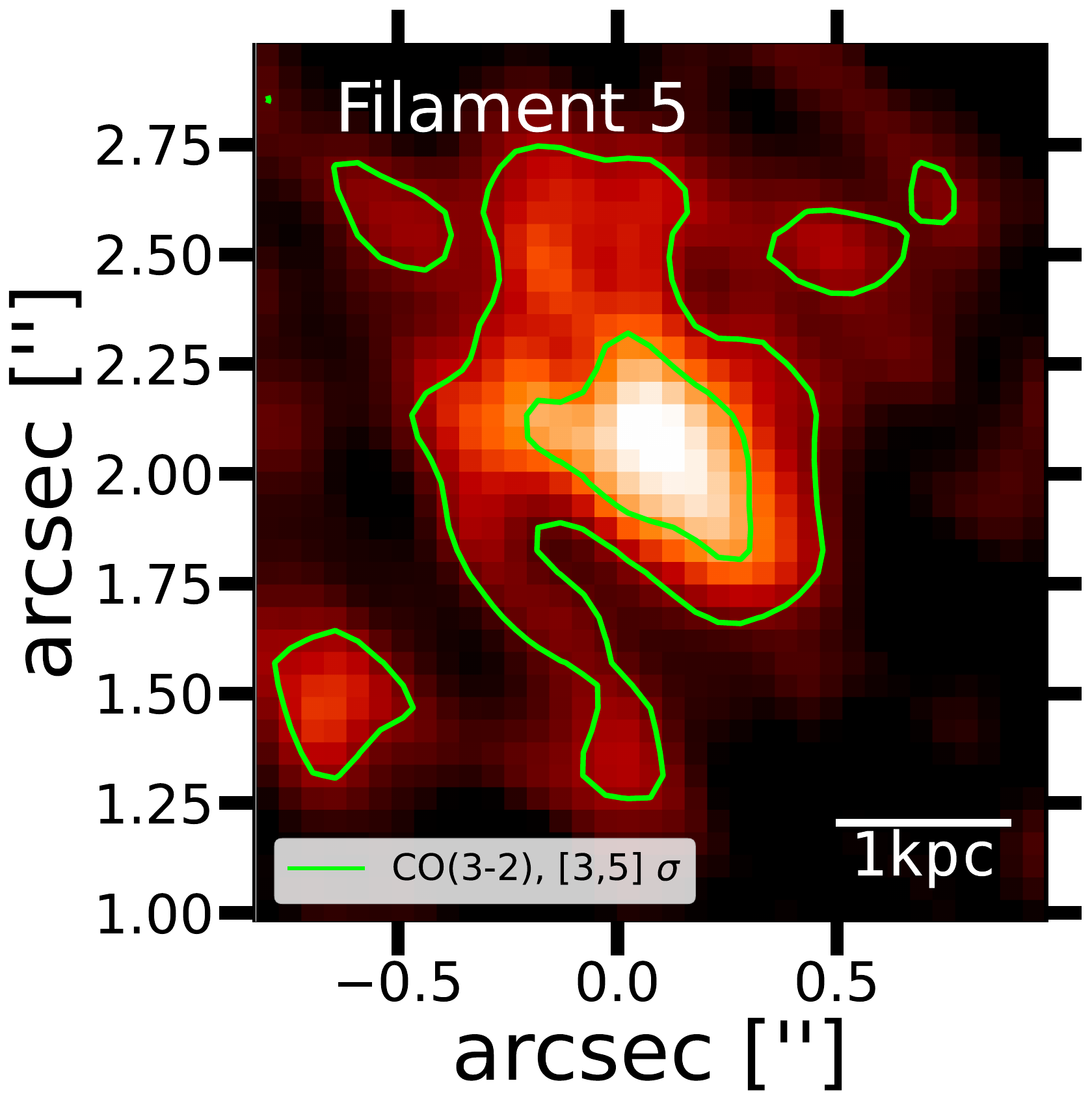} & 
\includegraphics[width=.22\textwidth,height=3.5cm]{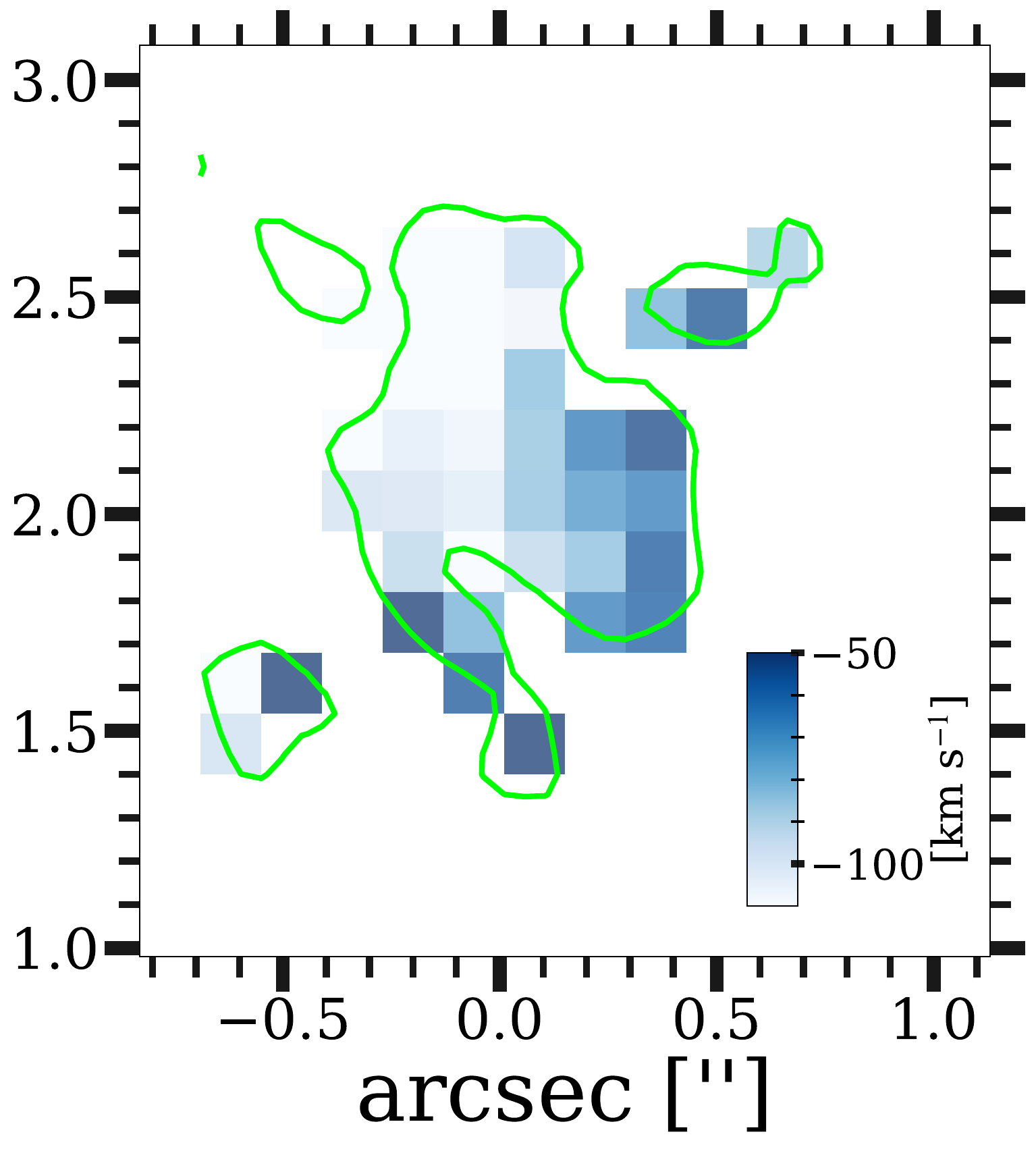} & 
 \includegraphics[width=.22\textwidth,height=3.5cm]{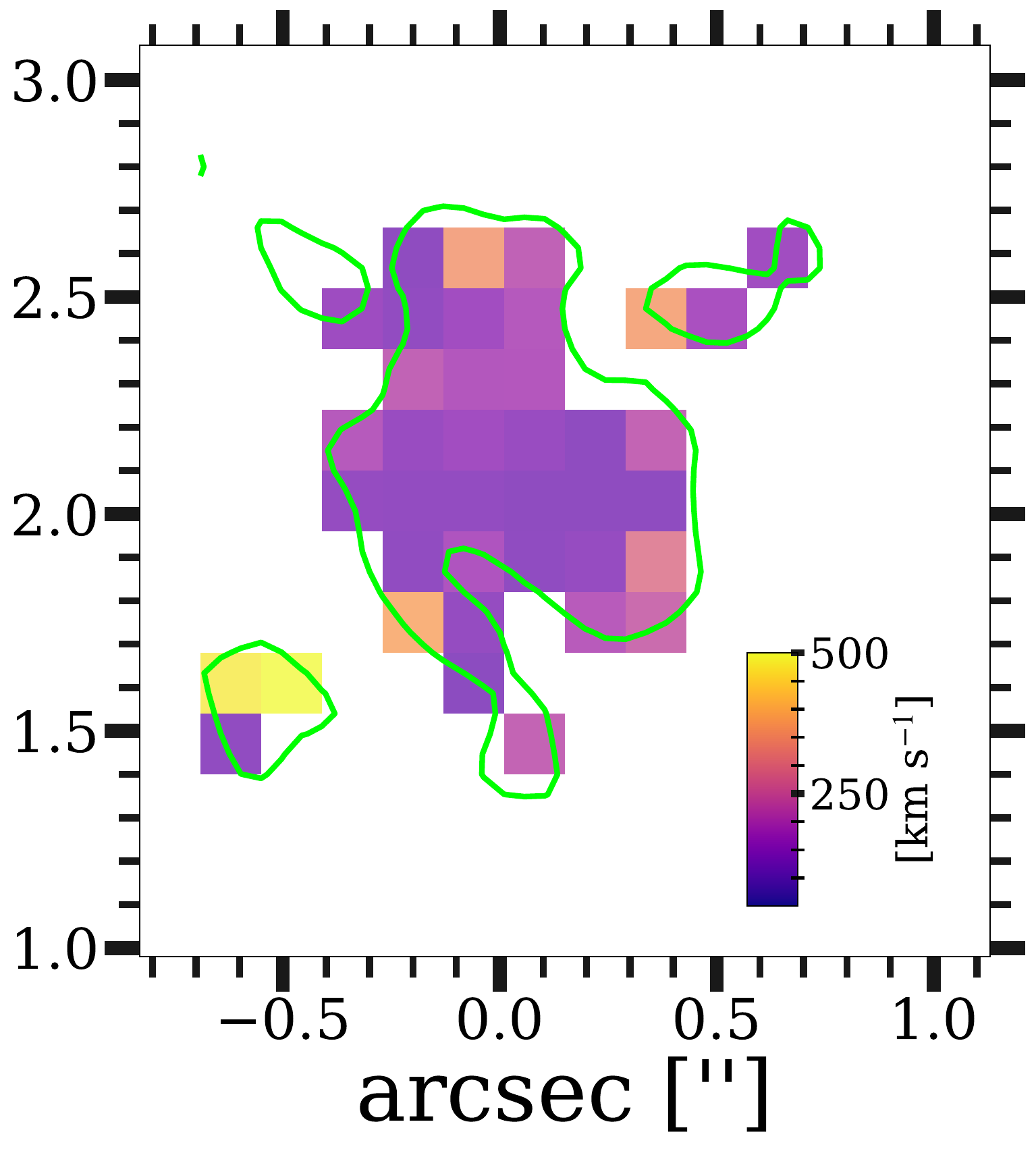} &
 \includegraphics[width=.3\textwidth,height=3cm]{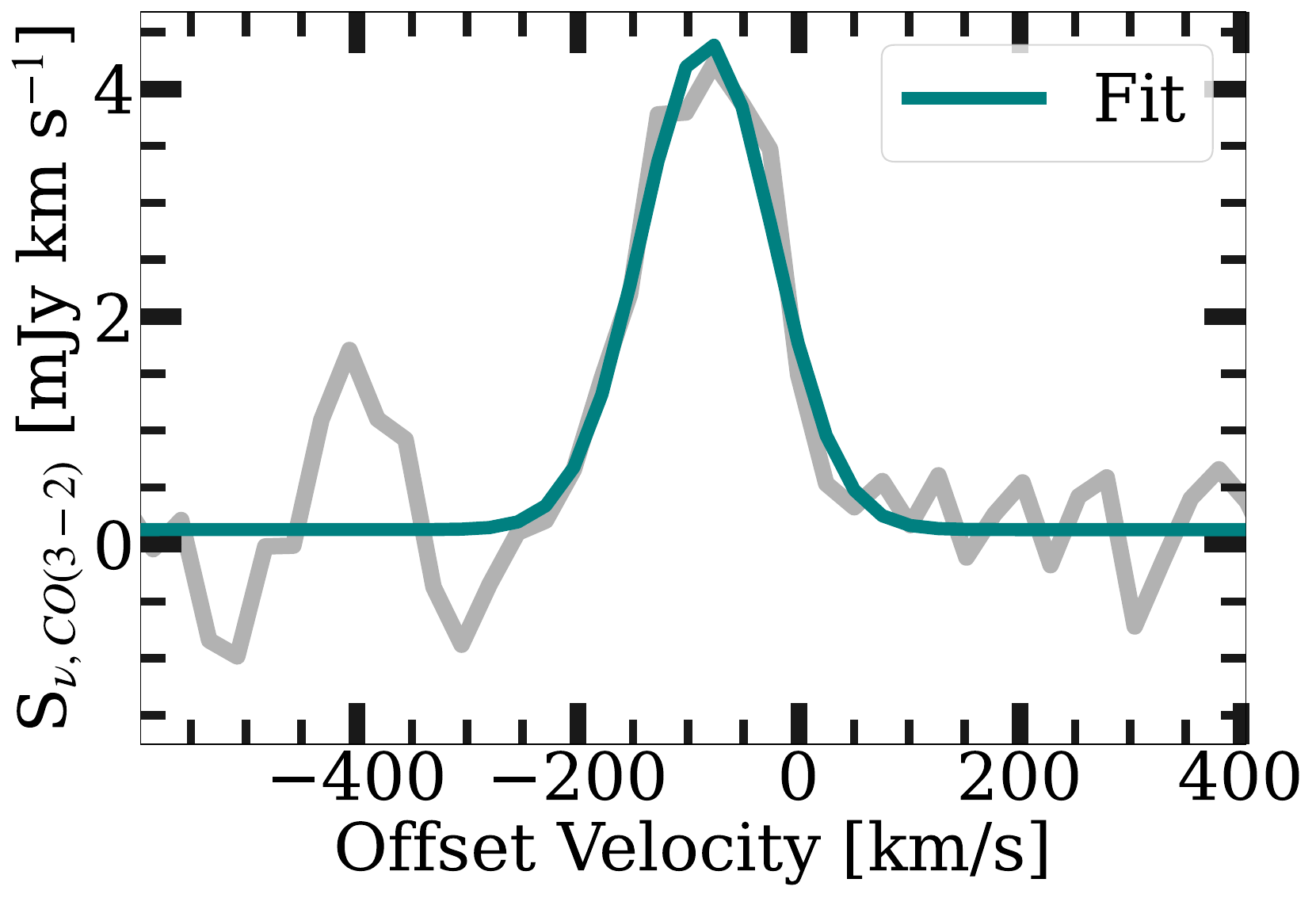} \\
\end{tabular}
\caption{Same as Figure \ref{appfig2: filamentsZoom1} but for J1000+1242 with its 5 identified filamentary molecular gas structures.}
\label{appfig3: filamentsZoom2}
\end{figure*}

The velocity and velocity width maps obtained following the methods outlined in Section \ref{sec: filProps}, are shown in Figure \ref{appfig2: filamentsZoom1} and Figure \ref{appfig3: filamentsZoom2} for each of the filaments of J1000+1242 and J1010+1413, respectively. From left to right, each of the panels in the figure represents (a) CO(3--2) emission narrow band image; (b) a map of the median velocity values in a group of (3$\times$3) spaxels; (c) a map of the velocity width (W$_{80}$); and (d) emission-line profile extracted over the entire filamentary regions. The CO (3--2) emission map in panel (a) shows a zoom-in of each of the filaments identified following the definition in Section \ref{sec: filDefine} and shown in Figure \ref{fig3: filaments_overview}. The green and cyan contours show the emission at 3$\sigma$ and 5$\sigma$ respectively. The two subsequent panels (b) and (c) represent the spatially resolved kinematics for each of the filaments with the 5$\sigma$ contour highlighting the edges and the colour bars at the bottom. The final panel (d) shows the integrated emission line profile over the entire filamentary region enclosed within the 5$\sigma$ contour. 

\end{document}